\pdfoutput=1
\documentclass[11pt,a4paper]{article}
\usepackage{jheppub,color,amsfonts,mathrsfs}
\usepackage[caption=false]{subfig}
\usepackage{alphalph}
\usepackage{floatrow}
\usepackage{longtable}
\usepackage{multirow}
\usepackage[titletoc,title]{appendix}
\usepackage{caption}
\usepackage{floatrow}
\newcommand{\bea}{\begin{eqnarray}}
\newcommand{\eea}{\end{eqnarray}}
\newcommand{\ignore}[1]{}

\newcommand{\cd}{\partial}

\newcommand{\R}{\mathbb{R}}

\newcommand{\I}{\mathbb{I}}
\newcommand{\ra}{\rightarrow}
\newcommand{\ip}[1]{\left\langle #1 \right\rangle}
\newcommand{\eps}{{\boldsymbol{\varepsilon}}}

\renewcommand{\d}{{\mathrm{d}}}

\newcommand{\g}{\mathfrak{g}}

\newcommand{\beq}{\begin{eqnarray}}
\newcommand{\eeq}{\end{eqnarray}}
\newcommand{\non}{\nonumber\\}

\newcommand{\p}{\partial}
\DeclareMathOperator{\U}{U}
\DeclareMathOperator{\SU}{SU}

\DeclareMathOperator{\SO}{SO}

\DeclareMathOperator{\diag}{diag}

\DeclareMathOperator{\tr}{tr}

\DeclareMathOperator{\MeV}{MeV}
\DeclareMathOperator{\GeV}{GeV}
\DeclareMathOperator{\fm}{fm}

\DeclareMathOperator{\grad}{grad}
\newcommand{\bphi}{{\boldsymbol{\phi}}}

\newcommand{\Lag}{\mathcal{L}}

\renewcommand{\i}{\mathrm{i}}

\newcommand{\red}[1]{{\color{red}#1}}

\newtheorem{theorem}{Theorem}

\newtheorem{lemma}[theorem]{Lemma}

\title{Backreacted Coulomb energy in the Skyrme model} 

\author{Sven Bjarke Gudnason$^1$,}
\affiliation{$^1$Institute of Contemporary Mathematics, School of
  Mathematics and Statistics, Henan University, Kaifeng, Henan 475004,
  P.~R.~China}
\emailAdd{gudnason(at)henu.edu.cn}
\author{James Martin Speight$^2$}
\affiliation{$^2$School of Mathematics, University of Leeds, Leeds LS2
  9JT, England}
\emailAdd{j.m.speight(at)leeds.ac.uk}
\abstract{
The Skyrme model is extended with the Maxwell action and a source term
for the gauge field.
We consider the specialized case of vanishing isospin states, such that
only an electric potential is turned on and study the backreaction
onto the Skyrme fields.
In particular, we study Skyrmions with baryon numbers $B=4,8,12,16$
and $40$.
We find, in agreement with physical expectations, that the Coulomb
backreaction is most pronounced for large Skyrmions and find
furthermore that the dynamics of the theory is more sensitive to the
backreaction than the ground states (global minimizers of the energy).
Calibrating the model to Carbon-12, we find excellent agreement of the
masses of the studied Skyrmions -- within $1.86\%$ of experimental
data.
The Coulomb energies are slightly larger than phenomenological fits
suggest, but only by about $3-22\%$, whereas the radii are within
$15\%$ errors, with the largest errors on the smallest baryon number
($B=4$) and the smallest errors on the large baryon numbers. 
}

\begin{document}
\maketitle

\section{Introduction}

Symmetries are fundamentally important in formulating physical
theories of any system.
In the strong sector, it so happens that the two lightest quarks are
close enough to being massless (compared with the energy scale of the
strong interactions) that $\SU(2)$ flavour symmetry is a good
approximate symmetry.
In fact, it is a good approximate symmetry for quarks both with
left-handed chirality and right-handed chirality, separately.
This fact is called chiral symmetry and the group is
$\SU(2)_L\times\SU(2)_R$.
More precisely, the flavour symmetry is classically
$\U(2)_L\times\U(2)_R$ but due to the ABJ anomaly
$\U(1)_{L-R}$ is anomalous, so the anomaly free symmetry is
$\SU(2)_L\times\SU(2)_R\times\U(1)_V$, where the Abelian
vector symmetry $V=L+R$ corresponds to the baryon current.
Chiral symmetry is broken at, perhaps, the same scale that physics of
the strong interactions confines the quarks and all other colour degrees of
freedom (the charge or free indices of quarks and gluons).
Pions composed by the two lightest quarks, up and down, would be
exactly massless if chiral symmetry were an exact symmetry of Nature.
Chiral symmetry breaking, nevertheless, is a small effect, leaving the
pions as the lightest particles of the strong sector.

The pions can be neatly arranged in a matrix form
\beq
U = e^{\i\pi^a\tau^a F_\pi^{-1}}, \qquad
U \to V_L U V_R^\dag, \qquad
V_L\in\SU(2)_L, \qquad
V_R\in\SU(2)_R,
\eeq
with $F_\pi$ called the pion decay constant, $\pi^a$ the three pions,
$\tau^a$ the three Pauli matrices, $V_L$ the left flavour
transformation matrices and $V_R$ the right flavour transformation
matrices.
The left- and right-invariant chiral currents, $L_\mu=U^\dag\p_\mu U$
and $R_\mu=\p_\mu U U^\dag$ are the building blocks of the mesonic
sector of chiral perturbation theory (ChPT) \cite{Scherer:2002tk}.
The pure pion operators in ChPT up to $p^4$ (i.e.~a common notation
for operators including no more than 4 derivatives) consist of 3
terms, with the lowest-order term being simply the kinetic term for
the pions.
A particular combination of the two fourth-order derivative terms, that has
no more than 2 time derivatives, is called the Skyrme term and is the
foundation of the Skyrme model \cite{Skyrme:1961vq,Skyrme:1962vh}.
The Skyrme term is crucial for the simplest model, as it allows for
simple Hamiltonian quantization of zero modes and it prevents the
soliton of the theory -- the Skyrmion -- from collapsing.

It is important to notice that the flavour symmetries until now are
global symmetries. That is, they are not gauged. 
What are gauged symmetries then?
They are related to and responsible for all known fundamental forces
(with the exception of gravity, where it is not known how or if a
quantum mechanical fundamental theory lies behind the ``classical''
gravitational attractive force)\footnote{Some ideas in the literature
propose that gravity is the gauge theory of diffeomorphisms
\cite{Krasnov:2011up}.}. 
The three well-tested fundamental forces in the standard model
correspond to $\U(1)$, $\SU(2)$ and $\SU(3)$ gauge groups, which
represent electromagnetism, the weak nuclear force and strong nuclear
force, respectively.

One could naively think that gauging the $\U(1)_V$ part of the chiral
symmetry would correspond to including electromagnetism in the model.
If the generator is simply the unit matrix, this gives all the
components and hence all the ``quarks'' the same charge, which is
known from phenomenology to be wrong.
Indeed gauging $\U(1)_V$ would correspond to gauging the baryon
symmetry, which has been considered in some
Beyond-Standard-Model (BSM) physics \cite{Pais:1973mi}.
The fact that the quarks have different electric charges can easily
be accommodated by changing the generator of the $\U(1)$ symmetry such
that $U\to U+\i e\alpha[Q,U]$ and $A_\mu\to A_\mu+\p_\mu\alpha$ with
$Q=\diag(\tfrac23,-\tfrac13)$.
The standard Skyrme model with the replacement of $\p_\mu$ by
$D_\mu=\p_\mu-\i e A_\mu[Q,\cdot]$ remains gauge invariant, although
the topological charge density does not and furthermore, it was
pointed out by Callan and Witten that this naive gauging of the Skyrme
model is not consistent with the anomalies of QCD \cite{Callan:1983nx}.
The topological charge is not simply fixed by replacing partial
derivatives with covariant derivatives, but instead a certain total
derivative term must be added, making the entire expression
gauge invariant and topological \cite{Callan:1983nx}.
The Lagrangian, on the other hand, must be amended with some anomalous
terms akin to (gauged) Wess-Zumino-Witten terms that are differential
4-forms \cite{Callan:1983nx}.

The seminal paper \cite{Piette:1997ny} by Piette and Tchrakian studies
the minimal $\U(1)$-gauged Skyrme model, which is essentially the
Skyrme model with partial derivatives replaced by covariant
derivatives and the addition of the Maxwell term in the Lagrangian.
The model is self-consistent and gauge invariant, but does not
reproduce the anomalies of QCD as pointed out by Callan and Witten
\cite{Callan:1983nx}\footnote{$B=1$ gauged Skyrmions with the
Callan-Witten anomaly terms taken into account were studied recently
in ref.~\cite{Navarro-Lerida:2023hbv}. }.
The gauge prescription of coupling electromagnetism (EM) to the Skyrme
model also does more than including just the Coulomb energy in the
model.
In particular, it turns on a magnetic field that cannot be turned off
-- even in the neutral solution (corresponding to the neutron).
The magnetic field explains also a physical effect, namely the
anomalous magnetic moment of the nucleon\footnote{The inclusion of the
nucleon's spin in this model was done in ref.~\cite{Radu:2005jp} and
the inclusion of the pion mass was done in
ref.~\cite{Livramento:2023keg} also for the nucleon ($B=1$).
Skyrmions of higher baryon numbers, i.e.~$B=1$ through 5 were studied
in ref.~\cite{Livramento:2023tmm}.
Gravitating $B=1$ $\U(1)$-gauged Skyrmions were studied in
ref.~\cite{Kirichenkov:2023omy}. }.
On the other hand, the inclusion of the gauge field -- i.e.~turning on
the magnetic field -- reduces the energy of the 1-Skyrmion and it
turns out to exacerbate the binding energy problem of the Skyrmions.
The worsened binding energy problem for the gauged Skyrmions -- a
problem that is already severe in the standard Skyrme model -- has
been considered recently in ref.~\cite{Cork:2021ylu}.
In ref.~\cite{Cork:2021ylu} the energy functional is derived from
Yang-Mills (YM) calorons, which are YM instantons on
$S^1\times\mathbb{R}^3$, giving rise to two extra terms that are
essentially the inner product between the YM field strength as well as
its Skyrme field conjugated counterpart and the curvature term
$L_\mu L_\nu\d x^\mu\wedge\d x^\nu$.
It is not clear to us, that this is not related to the
Wess-Zumino-Witten term of ref.~\cite{Callan:1983nx} by a total
derivative term.
Nevertheless, in ref.~\cite{Cork:2021ylu} the couplings and fields are
numerically optimized to obtain a much lower energy compared to its
lower bound, than was found in the model of
ref.~\cite{Piette:1997ny}.
A detailed analysis is needed to conclude whether the model of
ref.~\cite{Cork:2021ylu} reproduces the QCD anomalies or not, which
could be the case only if there exists a boundary term making this
model equal to that of Callan and Witten.
A geometric formulation of gauged Skyrmion was later put forward in
ref.~\cite{Cork:2023pft}, giving rise to BPS equations whose solutions
have vanishing classical binding energies.
The incorporation of the pion mass as well as solving the quantum
binding energy problem \cite{Gudnason:2023jpq} are still issues that
need to be tackled in this model of gauged Skyrmions.

$\U(1)$ gauged Skyrmions have been studied further in the
literature\footnote{There is a volume of literature on $\SO(3)$-gauged
Skyrmions and $\U(1)$-gauged baby-Skyrmions (in $2+1$ dimensions),
which we will not discuss here.}.
Refs.~\cite{Ohtani:2004aw,Ohtani:2005xv} like
ref.~\cite{Piette:1997ny} computed the anomalous magnetic moment of
the nucleon, but including the Wess-Zumino term that contains the
baryon current, hence providing a more physical source for EM.
The computation is done for a single $B=1$ Skyrmion with spherical
symmetry, under the assumption that the deformation of the nucleon
will be negligible -- an assumption not supported by the recent
results of ref.~\cite{Cork:2021ylu}.
Refs.~\cite{Ohtani:2004aw,Ohtani:2005xv} use the WZ term in the
5-dimensional formulation and it is hence not straightforward to
compare with the proposed terms of Callan and Witten
\cite{Callan:1983nx}, although we expect them to be similar, if not
identical.
The same model has been utilized in ref.~\cite{He:2016oqk} to compute
the masses and magnetic moments of the neutron and proton under the
influence of a strong external magnetic field.
The computation is performed with a ``spherically symmetric'' Ansatz
$F(r)$, but in a coordinate system of an ellipsoid, which according to
ref.~\cite{Cork:2021ylu} will not suffice for finding true minimizers.
Analytic solutions to the massless $\U(1)$-gauged Skyrme equations on a
flat box are given in
refs.~\cite{Aviles:2017hro,Canfora:2018clt,Canfora:2019asc},
although whether such integrable solutions can be made to satisfy
physical boundary conditions of a box or whether the solutions
represent the absolute minimum of the energy functional are issues
that should be considered with care.

The Skyrme model also comes out as the low-energy effective theory of
the holographic Sakai-Sugimoto model \cite{Sakai:2004cn} (for a
review, see ref.~\cite{Rebhan:2014rxa}), more
specifically if all massive vector bosons are decoupled and the
effective action is integrated over the holographic direction. This
corresponds to taking the holonomy of the instanton as envisioned by
Atiyah and Manton \cite{Atiyah:1989dq} and the instanton in the 5-dimensional
low-energy effective action is also known as the Sakai-Sugimoto
soliton \cite{Bolognesi:2013nja}.
So one may ask the question: how to take electromagnetism into account
in the Sakai-Sugimoto model?
To answer this, we need a few more details on the construction.
The model is invented to describe the strong interactions, where
chiral symmetry and its breaking is geometrically encoded in string
theory.
That is, the left-handed and right-handed flavours of quarks each have
their own D8-brane (i.e.~8 spatial dimensions in its world volume) and
these two branes intersect with a D4-brane describing colour degrees of
freedom, that is, a string from the D4-brane to the left D8-brane
represents a left-handed quark and so on.
One dimension of the D4-brane is compactified in order to break the
would-be supersymmetry of the superstring theory and the 5 extra
dimensions of the D8-branes are assumed to have an $\SO(5)$ symmetry.
Considering the large-$N_c$ limit, the D8-branes can be viewed as
probe branes in the background of the heavy stack of D4-branes,
described by the Witten background \cite{Witten:1998qj}.
The resulting low-energy effective theory is now a 5-dimensional
Yang-Mills action for the flavour gauge fields coupled to a
5-dimensional Chern-Simons term, which has the prefactor of the flux
of the 5-sphere -- the number of colours, $N_c$.
External electromagnetic fields have been considered in such a
framework and they are gauge fields living on the D8 flavour branes
\cite{Johnson:2008vna,Bergman:2008sg}. 
It would be very interesting to include dynamics of the electromagnetic
fields and reduce the model to the low-energy effective action of the
Skyrme model and see whether it would coincide with the gauged Skyrme
model of Callan and Witten, especially including the specific
gauged Wess-Zumino-Witten-like term.

As evident from the above discussion, the gauging approach of
including electromagnetism into the Skyrme model, requires not only a
consistent minimal gauging of the Lagrangian, but also the inclusion
of a Wess-Zumino-Witten term, that will reproduce the
Gell-Mann-Nishijima (GMN) formula for charge, baryon number and isospin.
The gauging procedure further turns on a mandatory magnetic flux
\cite{Piette:1997ny} and strongly deforms the known Skyrmion solutions
\cite{Cork:2021ylu}, not to mention the large number of extra
couplings between the gauge field and the Skyrme fields.
For this reason, we consider here the absolute minimal coupling of the
Skyrme model with the Maxwell gauge field with a source dictated by
the phenomenological GMN formula.
The backreaction to the Skyrme field in this formulation thus happens
through the currents of the Skyrme field interacting with the $\U(1)$
gauge field. 
Studying Skyrmions of relatively large baryon number and their
deformations due to the backreaction of the Maxwell gauge field is the
main purpose of this paper.
It is worth noting that this minimal model can be obtained from the
Callan-Witten model simply by truncating the Lagrangian to order
$\mathcal{O}(e^1)$ (linear order in the electromagnetic coupling).
This implies that we neglect the $\mathcal{O}(e\p A)$ terms in the
source for the gauge fields (i.e.~$\mathcal{O}(e^2)$ in the
Lagrangian) as well as their self interactions,
which due to the smallness of the electromagnetic gauge coupling is
physically a quite good approximation.

Studies more similar to our simplistic approach to the problem of
Coulomb energy are given in refs.~\cite{Bonenfant:2012kt,Adam:2013tda,Adam:2013wya,Ma:2019fvk}.
In these cases, the Coulomb energy is computed from the Skyrmion via
its baryon charge density and its isospin current.
Then the classical technique of expanding the computed charge density
in spherical harmonics and then calculating the Coulomb energy as a
sum of multipole moments \cite{10.1119/1.1969367}, is carried out.
In all these papers, no backreaction from the gauge field onto the
Skyrme field is taken into account.
For small baryon numbers ($B$ of order one), this makes physical
sense as the Coulomb energy is small compared to the total mass and
hence the backreaction is expected to be small too.
This may not be the case when the $\U(1)$-gauged approach is
considered though \cite{Cork:2021ylu}.

As evident from the GMN formula, the situation simplifies for two
reasons in the isospin-0 cases: The isospin current is not needed and
the baryon current becomes simply the baryon density (topological
density) of the Skyrmions with all the spatial components vanishing.
Two complications arise once a nonvanishing isospin is turned on. The
isospin current needs to be normalized, turning the PDEs (partial
differential equations) into integro-differential PDEs. Secondly,
time-dependence of the Skyrmion turns on nonvanishing spatial
components in the baryon current ($B^i$), which in turn require
quantization to be implemented in the equations that need be
solved.
This thus also turns on a magnetic flux, just as was found in the
gauged Skyrme models.
We will thus limit ourselves to the isospin-0 case in this paper, but
we will take the full backreaction of the Coulomb force into account.
We will calibrate the model to physically reasonably chosen
observables: the fine-structure constant, the pion mass, the mass and
radius of the ${}^{12}$C nucleus, treating $F_\pi$ and the Skyrme term
coupling as free parameters.
Finally, since we have fitted the model to physical observables, we
expect the backreaction of the Coulomb force to become important near
the largest baryon numbers that are stable.
Due to our simplification of the problem of treating only the
isospin-0 cases, we study Skyrmions with baryon numbers up to $B=40$,
as Calcium-40 is the largest stable $B=4n$ ($n\in\mathbb{N}$)
nucleus.
A large number of Skyrmion solutions have been found in
ref.~\cite{Gudnason:2022jkn}, so we will use the $B=4,8,12,16$
solutions as initial conditions for the computations of this paper.
For the $B=40$ Skyrmions, we adopt the same strategy as in
ref.~\cite{Gudnason:2022jkn} i.e.~we generate a ``large'' number of
initial conditions that are made of random constellations of $B=1$
Skyrmions with random orientations in a product Ansatz.

\section{The Skyrme model with Coulomb energy}\label{sec:model}

\subsection{The field theory}\label{sec:ft}

The Lagrangian on base manifold $M=\mathbb{R}^3$ equipped with
Minkowski metric $\eta$, consisting of the Maxwell term, the Dirichlet
(kinetic) term, the Skyrme term, the pion mass term and the minimal
coupling of the gauge field to the electric charge density, reads
\begin{align}
  \Lag &=
  -\frac{1}{4}F_{\mu\nu}F^{\mu\nu}
  +\frac{F_\pi^2}{16}\tr(R_\mu R^\mu)
  +\frac{1}{32g^2}\tr\left([R_\mu,R_\nu][R^\mu,R^\nu]\right)
  +\frac{F_\pi^2m_\pi^2}{8}\tr(U - \mathbf{1}_2)\non
  &\phantom{=\ }
  -e A_\mu J^\mu,
\label{eq:Lfull}
\end{align}
with the electric, baryon, isospin and vectorial currents (all divergenceless)
\begin{align}
  J^\mu &= \frac12 B^\mu + I^\mu\label{eq:Jmu}\\
  B^\mu &= -\frac{1}{24\pi^2}\epsilon^{\mu\nu\rho\sigma}\tr(R_\nu R_\rho R_\sigma),\label{eq:Bmu}\\
  I^\mu &= \frac{(Z-N)J_V^{\mu 3}}{2\int_M J_V^{03}\d^3x},\\
J_V^{\mu a} &= \frac{\i F_\pi^2}{16}\tr\left[(R^\mu-L^\mu)\tau^a\right]
+\frac{\i}{16g^2}\tr\left[\left([R_\nu,[R^\mu,R^\nu]]-[L_\nu,[L^\mu,L^\nu]]\right)\tau^a\right],\label{eq:JVmu}
\end{align}
the Maxwell field strength
$F_{\mu\nu} = \p_\mu A_\nu - \p_\nu A_\mu$,
the right-invariant and left-invariant chiral currents
\beq
R_\mu = \p_\mu U U^\dag, \qquad
L_\mu = U^\dag \p_\mu U,
\eeq
$F_\pi$ is the pion decay constant, $g$ is the Skyrme coupling
constant, $m_\pi$ is the pion mass, $e$ is the electric charge
related to the fine structure constant by
$\alpha=\frac{e^2}{4\pi}$ in Heaviside-Lorentz conventions, $Z$ is the
number of protons and $N$ is the number of neutrons, and the chiral
Lagrangian or Skyrme field $U$ is related to the pions via
\beq
U = \mathbf{1}_2\sigma + \i\tau^a\pi^a, \qquad
a = 1,2,3,
\eeq
where $\tau^a$ are the standard Pauli spin matrices.
The total (integral) electric charge is given by the
Gell-Mann-Nishijima formula
\beq
Q = \frac12\int_M(B^0 + 2I^0)\;\d^3x,
\eeq
which is the time component of the electric charge current that is
coupled to the electromagnetic potential $A_\mu$. 
Finally, we set the speed of light $c=1$ and the reduced Planck constant
$\hbar=1$, and use the mostly-positive metric signature.

Although the minimal coupling of the gauge field $A_\mu$ to the
electric charge current looks gauge variant, it is indeed gauge
invariant due to current conservation of the baryon charge current and
the isospin charge current
\beq
A_\mu J^\mu 
\to (A_\mu - \p_\mu\lambda)J^\mu
= A_\mu J^\mu - \p_\mu(\lambda J^\mu),
\eeq
up to a total derivative, because
\beq
\p_\mu J^\mu = 0, \qquad
J^\mu := \frac12B^\mu + I^\mu,
\eeq
is conserved.

The baryon number is the spatial integral of the baryon charge
density, which is also the topological degree of the Skyrme field $U$:
\beq
B = \int_{M} B^0\;\d^3x.
\eeq
The number of protons $Z$ and neutrons $N$ in a baryon are related by 
\beq
B = Z + N.
\eeq
On the other hand, the isospin $I$ of a nucleus is given by
\beq
2I = Z - N,
\eeq
with the isospin charge
\beq
I = \int_M I^0\;\d^3x.
\eeq

The Maxwell equations read
\beq
\p_\mu F^{\mu\nu} - \frac{e}{2}\left(B^\nu + 2I^\nu\right) = 0,
\eeq
and it is well known that for static electric charges, the magnetic gauge
potential decouples, or in other words: for $B^i=I^{i}=0$ we can
set $A_i:=0$.
For a static Skyrme field ($\p_0U=0$) we have that $B^i=0$, but
$J_V^{i3}\neq 0$.
Therefore, the situation drastically simplifies if $Z=N$ yielding
$I^i=0$: that is, the isospin zero case.
This case is furthermore simplified, because we do not have to deal
with the quantization of the isospin zeromode.

We will focus on the isospin-zero case in the remainder of this paper.
Since we have decoupled the magnetic gauge potential ($A_i$) and work with
static Skyrme fields ($\p_0U=0$), we can now simplify the field theory
model to 
\begin{align}
  \Lag &=
  \frac12(\p_i A_0)^2
  +\frac{F_\pi^2}{16}\tr(R_i^2)
  +\frac{1}{32g^2}\tr\left([R_i,R_j]^2\right)
  +\frac{F_\pi^2m_\pi^2}{8}\tr(U - \mathbf{1}_2)
  -\frac{e}{2}A_0 B^0.
\label{eq:L_I=0}
\end{align}
Since the model is static\footnote{The model is static if we treat the
electric potential as a scalar field. If we treat it instead as the
time-component of a vector field, the Legendre transform will modify
the Hamiltonian, but only by a total derivative.},
the Hamiltonian (energy) is simply minus the Lagrangian
\begin{align}
  E &= \int_{M}
  \bigg[
    -\frac12(\p_i A_0)^2
    -\frac{F_\pi^2}{16}\tr(R_i^2)
    -\frac{1}{32g^2}\tr\left([R_i,R_j]^2\right)
    +\frac{F_\pi^2m_\pi^2}{8}\tr(\mathbf{1}_2-U) \non
    &\phantom{=\int_{\mathbb{R}^3}\bigg[\ }
    +\frac{e}{2}A_0 B^0
  \bigg]\;\d^3x.
\end{align}
It will now be convenient to switch to Skyrme units, by rescaling
lengths and energies by $x^i\to\lambda x^i$ and $E\to\mu E$,
respectively, for which we get the Skyrme units with energies and
lengths measured in units of
\beq
\mu = \frac{F_\pi}{4g},\qquad
\lambda = \frac{2}{g F_\pi},
\label{eq:units}
\eeq
and we are left with the dimensionless energy functional
\begin{align}
  E &= \int_{M}\!
  \bigg[
    -\frac{\kappa}{2}(\p_i V)^2
    -\frac12\tr(R_i^2)
    -\frac{1}{16}\tr\left([R_i,R_j]^2\right)
    +m^2\tr(\mathbf{1}_2 - U)
    +\kappa V B^0
  \bigg]\;\d^3x,
\end{align}
where we have defined
\beq
m \equiv \frac{2m_\pi}{g F_\pi}, \qquad
V \equiv \frac{4}{e g F_\pi} A_0, \qquad
\kappa\equiv \frac{e^2g^2}{2}.
\label{eq:m_V_kappa}
\eeq
The energy functional and hence the (static) theory depends only on two
parameters: $m$ and $\kappa$ (after fixing length and energy units). 

For numerical calculations, it will be more convenient to use a
4-vector field $\bphi$
\beq
U = \mathbf{1}_2 \phi_0 + \i\tau^a\phi_a, \qquad
a=1,2,3,
\eeq
instead of the SU(2) matrix-valued field $U$.
In terms of the $\bphi=\{\phi_0,\phi_1,\phi_2,\phi_3\}$ field, the
energy functional reads 
\begin{align}
  E &= \int_{M}
  \bigg[
    -\frac{\kappa}{2}(\p_i V)^2
    +\p_i\bphi\cdot\p_i\bphi
    +\frac12(\p_i\bphi\cdot\p_i\bphi)^2
    -\frac12(\p_i\bphi\cdot\p_j\bphi)^2
    +2m^2(1 - \phi_0)\non
    &\phantom{=\int_{\mathbb{R}^3}\bigg[\ }
    +\kappa V B^0
  \bigg]\;\d^3x,\label{eq:E_phi}
\end{align}
the baryon charge density now reads
\beq
B^0 = \frac{1}{12\pi^2}\epsilon_{i j k}\epsilon_{a b c d}
\p_i\phi_a\p_j\phi_b\p_k\phi_c\phi_d,
\eeq
where we adopt the conventions $\epsilon_{123}=\epsilon_{0123}=+1$ and
the vector indices $a,b,c,d=0,1,2,3$. 

The equations of motion read
\begin{align}
  \p_i^2\phi_a
  -(\bphi\cdot\p_i^2\bphi)\phi_a
  +(\p_j\bphi)^2\p_i^2\phi_a
  -(\p_j\bphi)^2(\bphi\cdot\p_i^2\bphi)\phi_a
  +(\p_i\p_j\bphi\cdot\p_j\bphi)\p_i\phi_a\non
  \mathop-(\p_i^2\bphi\cdot\p_j\bphi)\p_j\phi_a
  -(\p_i\bphi\cdot\p_j\bphi)\p_i\p_j\phi_a
  +(\p_i\bphi\cdot\p_j\bphi)(\bphi\cdot\p_i\p_j\bphi)\phi_a\non
  \mathop+\frac{\kappa}{8\pi^2}\epsilon_{a b c d}\epsilon_{i j k}
  \p_i V\p_j\phi_b\p_k\phi_c\phi_d
  +m^2\left(\delta_{a0} - \phi_0\phi_a\right) = 0, \label{eq:phi}\\
  \Delta V = -\p_i^2V = B^0.\label{eq:V}
\end{align}
Notice that standard variational approaches to minimizing the energy
will fail due to the ``wrong sign'' of the kinetic energy for $V$.
For this reason, we have to use a constrained variational method that
we developed in ref.~\cite{Gudnason:2020arj}.

It will prove convenient to rewrite the Coulomb part of the energy
\begin{align}
  E_C &= \int_{M}
  \bigg[
    -\frac{\kappa}{2}(\p_i V)^2
    +\kappa V B^0
    \bigg]\;\d^3x\non
  &= \frac{\kappa}{2}\int_{M}
  V(\p_i^2 V + 2B^0)\;\d^3x\non
  &= \frac{\kappa}{2}\int_{M} (\p_i V)^2\;\d^3x\non
  &= \frac{\kappa}{2}\int_{M}
  VB^0\;\d^3x,\label{eq:EC}
\end{align}
where we have integrated the kinetic term for $V$ by parts and used
the equation of motion. 
The last line in the above equation is numerically easier to evaluate,
because for $m\neq 0$, the $\bphi$ fields tend to $(1,0,0,0)$
exponentially, whereas $V$ tends to zero polynomially at spatial
infinity.
Importantly, the second last line proves that the Coulomb energy is
positive semi-definite, which is not \emph{a priori} clear from the
energy functional \eqref{eq:E_phi}.

\subsection{Geometric approach to the variational problem}\label{sec:geometry}

The Skyrme energy of a smooth map $\phi:M\ra G$,  where $M=\R^3$ is physical space, and $G=\SU(2)$ is target space, is
\beq\label{melwyl}
E_{\rm Skyrme}(\phi)=\int_M\left\{|\phi^*\mu|^2+\frac14|\phi^*\omega|^2+f(\phi)\right\},
\eeq
where $\mu\in\Omega^1(G)\otimes \g$ is the right Maurer-Cartan form on $G$,
$\omega\in\Omega^2(G)\otimes\g$ is the associated 2-form
$\omega(X,Y)=[\mu(X),\mu(Y)]$, $f:G\ra\R$ is a potential function (required to give the pions mass) and we have chosen an $Ad(G)$ invariant inner product on $\g$ (namely $\ip{X,Y}_\g=-\frac12\tr(XY)$). If we interpret $\phi$ as a static Skyrme field, we should ascribe to it the electric charge density 
\beq
\rho=\frac12*\phi^*\Omega,
\eeq
where $\Omega$ is the volume form on $G$ normalized so that $\int_G\Omega=1$, and $*$ is the Hodge isomorphism on $M$. Note that we have implicitly chosen $e$, the charge of a proton, as our unit of electric charge in making this assertion, and that the total electric charge is $\int_M\rho=B/2$, so that our interpretation of $\phi$ is consistent only if $B$ is even. Since the field has electric charge, it induces an electrostatic potential $V:M\ra\R$ which is, by definition, the solution of
\beq\label{gauss}
\Delta V=\frac{\rho}{\eps_0},
\eeq
satisfying the boundary condition $V(\infty)=0$. Here $\Delta=-\cd_i^2$ is the  Laplacian in the geometer's sign convention and $\eps_0$ is the permittivity of free space, a universal physical constant whose numerical value in our coordinate system will depend on our choice of calibration (see later). It follows that the field $\phi$ induces a Coulomb energy
\beq
E_{C}(\phi)=\frac12\int_M V\rho,
\label{eq:ECgeom}
\eeq
and hence that the map $\phi:M\ra G$ corresponding to an isospin $0$ nucleus of even baryon number $B$ (consisting of $B/2$ protons and $B/2$ neutrons) is not (as is usually taken) the degree $B$ field that minimizes $E_{\rm Skyrme}(\phi)$, but rather the degree $B$ field that minimizes
\beq
E(\phi)=E_{\rm Skyrme}(\phi)+E_C(\phi). 
\eeq
The purpose of this section is to derive the first variation formula for this variational problem.

Let $\phi_t$ be a smooth variation of $\phi=\phi_0:M\ra G$ of compact support (meaning that $\phi_t(x)=\phi(x)$ for all $x$ outside some compact subset of $M$), and $\eps=\cd_t\phi_t|_{t=0}\in \Gamma(\phi^{-1}TG)$ its infinitesimal generator (necessarily also of compact support). We seek a formula for the section
$\grad E(\phi)\in \Gamma(\phi^{-1}TG)$ which, by definition, for all such variations satisfies
\beq
\frac{\d\: }{\d t}E(\phi_t)\bigg|_{t=0}=\ip{\eps,\grad E(\phi)}_{L^2}.
\eeq
A map $\phi$ is then a critical point of $E$ if $\grad E(\phi)=0$. 

The calculation of $\grad E(\phi)$ is nontrivial principally because $E_C(\phi)$ is a nonlocal functional of $\phi$. Indeed, the first variation of
$E_{\rm Skyrme}(\phi)$ is well known \cite{Manton:2004tk}:
\beq
\frac{\d\: }{\d t}E_{\rm Skyrme}(\phi_t)\bigg|_{t=0}=
\ip{\mu(\eps),2\delta\phi^*\mu+\frac12\delta\xi_\phi+\mu(\nabla f)}_{L^2},
\eeq
where $\xi_\phi\in\Omega^1(M)\otimes\g$ is the $\g$-valued one-form
\beq
\xi_\phi(X)=\sum_{i}\left[\phi^*\mu(e_i),\phi^*\omega(X,e_i)\right],
\eeq
$\{e_i\}$ is a local orthonormal frame on $M$ and $\delta=-*\d*:\Omega^1(M)\ra\Omega^0(M)$ is the coderivative adjoint to $\d$. Hence
\beq\label{gradsky}
\grad E_{\rm Skyrme}(\phi)=\d R_\phi\left\{\delta\left(2\phi^*\mu+\frac12\xi_\phi\right)\right\}+
 (\nabla f)\circ \phi
 \eeq
where $R_\phi:G\ra G$ is the right multiplication map $g\ra g\phi$. We note in passing that the formula \eqref{gradsky} for $\grad E_{\rm Skyrme}$ is valid on any oriented Riemannian 3-manifold $M$, for any compact semi-simple Lie group $G$.

It remains to compute the gradient of 
\beq
E_C(\phi)=\frac12\int_{\R^3}V\rho
=\frac{\eps_0}{2}\int_{\R^3}V\Delta V.
\eeq
 Here we must be careful: although the variation of $\phi$ (and hence of $\phi^*\Omega$) has compact support, $V$ and its induced variation do not. Boundary terms must be treated with care, therefore, and our analysis is
 restricted to the case of primary interest ($M=\R^3$, $G=SU(2)$). 
 Let us assume that $\phi:M\ra G$ is exponentially spatially localized, in the sense that both $|\phi-\I_2|$ and $|\d\phi|$ are exponentially localized, that is, there exists $C>0$ such that, for all $x\in M$,
\beq
|\phi(x)-\I_2|,\quad |\d\phi_x|\leq Ce^{-|x|/C}.
\eeq
This condition is natural since the underlying model has massive pions, so we expect the Skyrme field to decay like $e^{-m_\pi|x|}$. Since our variation has compact support, it follows immediately that $\phi_t$ is also localized in the same sense, and that the electric charge density of $\phi_t$, $\rho_t=\frac12*\phi_t^*\Omega$ is exponentially spatially localized. A key observation is that the electrostatic potential induced by such a charge distribution is (at least) $1/|x|$ localized:

\begin{lemma}\label{loclem}
Let $C>0$ and $\rho:\R^3\ra\R$ be any smooth function such that $|\rho(x)|\leq Ce^{-|x|/C}$ for all $x$. Let $V:\R^3\ra\R$ be the electrostatic potential induced by $\rho$. Then there exists $K>0$ such that, for all $x\in\R^3$,
$$
|V(x)|\leq \frac{K}{|x|},\qquad\left|\frac{\cd V}{\cd |x|}(x)\right|\leq \frac{K}{|x|^2}.
$$
\end{lemma}

This follows from elementary estimates on the integral formula for $V$
obtained by Green's function methods, which we present in appendix \ref{app:loclem_proof}. 

Let $V_t$ be the potential induced by $\rho_t$ and $\dot{V}=\cd_t\rho_t|_{t=0}$. Denote by $B_R$ the ball in $\R^3$ centred at $0$ of radius $R$. Then
\bea
\frac{\d\: }{\d t}E_C(\phi_t)\bigg|_{t=0}&=&\frac{\eps_0}{2}\int_{\R^3}(\dot V\Delta V+V\Delta\dot{V})\nonumber \\
&=&\lim_{R\ra\infty}\frac{\eps_0}{2}\int_{B_R}(\dot V\Delta V+V\Delta\dot{V})\nonumber \\
&=&\lim_{R\ra\infty}\frac{\eps_0}{2}\left\{2\int_{B_R}V\Delta\dot V
+\int_{\cd B_R}(V*\d\dot V-\dot V*\d V)\right\},
\eea
where we have used Stokes's theorem. 
Now $V$ is induced by $\rho$, which is exponentially localized, and $\dot{V}$ is induced by $\dot\rho$, which has compact support, and hence is also exponentially localized. It follows from Lemma \ref{loclem} that
\beq
\lim_{R\ra\infty}\int_{\cd B_R}(V*\d\dot V-\dot V*\d V)=0,
\eeq
and hence
\beq
\frac{\d\: }{\d t}E_C(\phi_t)\bigg|_{t=0}=\eps_0\int_{\R^3}V\Delta \dot{V}=\int_{\R^3}V\dot\rho.
\eeq
By the Homotopy Lemma, 
\beq
\dot\rho=\frac12\cd_{t}*\phi_t^*\Omega\big|_{t=0}=\frac12*\d(\phi^*\iota_\eps\Omega),
\eeq
and hence
\bea
\frac{\d\: }{\d t}E_C(\phi_t)\bigg|_{t=0}&=&\lim_{R\ra\infty}\frac12\int_{B_R}
V\d\phi^*\iota_\eps\Omega\nonumber\\
&=&\lim_{R\ra\infty}\frac12\left\{\int_{\cd B_R}V\phi^*\iota_\eps\Omega
-\int_{B_R}\d V\wedge\phi^*\iota_\eps\Omega\right\}\nonumber \\
&=&-\frac12\int_{\R^3}\d V\wedge \phi^*\iota_\eps\Omega,\label{chbr}
\eea
since $\eps$ has compact support. 

To extract $\grad E_C$ from eq.~\eqref{chbr} we note that the normalized volume form at $\I_2\in G$ is the totally skew-symmetric map
\beq
\Omega_{\I_2}:\g\times\g\times\g\ra\R,\qquad
\Omega_{\I_2}(X,Y,Z)=-\frac{1}{4\pi^2}\ip{X,[Y,Z]}_\g.
\eeq
But $\Omega$ is right invariant, so
\beq
\Omega_g(X,Y,Z)=-\frac{1}{4\pi^2}\ip{\mu(X),\omega(Y,Z)}_\g.
\eea
Hence
\bea
*\d V\wedge \phi^*\iota_\eps\Omega&=&(\d V\wedge \phi^*\iota_\eps\Omega)(e_1,e_2,e_3) \nonumber \\
&=&e_1[V]\Omega(\eps,\d\phi(e_2),\d\phi(e_3))+\mbox{cyclic perms}
\ignore{e_2[V]\Omega(\eps,\d\phi(e_3),\d\phi(e_1))+
e_3[V]\Omega(\eps,\d\phi(e_1),\d\phi(e_2))}\nonumber \\
&=&\frac{-1}{4\pi^2}\left\{e_1[V]\ip{\mu(\eps),\phi^*\omega(e_2,e_3)}_\g
+\mbox{cyclic perms}
\ignore{+e_2[V]\ip{\mu(\eps),\phi^*\omega(e_3,e_1)}_\g
+e_3[V]\ip{\mu(\eps),\phi^*\omega(e_1,e_2)}_\g}\right\}\nonumber \\
&=&\ip{\mu(\eps),\frac{-1}{4\pi^2}*\d V\wedge\phi^*\omega}_\g.
\eea
It follows that
\bea
\frac{\d\:}{\d t}E_C(\phi_t)\bigg|_{t=0}&=&
\ip{\mu(\eps),\frac1{8\pi^2}*\d V\wedge\phi^*\omega}_{L^2},
\eea
and hence that
\beq
\grad E_C(\phi)=\frac1{8\pi^2} \d R_\phi(*\d V\wedge\phi^*\omega).
\eeq

In conclusion, a smooth exponentially localized map $\phi:M\ra G$ is a critical point of $E=E_{\rm Skyrme}+E_C$ with respect to all smooth variations of compact support if and only if
\beq
\grad E(\phi)=
\d R_\phi\left\{\delta\left(2\phi^*\mu+\frac12\xi_\phi\right)
+\frac1{8\pi^2}*\d V\wedge\phi^*\omega
\right\}
+(\nabla f)\circ \phi=0,
\label{eq:gradE}
\eeq
where $V:M\ra\R$ is the unique solution of eq.~\eqref{gauss}.
Note that this formula for $\grad E(\phi)$ coincides with (minus two
times) the left hand side of \eqref{eq:phi} in the case where
$f(\phi)=m^2\tr(\I_2-\phi)=2m^2(1-\phi_0)$ is the usual pion-mass
potential and up to the correct normalization of $V$.

In order to normalize $V$ correctly, we recall that the
energy functional for $\phi$ \eqref{melwyl} is defined in ``Skyrme
units'', whereas Gauss' law is defined with $\rho$ normalized by the
charge of the proton.
The correct rescaling of $V$ is thus $V\to\frac{e}{2\lambda}V$, with
$e$ being the charge of the proton and $\lambda$ being the length unit.
In order to compute the coupling between the Skyrme field $\phi$ in
Skyrme units and the field $V$, we need to compare the energy units of
the Coulomb energy and the Skyrme energy functional.
This amounts to multiplying the Coulomb energy by the length scale and
dividing by the energy scale.
Combining it all and using that the Coulomb energy contains two $V$'s,
we have
\beq
\kappa = \left(\frac{e}{2\lambda}\right)^2\frac{\lambda}{\mu}
= \frac{e^2}{4\mu\lambda}
= \frac{e^2g^2}{2},
\eeq
where in the last step we have used the definitions of the length and
energy units of eq.~\eqref{eq:units}.
Notice that we have recovered the coupling $\kappa$ of
eq.~\eqref{eq:m_V_kappa}, which should be inserted in front of the
last term in the curly braces of the gradient \eqref{eq:gradE}.

\subsection{Derrick scaling}

It is instructive to consider how the Coulomb energy behaves under isotropic dilation of the Skyrme field \cite{Derrick:1964ww}. That is, for given $\phi:M\ra G$, consider the one-parameter family $\phi_\lambda:M\ra G$, $\phi_\lambda(x)=\phi(\lambda x)$ where $\lambda\in(0,\infty)$. The associated electric charge density is $\rho_\lambda(x)=\lambda^3\rho(\lambda x)$, so the electrostatic potential $V_\lambda$ induced by $\phi_\lambda$ satisfies
\beq
-\lambda^2\sum_{i=1}^3\frac{\cd^2\:}{\cd(\lambda x_i)^2}V_\lambda=\frac{\lambda^3}{\eps_0}\rho(\lambda x),
\eeq
whence we deduce that $V_\lambda(x)=\lambda V(\lambda x)$. Hence, the Coulomb energy of $\phi_\lambda$ is
\bea
E_C(\phi_\lambda)&=&\frac12\int_{\R^3}V_\lambda(x)\rho_\lambda(x)\;\d^3x
=\frac\lambda2\int_{\R^3}V(\lambda x)\rho(\lambda x)\;\d^3(\lambda x)
=\lambda E_C(\phi).
\eea
Defining, as usual, the individual contributions to $E_{\rm Skyrme}$,
\beq
E_2(\phi)=\|\phi^*\mu\|_{L^2}^2,\quad
E_4(\phi)=\frac14\|\phi^*\omega\|_{L^2}^2,\quad
E_0(\phi)=\int_M *f\circ\phi,
\eeq
we note
that $E_C$ has the same scaling behaviour as $E_4$. 

Since any static solution of the model is a critical point of $E(\phi)$ for all smooth variations of $\phi$, including $\phi_\lambda$, we conclude that such a solution must satisfy the virial identity
\bea
\frac{\d\:}{\d\lambda}
E(\phi_\lambda)\bigg|_{\lambda=1}=\frac{\d\:}{\d\lambda}\left\{
\frac1\lambda E_2(\phi)+\lambda(E_4(\phi)+E_C(\phi))+\frac1{\lambda^3}E_0(\phi)\right\}\bigg|_{\lambda=1}&=&0, \nonumber \\
\Rightarrow\quad
-E_2(\phi)+E_4(\phi)+E_C(\phi)-3E_0(\phi)&=&0. \label{virial}
\eea
This provides a useful check on our numerical results. In particular, for fields defined on a bounded box, the variation $\phi_\lambda$ is only well-defined for $\lambda\in[1,\infty)$, and minimality of $E$ implies only that $\d E(\phi_\lambda)/\d\lambda|_{\lambda=1}\geq 0$. Hence, on a bounded domain, the virial identity is replaced by the condition 
\beq 
\frac{1}{E(\phi)}\left\{-E_2(\phi)+E_4(\phi)+E_C(\phi)-3E_0(\phi)\right\}\geq 0,
\eeq
and the value of this quantity measures the pressure exerted by the
box boundary. Of course, one aims to choose the computation domain
large enough that this pressure is small.
Typical values for the
numerical solutions presented below are 0.01-0.08\% for ``light''
Skyrmions, i.e.~with baryon numbers in the range $[4,16]$ and around
0.35\% for the large $B=40$ solutions.
The latter rather large pressure suggests that the computational box
we have used is on the limit of being big enough, but due to the heavy
computational cost of those computations, we have chosen this
compromise.

Since $E_C$ scales similarly to $E_4$, it is natural to wonder whether
Skyrmions can be stabilized by $E_C$ alone, dropping $E_4$
entirely. Simulations in the $B=4$ sector suggest that they can, but
the model with no Skyrme term is unlikely to be phenomenologically
competitive.
Our simulations suggest that the electromagnetic coupling (charge)
should be about two orders of magnitude larger than experimentally
known values, in order to reproduce phenomenologically viable sizes of
the nuclei.
The stabilization of Skyrmions through electromagnetic interactions
alone, with no Skyrme term, was also observed in the gauge-theoretic
context in ref.~\cite{Navarro-Lerida:2023hbv}.

\subsection{Topological energy bound}

The Skyrme part of the energy has the topological energy bound given
in refs.~\cite{Harland:2013rxa,Adam:2013tga,Gudnason:2022jkn}:
\beq
E_{\rm Skyrme} \geq 12\pi^2\left(\sqrt{\alpha} + \frac{128\sqrt{m}(1-\alpha)^{\frac34}\Gamma^2(\tfrac34)}{45\pi^{\frac32}}\right)|B|,
\label{eq:Ebound}
\eeq
with $\alpha\in[0,1]$:
\beq
\alpha = \frac{a^2}{2}\left(\sqrt{1+\frac{4}{a^2}}-1\right),
\eeq
and finally
\beq
a = \frac{225\pi^3}{4096m\Gamma^4(\tfrac34)}.
\eeq
We will shortly calibrate the model and obtain $m\simeq0.650$, for
which $a\simeq1.162$, $\alpha\simeq0.6688$ and
\beq
E_{\rm Skyrme} \geq 1.088\times 12\pi^2|B|.
\eeq
Since the Coulomb energy is positive semi-definite, we can infer that
the total energy has an energy bound $E\geq E_{\rm Skyrme}$.
It would be interesting to know whether this bound can be improved.

\subsection{Numerical algorithm}

The numerical algorithm that will be used in this paper is basically
that developed in ref.~\cite{Gudnason:2020arj}.
That is, we perform an arrested Newton flow for the Skyrme field $\bphi$:
\beq
P_\phi(\ddot\phi) = -\grad E(\phi),
\eeq
where $P_\phi:\mathbb{R}^4\to T_\phi N$ denotes the orthogonal
projection defined by the isometric embedding $N\subset\mathbb{R}^4$,
with $N=S^3$ here, and $\grad E(\phi)$ is given in eq.~\eqref{eq:gradE}.
The ``arrested'' part of the Newton flow consists of monitoring the
total energy $E_\bphi$ at every step and restarting the flow with
vanishing kinetic energy $\dot\bphi=0$, if the energy increases.
The Coulomb potential $V$, on the other hand, cannot be solved with
the same variational methods, due to its ``wrong sign'' in the
action.
For this reason, the numerical algorithm developed in
ref.~\cite{Gudnason:2020arj}, solves the eq.~\eqref{gauss} or
\eqref{eq:V} completely, at each step of the arrested Newton flow,
using the conjugate gradient method.
In this sense, the method is a constrained arrested Newton flow where
the flow only takes place on trajectories where $V$ is a solution to
its governing equation.

An important difference and technical detail, is that the $\omega_0$
field in ref.~\cite{Gudnason:2020arj} was massive and hence enjoyed
exponential spatial localization, whereas $A_0$ and hence $V$
in the present paper is massless and obeys a power-law falloff.
This makes the accurate evaluation of the total energy at each step of
the arrested Newton flow difficult.
The solution is simply to use the Coulomb energy functional written in
the last line of eq.~\eqref{eq:EC} (see also eq.~\eqref{eq:ECgeom}),
as opposed to that written in the second to last line.
Since the baryon charge $B^0$ {\em is} exponentially spatially localized, so is the
integrand $V B^0$.

The boundary condition on $U$:
$\lim_{|x|\to\p{\rm box}}U(x)=\mathbf{1}$ is simple to impose for a
sufficiently large box size.
In practice, we will instead adopt a Neumann condition on the boundary
of the box, since it induces a smaller numerical error in energy and
topological charge computations.
On the other hand, the boundary condition $\lim_{|x|\to\infty}V=0$ is
difficult to impose on a finite box, since the power-law falloff is
too slow for any manageably sized box.
One could attempt to use a complicated Robin boundary condition:
$|x|V+\hat{x}\cdot\nabla V=0$, where the charge needs not be
specified.
The unspecificity of the charge allows for complicated charge
distributions, but these Robin boundary conditions are somewhat tricky
to work with for the conjugate gradients method.
A much more stable boundary condition, is to assume that the box size
is sufficiently large, so that it induces only a small inaccuracy to
impose the Dirichlet boundary condition
$\lim_{|x|\to\p{\rm box}}V(x)=\frac{B}{4\pi|x|}$, where $B$ is the
total baryon number.\footnote{Notice that the total charge is
\emph{not} $B$, but $eB/2$. However, in the rescaled fields $V$ obeys
eq.~\eqref{eq:V}. The integral of the right-hand side is $B$ and hence
the approximation of $V$ with a point-charge $B$ source is $B/4\pi|x|$
in Heaviside-Lorentz units.}
This choice of boundary conditions is very stable, but gives an
inaccurate solution for $V$ near the boundary of the box.
This inaccuracy is negligible for the energy integral of $E_C$, due to
the strong localization of the electric charge distribution $\rho=e B^0/2$.
In this work, we thus work with larger box sizes, i.e.~$151^3$ lattice
sites (sometimes larger) and a lattice distance typically of the order
of $h_x\simeq0.13$, in order to minimize the inaccuracy of the Coulomb
energy. 

To summarize the numerical algorithm:
\begin{enumerate}
\item Perform Newton flow step using Runge-Kutta 4 (RK4).
\item Solve Gauss's equation for $V$.
\item Compute the total energy and compare with the value of the
  previous step: If larger, set all kinetic energy to zero.
\item If the residue of the Skyrme field is smaller than a threshold
  value $\|\grad E(\phi)\|_{L^2}^2<T$ ($T$ is typically chosen as
  $T=10^{-3}$), then stop the algorithm.
\item Continue to the next step (start over).
\end{enumerate}

Our code is implemented in CUDA C and run on a small cluster of NVIDIA
GPUs.

\subsection{Calibration}

Numerically, in Heaviside-Lorentz units, the electric charge is
related to the fine-structure constant \cite{Zyla:2020zbs}
\beq
\alpha = \frac{e^2}{4\pi} \approx \frac{1}{137.036},
\eeq
and thus equal to
\beq
e = 2\sqrt{\pi\alpha} \approx 0.302822.
\eeq
The pion masses are given by \cite{Zyla:2020zbs}
\beq
m_{\pi^\pm} \approx 139.570\ \MeV, \qquad
m_{\pi^0} \approx 134.977\ \MeV.
\eeq
Since we work with unbroken isospin symmetry, we will take the
geometric average
\beq
m_\pi = \sqrt[3]{m_{\pi^\pm}^2m_{\pi^0}} \approx 138.022\ \MeV.
\eeq
In order to calibrate the model, we need two data points of energy and
length dimensions, respectively.
We will consider the following calibration scheme: We fit the mass and
size of the lowest energy $B=12$ Skyrmion to those of the Carbon-12
nucleus.

\begin{figure}[!htp]
  \begin{center}
    \mbox{\subfloat[]{\includegraphics[width=0.33\linewidth]{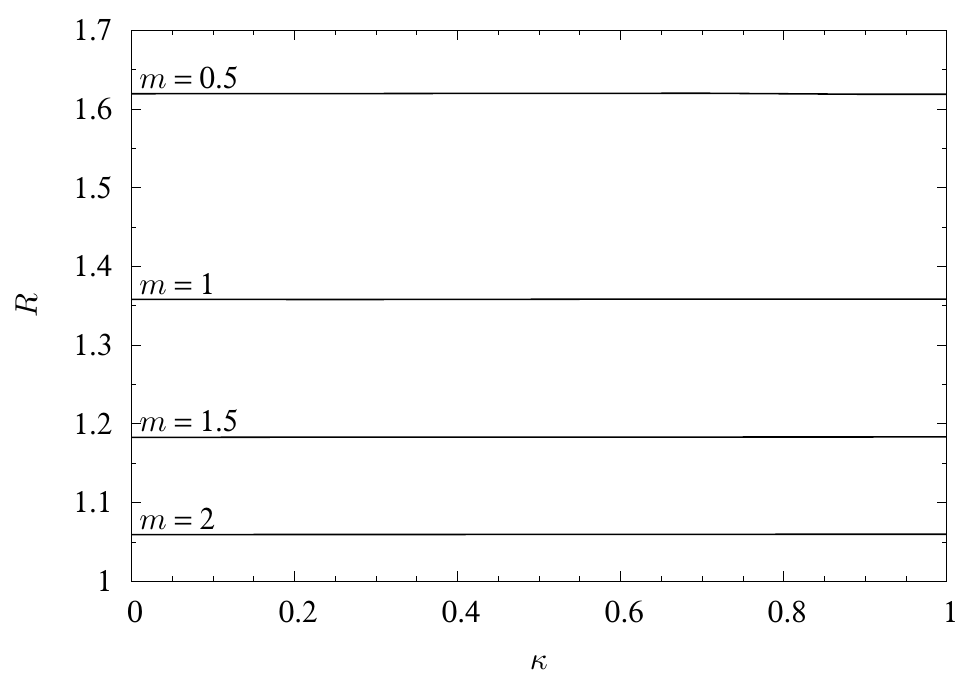}}
      \subfloat[]{\includegraphics[width=0.33\linewidth]{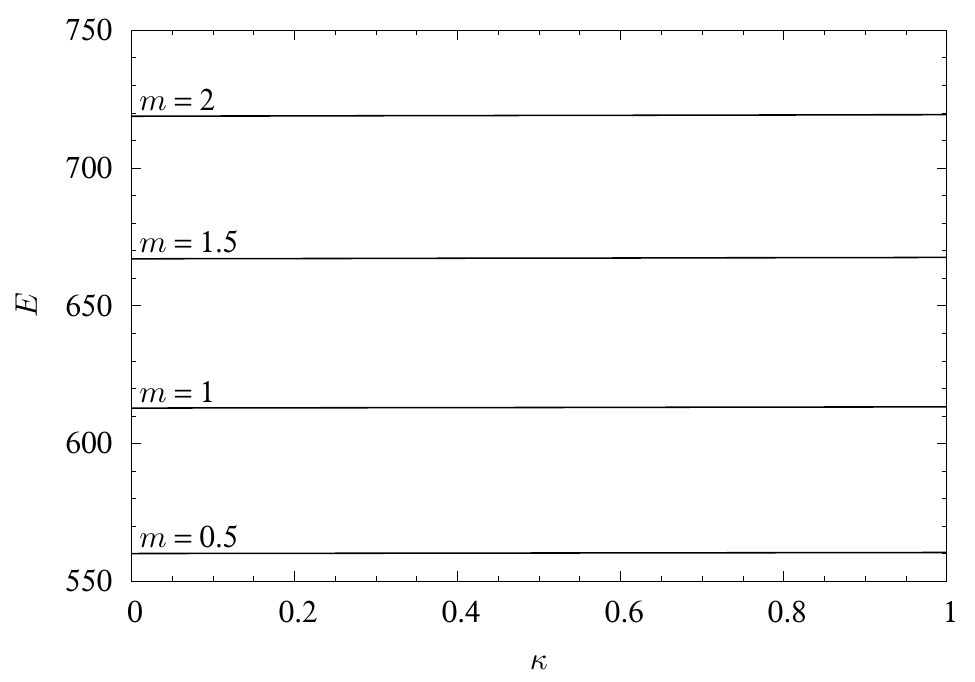}}
      \subfloat[]{\includegraphics[width=0.33\linewidth]{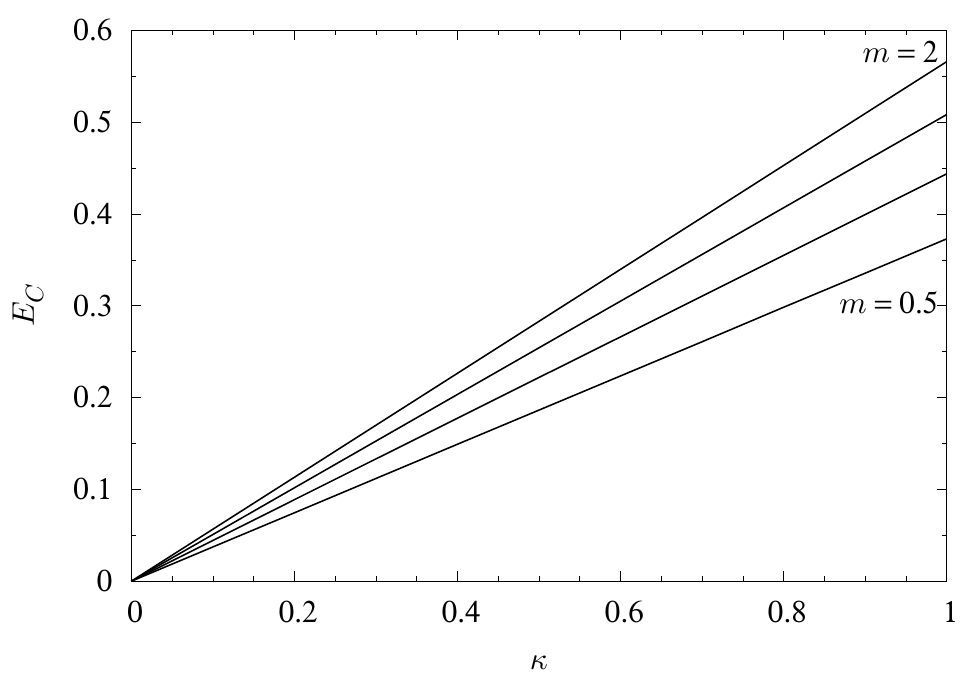}}}
    \caption{Dependence of the (a) radius, (b) total energy and (c) Coulomb
      energies on $\kappa$ in Skyrme units for the $B=4_a$ cube. The
      four curves correspond to $m=0.5, 1, 1.5, 2$, respectively. }
    \label{fig:kappa4}
  \end{center}
\end{figure}

\begin{figure}[!htp]
  \begin{center}
    \mbox{\subfloat[]{\includegraphics[width=0.33\linewidth]{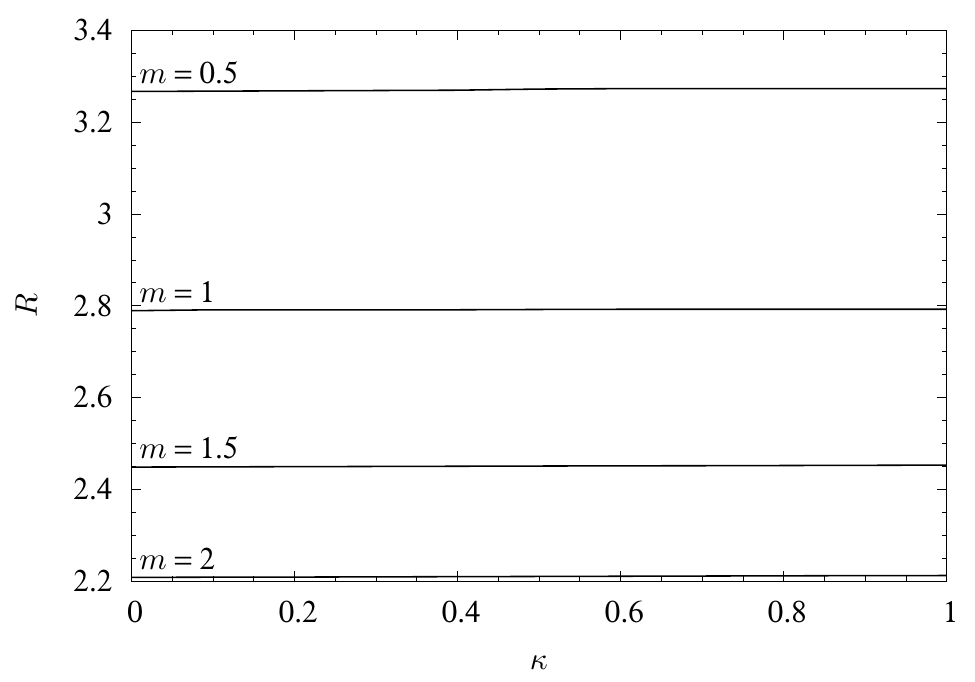}}
      \subfloat[]{\includegraphics[width=0.33\linewidth]{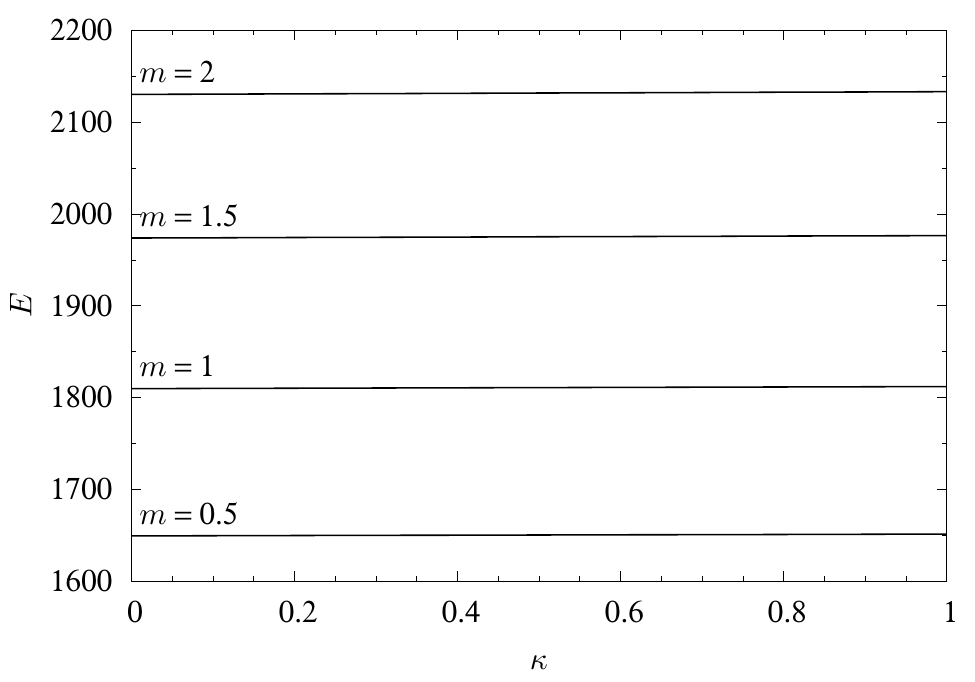}}
      \subfloat[]{\includegraphics[width=0.33\linewidth]{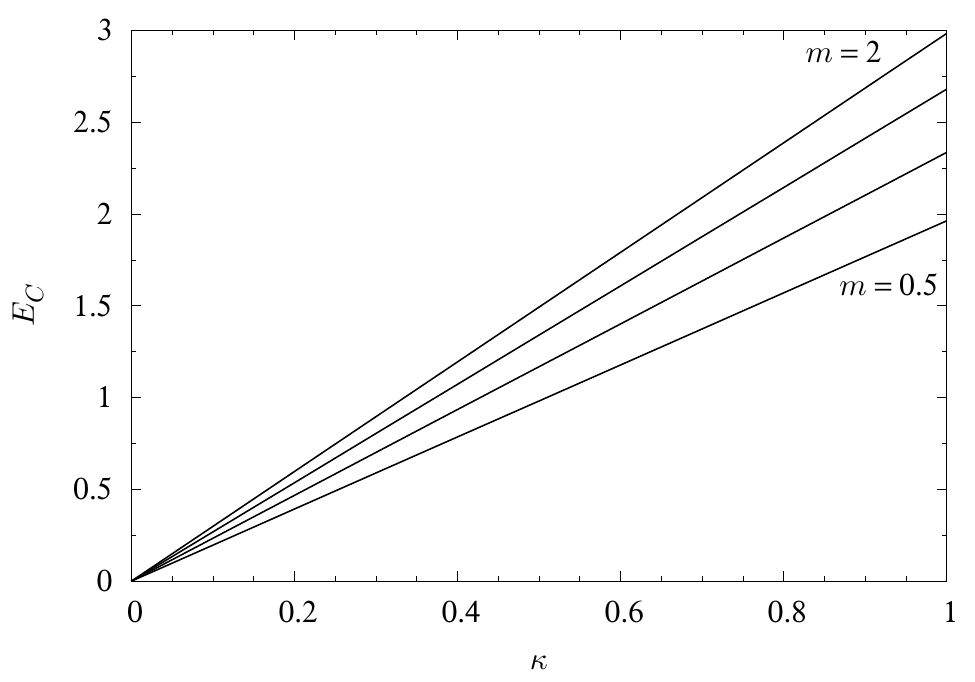}}}
    \caption{Dependence of the (a) radius, (b) total energy and (c) Coulomb energies on $\kappa$ in Skyrme units for the $B=12_a$ chain. The
      four curves correspond to $m=0.5, 1, 1.5, 2$, respectively.}
    \label{fig:kappa12a}
  \end{center}
\end{figure}

In order to see some qualitative dependence on the parameters $m$,
$\kappa$ for the Skyrmion observables, we plot the radius, the total
energy and the Coulomb energy as a function of $\kappa$ for various
$m=0.5,1,1.5,2$ in figs.~\ref{fig:kappa4} and \ref{fig:kappa12a}
for the $B=4_a$ (cube) and $B=12_a$ (chain of cubes) Skyrmions,
respectively.
Clearly the strongest impact on both the radius and the total energy
comes from any change in the pion mass parameter, $m$, whereas the
dependence on $\kappa$ is very mild.

\def\scaletwelf{0.23}
\begin{figure}[!tp]
  \centering
  \mbox{
    \subfloat[]{\includegraphics[scale=\scaletwelf]{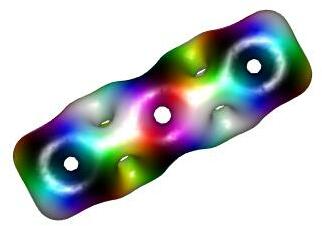}}
    \subfloat[]{\includegraphics[scale=\scaletwelf]{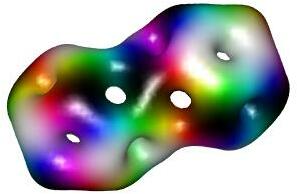}}
    \subfloat[]{\includegraphics[scale=\scaletwelf]{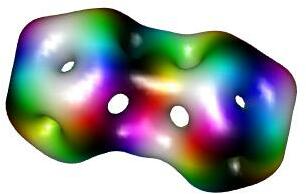}}
    \subfloat[]{\includegraphics[scale=\scaletwelf]{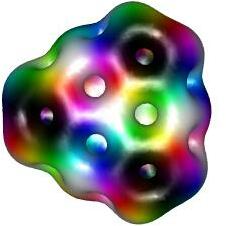}}
    \subfloat[]{\includegraphics[scale=\scaletwelf]{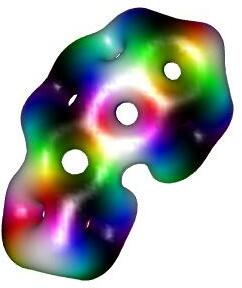}}
    \subfloat[]{\includegraphics[scale=\scaletwelf]{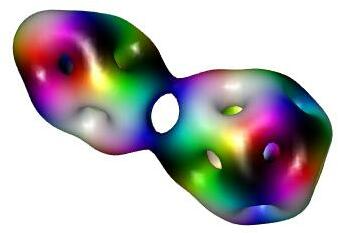}}
  }
  \mbox{
    \subfloat[]{\includegraphics[scale=\scaletwelf]{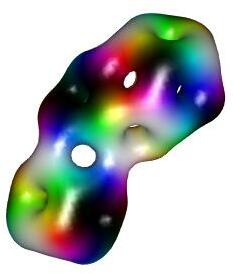}}
    \subfloat[]{\includegraphics[scale=\scaletwelf]{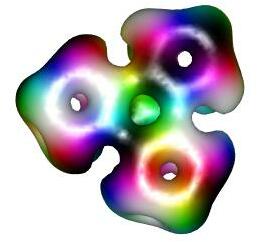}}
    \subfloat[]{\includegraphics[scale=\scaletwelf]{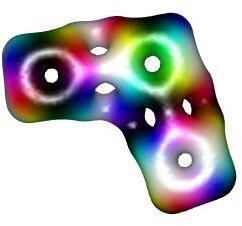}}
    \subfloat[]{\includegraphics[scale=\scaletwelf]{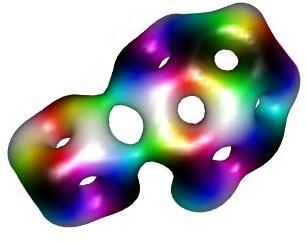}}
    \subfloat[]{\includegraphics[scale=\scaletwelf]{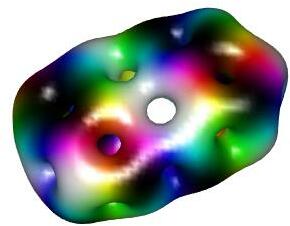}}
    \subfloat[]{\includegraphics[scale=\scaletwelf]{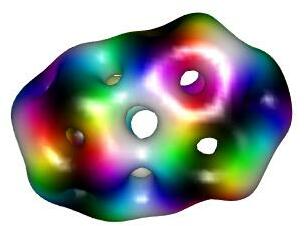}}
  }
  \mbox{
    \subfloat[]{\includegraphics[scale=\scaletwelf]{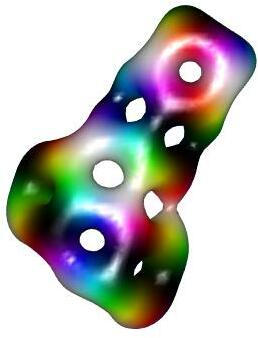}}
    \subfloat[]{\includegraphics[scale=\scaletwelf]{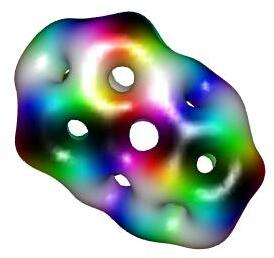}}
    \subfloat[]{\includegraphics[scale=\scaletwelf]{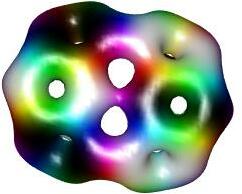}}
    \subfloat[]{\includegraphics[scale=\scaletwelf]{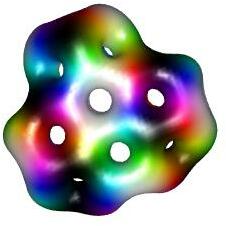}}
    \subfloat[]{\includegraphics[scale=\scaletwelf]{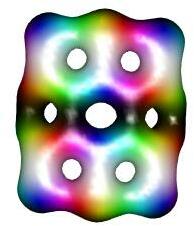}}
    \subfloat[]{\includegraphics[scale=\scaletwelf]{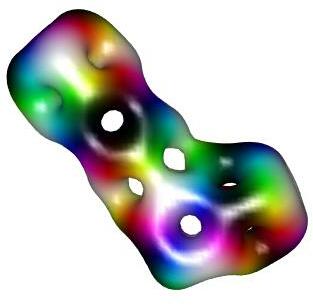}}
  }
  \mbox{
    \subfloat[]{\includegraphics[scale=\scaletwelf]{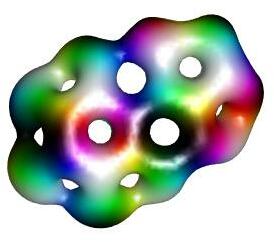}}
    \subfloat[]{\includegraphics[scale=\scaletwelf]{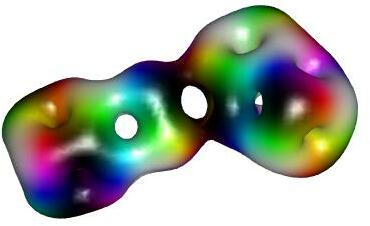}}
    \subfloat[]{\includegraphics[scale=\scaletwelf]{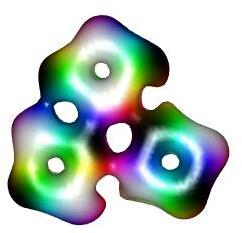}}
    \subfloat[]{\includegraphics[scale=\scaletwelf]{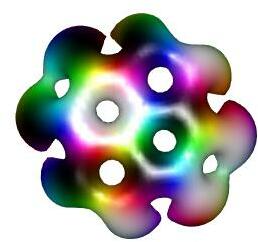}}
    \subfloat[]{\includegraphics[scale=\scaletwelf]{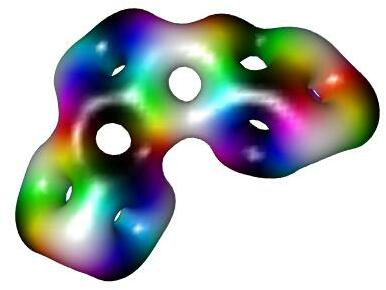}}
    \subfloat[]{\includegraphics[scale=\scaletwelf]{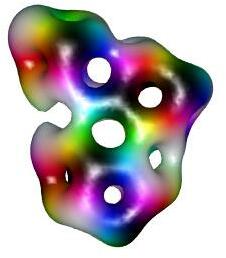}}
  }
  \mbox{
    \subfloat[]{\includegraphics[scale=\scaletwelf]{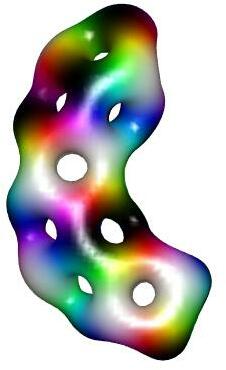}}
    \subfloat[]{\includegraphics[scale=\scaletwelf]{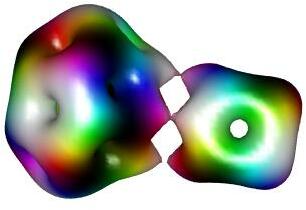}}
    \subfloat[]{\includegraphics[scale=\scaletwelf]{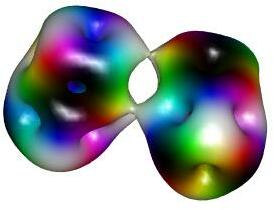}}
    \subfloat[]{\includegraphics[scale=\scaletwelf]{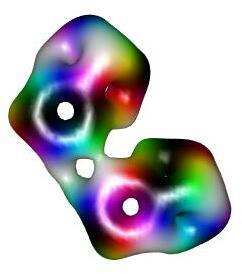}}
  }
  \caption{$B=12$ Skyrmion solutions in order of increasing mass
    (energy), for $\kappa=0$ and pion mass $m=1$.
    These figures are taken from ref.~\cite{Gudnason:2022jkn}
  }
  \label{fig:B12s}
\end{figure}
\begin{figure}[!htp]
  \centering
  \includegraphics[width=\linewidth]{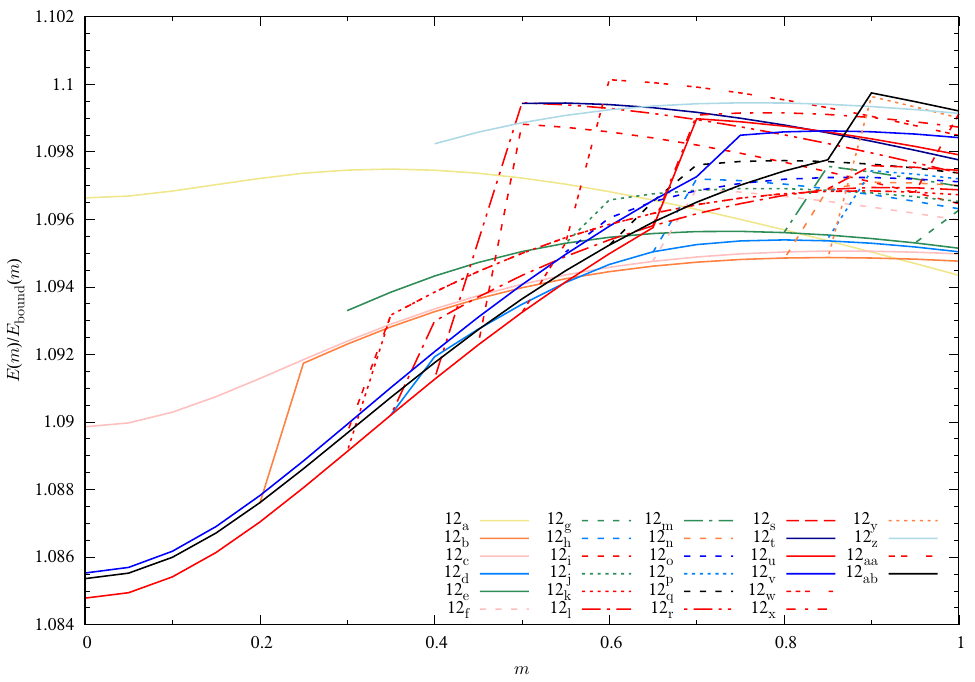}
  \caption{$B=12$ Skyrmion solutions as functions of the pion mass
    $m$ for $\kappa=0$.
    Since the energy grows drastically with $m$, we display the
    energies divided by their topological energy bound, see
    eq.~\eqref{eq:Ebound}.
    When a curve stops, the solution ceases to exist, but when a curve
    drops to another existing curve, the solution decays or transforms
    itself to that solution.
  }
  \label{fig:energy}
\end{figure}
In order to understand which $B=12$ Skyrmion has the lowest energy as
a function of the pion mass, we take all the $B=12$ Skyrmion solutions of the
Sm\"org\aa rdsbord \cite{Gudnason:2022jkn} (they are displayed for
$m=1$ and $\kappa=0$ in fig.~\ref{fig:B12s}) and vary the pion mass
parameter in the range $m\in[0,1]$, which \emph{a posteriori} turns out to be
sufficient for calibrating the theory.
For Skyrmion number $B=12$ and in particular for the most stable
$B=12$ Skyrmions, the backreaction of the Coulomb energy has a 
nearly negligible impact, see fig.~\ref{fig:kappa12a}.
The energy of all the $B=12$ solutions of fig.~\ref{fig:B12s} are
shown in fig.~\ref{fig:energy} as functions of their pion mass
parameter, $m$, for $\kappa=0$.
When a curve stops, the solution ceases to exist, but when a curve
drops to another existing curve, the solution decays or transforms
itself to that solution.
Hysteresis is existing, but is negligible and comparable to the
slope of the curves transitioning between solutions.

\def\scaletwelflarge{0.35}
\begin{figure}[!htp]
  \centering
  \mbox{
    \subfloat[$0\leq m\leq 0.547$]{\includegraphics[scale=\scaletwelflarge]{{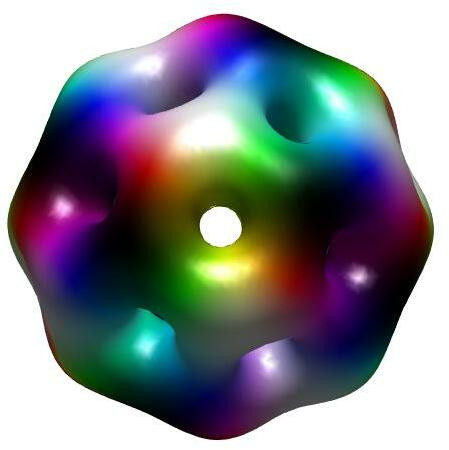}}}
    \subfloat[$0.547\leq m\leq 0.566$]{\includegraphics[scale=\scaletwelflarge]{{B12_121s8c29_crop}}\quad}
    \subfloat[$0.566\leq m\leq 0.928$]{\includegraphics[scale=\scaletwelflarge]{{B12_121s8c44_crop}}}
    \subfloat[$0.928\leq m\leq 1$]{\includegraphics[scale=\scaletwelflarge]{{B12_121s8c51_crop}}}
  }
  \caption{Stable $B=12$ Skyrmion solutions (global energy minimizers)
    for $\kappa=0$ and pion mass in the interval $m\in[0,1]$. }
  \label{fig:B12stable}
\end{figure}
Ignoring for the moment the backreaction of the Coulomb energy, we can
read off the stable $B=12$ Skyrmion solutions in the interval
$m\in[0,1]$, which we illustrate in fig.~\ref{fig:B12stable}.
It turns out that the chain of 3 alpha particles is only the ground
state of the $B=12$ sector for large pion mass, near $m\sim1$, whereas
the relevant Skyrmion for a more realistic pion mass is made of two
$B=7$ Skyrmions sharing a face.
We also checked explicitly, that turning on a finite $\kappa$ of order
one, does not change the ground states (global energy minimizers). 
\begin{figure}[!tp]
  \centering
  \sidesubfloat[]{\includegraphics[width=0.91\linewidth]{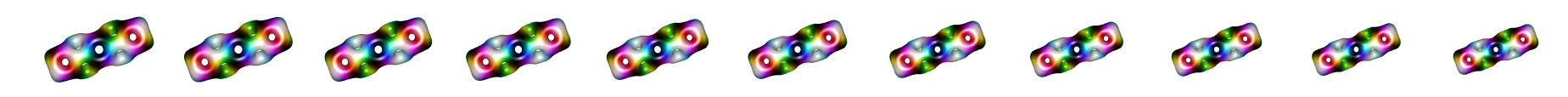}}\\
  \sidesubfloat[]{\includegraphics[width=0.91\linewidth]{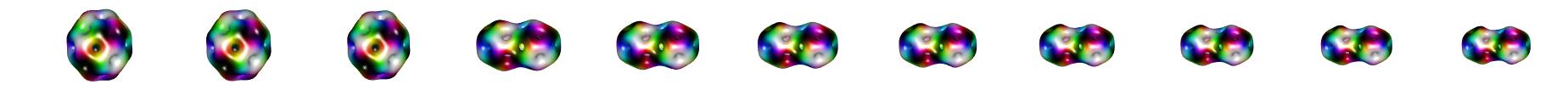}}\\
  \sidesubfloat[]{\includegraphics[width=0.91\linewidth]{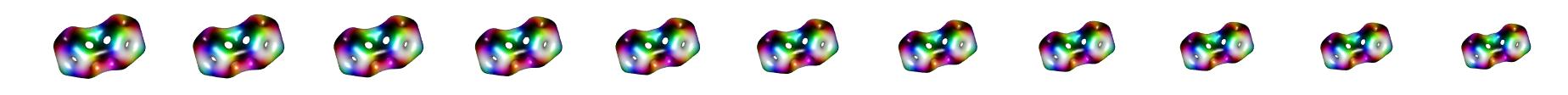}}\\
  \sidesubfloat[]{\includegraphics[width=0.91\linewidth]{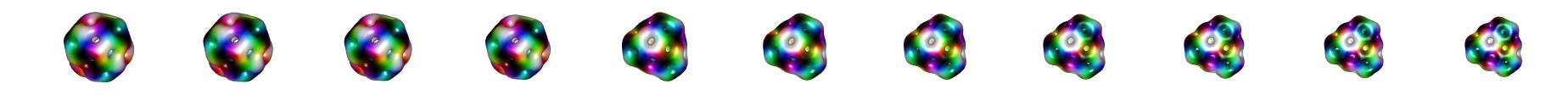}}\\
  \sidesubfloat[]{\includegraphics[width=0.91\linewidth]{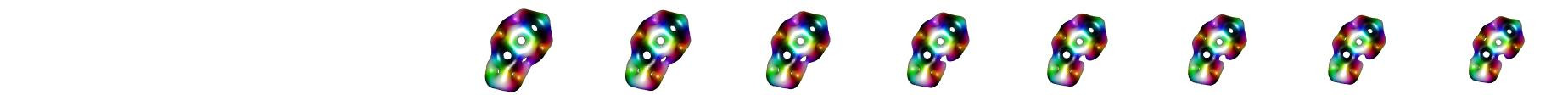}}\\
  \sidesubfloat[]{\includegraphics[width=0.91\linewidth]{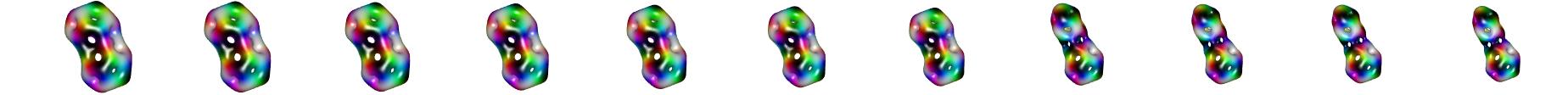}}\\
  \sidesubfloat[]{\includegraphics[width=0.91\linewidth]{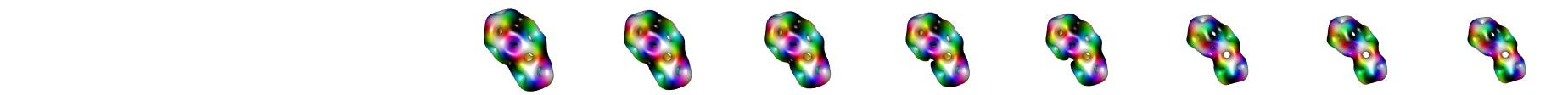}}\\
  \sidesubfloat[]{\includegraphics[width=0.91\linewidth]{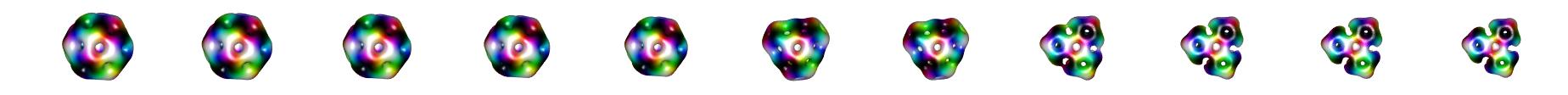}}\\
  \sidesubfloat[]{\includegraphics[width=0.91\linewidth]{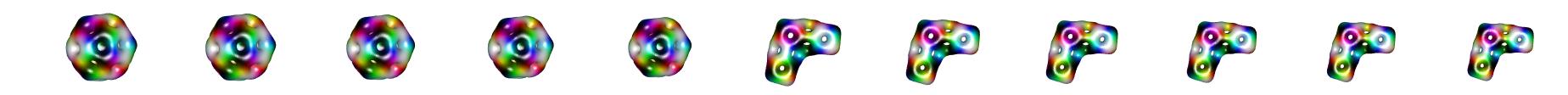}}\\
  \sidesubfloat[]{\includegraphics[width=0.91\linewidth]{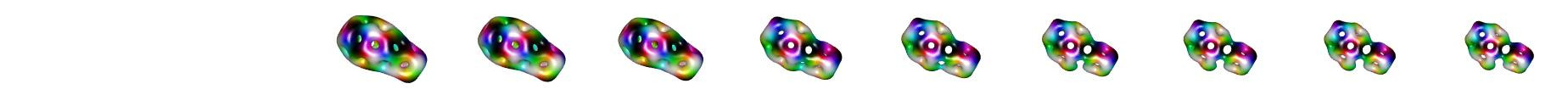}}\\
  \sidesubfloat[]{\includegraphics[width=0.91\linewidth]{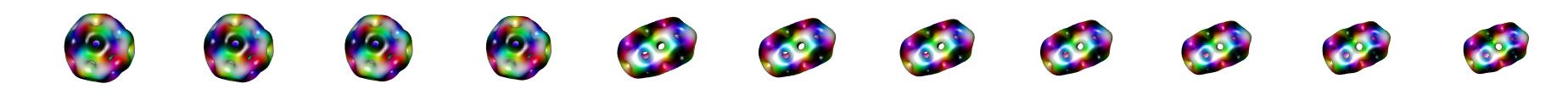}}\\
  \sidesubfloat[]{\includegraphics[width=0.91\linewidth]{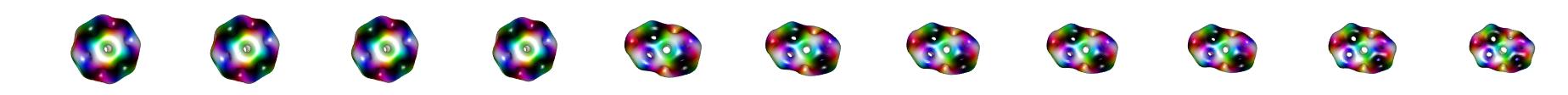}}\\
  \sidesubfloat[]{\includegraphics[width=0.91\linewidth]{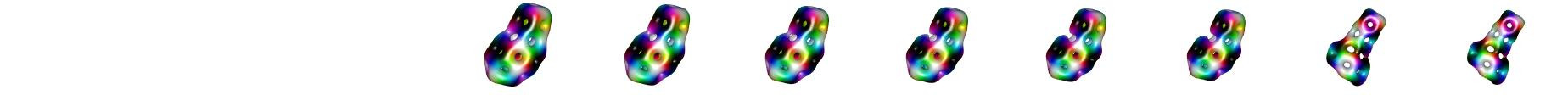}}\\
  \sidesubfloat[]{\includegraphics[width=0.91\linewidth]{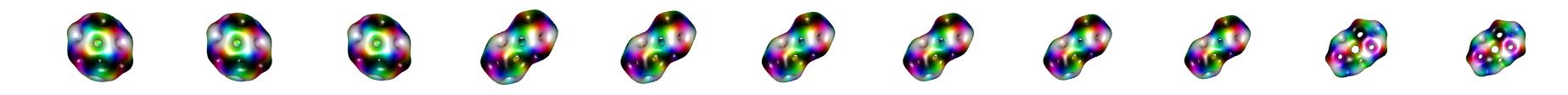}}\\
  \sidesubfloat[]{\includegraphics[width=0.91\linewidth]{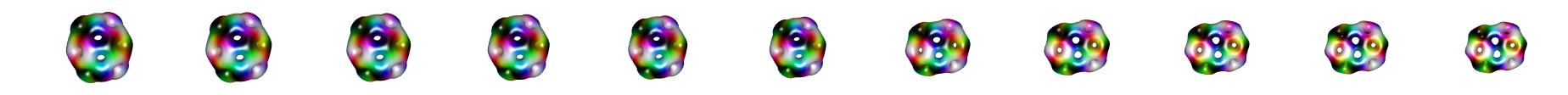}}\\
  \sidesubfloat[]{\includegraphics[width=0.91\linewidth]{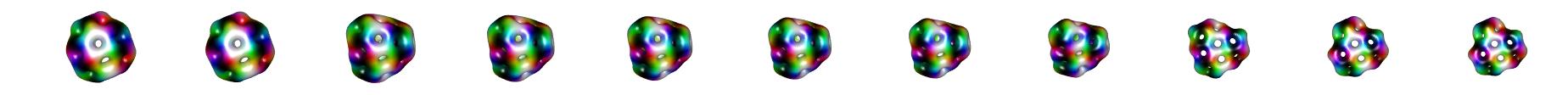}}\\
  \sidesubfloat[]{\includegraphics[width=0.91\linewidth]{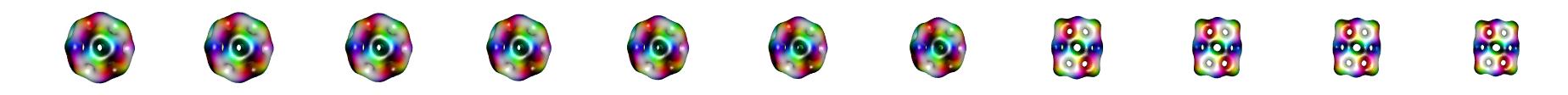}}\\
  \sidesubfloat[]{\includegraphics[width=0.91\linewidth]{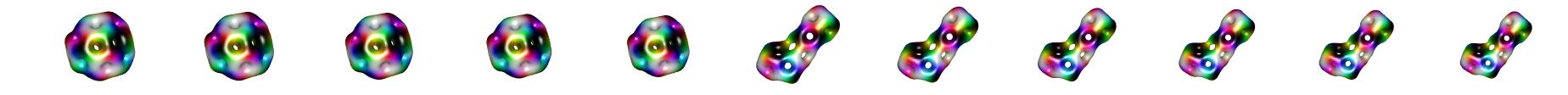}}\\
  \sidesubfloat[]{\includegraphics[width=0.91\linewidth]{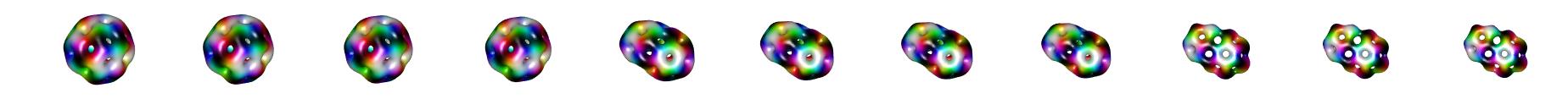}}\\
  \sidesubfloat[]{\includegraphics[width=0.91\linewidth]{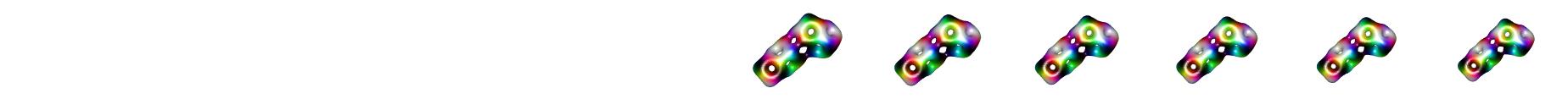}}\\
  \sidesubfloat[]{\includegraphics[width=0.91\linewidth]{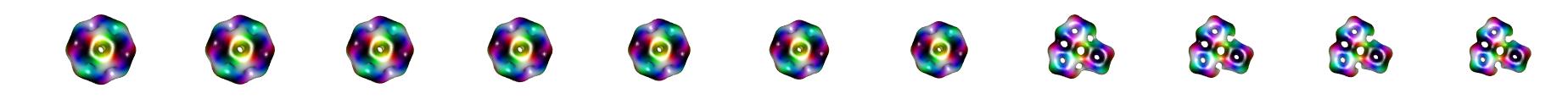}}\\
  \sidesubfloat[]{\includegraphics[width=0.91\linewidth]{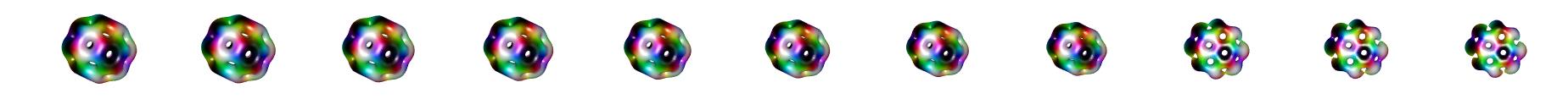}}\\
  \sidesubfloat[]{\includegraphics[width=0.91\linewidth]{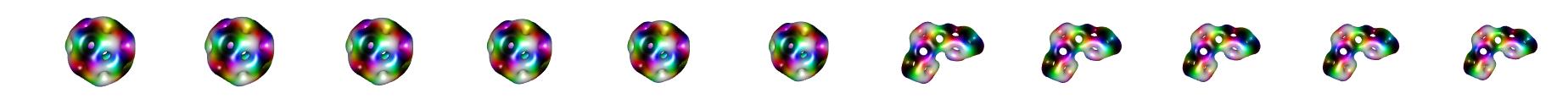}}
  \caption{The evolution of $B=12$ Skyrmion solutions of
    fig.~\ref{fig:B12s} as functions of the pion mass for $\kappa=0$,
    which is lowered from $m=1$ (right-most column) to $m=0$ (left-most
    column).
    The steps in $m$ between each column is twice as large as compared
    to those shown in fig.~\ref{fig:energy}. 
  }
  \label{fig:evolution}
\end{figure}
\begin{figure}[!tp]
  \ContinuedFloat
  \centering
  \sidesubfloat[]{\includegraphics[width=0.91\linewidth]{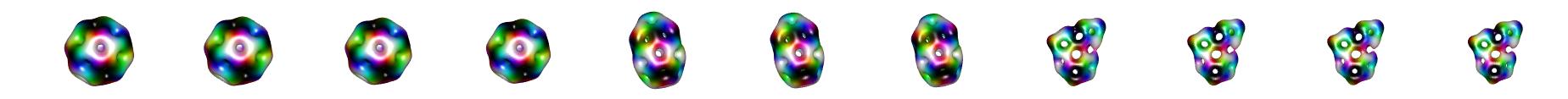}}\\
  \sidesubfloat[]{\includegraphics[width=0.91\linewidth]{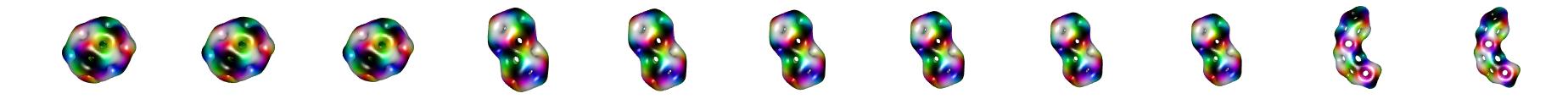}}\\
  \sidesubfloat[]{\includegraphics[width=0.91\linewidth]{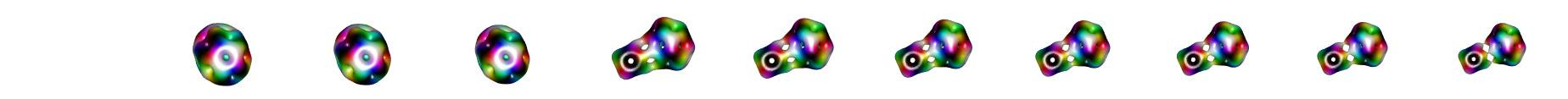}}\\
  \sidesubfloat[]{\includegraphics[width=0.91\linewidth]{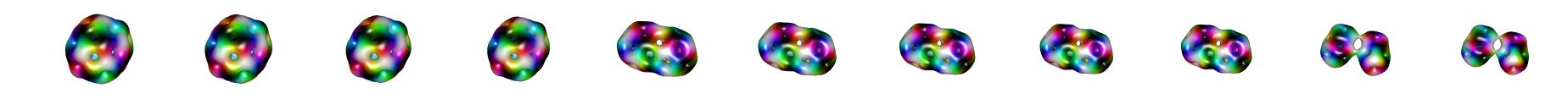}}\\
  \sidesubfloat[]{\includegraphics[width=0.91\linewidth]{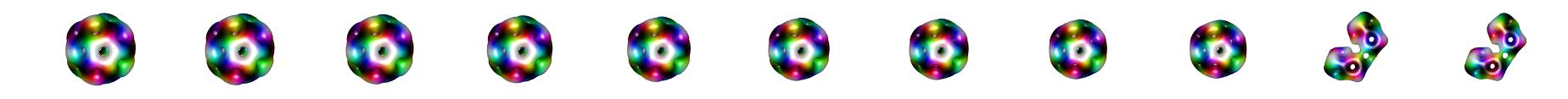}}
  \caption{This figure is continued from the previous page.}
  \label{fig:evolution2}
\end{figure}
The evolution of the solutions as functions of the pion mass parameter
$m$ for $\kappa=0$ is shown in fig.~\ref{fig:evolution}.

Fitting now the mass and radius of the $B=12$ ground state to the mass
and radius of the $^{12}$C nucleus, we have
\cite{Angeli:2013epw,Zyla:2020zbs}
\begin{align}
\frac{F_\pi}{4g} E_{12}(m,\kappa) &= M_{^{12}{\rm C}} \approx 11177.9\ \MeV, \non
\frac{2}{g F_\pi} R_{12}(m,\kappa) &= R_{^{12}{\rm C}} \approx 2.4702\ \fm \approx 1.2518\times 10^{-2}\ \MeV^{-1}.
\end{align}
The charge radius of a Skyrmion in Skyrme units is calculated as
\beq
R_B = \sqrt{\frac{1}{B}\int_{M} r^2 B^0\; \d^3x},
\eeq
with $r\equiv |x-x_0|$ being the radial distance from the centre of
the charge distribution, $x_0$, which is calculated as
\beq
x_0^i = \frac{1}{B}\int_{M} x^i B^0\; \d^3x, \qquad
i=1,2,3.
\eeq

We will now perform the calibration of the model, by computing the
mass as a function of $m$ and $\kappa$, and it suffices to consider
the $B=12_b$ Skyrmion of fig.~\ref{fig:B12stable}(c), which is the
union of two $B=7$ Skyrmions with icosahedral symmetry that share a
face.
The result of the calibration is
\begin{align}
  m \simeq 0.650, \qquad \kappa \simeq 0.737,
\end{align}
for which the Skyrme coupling constant and the pion decay constant
read
\beq
g \simeq 4.010,\qquad
F_\pi \simeq 105.9\MeV,
\eeq
both of which are smaller than those obtained by Adkins-Nappi-Witten
using a fit of the nucleon and the Delta resonance (both are $B=1$
Skyrmions) \cite{Adkins:1983ya}, but both values are larger than those
fitted to Lithium-6 in ref.~\cite[eq.~(77)]{Manton:2006tq}.

\subsection{Skyrmion solutions}

We will now present the numerical Skyrmion solutions.
\begin{figure}[!htp]
  \centering
  \includegraphics[scale=\scaletwelf]{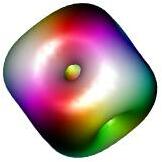}
  \caption{The $B=4$ Skyrmion. }
  \label{fig:B4}
\end{figure}
The 4-Skyrmion is unaltered and is still a cube, see
fig.~\ref{fig:B4}.
\begin{figure}[!htp]
  \centering
  (b)\includegraphics[scale=\scaletwelf]{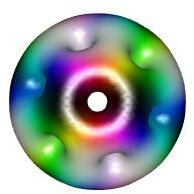}
  (a)\includegraphics[scale=\scaletwelf]{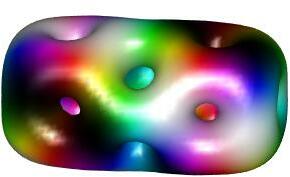}
  \caption{The two existing $B=8$ Skyrmion solutions at $m=0.65$.
    The $8_b$-Skyrmion is the lightest in our calibration. }
  \label{fig:B8}
\end{figure}
The 8-Skyrmions have reduced from 4 solutions in
ref.~\cite{Gudnason:2022jkn} to only the two lightest solutions, see
fig.~\ref{fig:B8}.
The lightest 8-Skyrmion at $m=1$ is the $8_a$ solution, which becomes
the next-lightest one at $m=0.65$, whereas the $8_b$ solution becomes
the lightest state.
No apparent effect is visible from the Coulomb energy with full
backreaction (for details on the impact of the Coulomb force, see the
next section); that 
is, the 2 heavier 8-Skyrmions that were a 7-Skyrmion with a 1-Skyrmion
and a 6-Skyrmion with a torus mounted on the side instead of at the
end, have both disappeared due to the lower pion mass, but with and
without the Coulomb backreaction taken into account they both decay to
the lightest $8_b$ Skyrmion.

\begin{table}[!htp]
  \begin{minipage}{0.42\linewidth}
  {\small
\begin{tabular}{c||cc}
  config & $\kappa=0$  & $\kappa=0.737$ \\
  ($m=1$) & ($m=0.65$) & ($m=0.65$)\\
  \hline\hline
  $12_f$ & \red{$12_c$} & $^*12_f$\\
  $12_g$ & \red{$12_e$} & \red{$12_e$}\\
  $12_h$ & \red{$12_d$} & \red{$12_d$}\\
  $12_m$ & \red{$12_e$} & \red{$12_e$}\\
  $12_n$ & \red{$12_b$} & \red{$12_b$}\\
  $12_p$ & \red{$12_d$} & \red{$12_d$}\\
  $12_q$ & \red{$12_o$} & \red{$12_o$}\\
\end{tabular}}
\end{minipage}
  \begin{minipage}{0.42\linewidth}
    {\small
  \begin{tabular}{c||cc}
    config & $\kappa=0$  & $\kappa=0.737$ \\
  ($m=1$) & ($m=0.65$) & ($m=0.65$)\\
  \hline\hline
  $12_s$ & \red{$12_\alpha$} & \red{$^*12_b$}\\
  $12_u$ & \red{$12_\beta$} & \red{$^*12_d$}\\
  $12_v$ & \red{$12_\gamma$} & \red{$12_\gamma$}\\
  $12_x$ & \red{$12_l$} & \red{$12_l$}\\
  $12_y$ & \red{$12_b$} & \red{$12_b$}\\
  $12_{aa}$ & \red{$12_k$} & \red{$^*12_e$}\\
  $12_{ab}$ & \red{$12_\delta$} & \red{$12_\delta$}\\
      \end{tabular}}
  \end{minipage}
\caption{$B=12$ Skyrmions' change from $m=1$, $\kappa=0$ to
  the physical fit of $m=0.65$ with and without the physical EM
  coupling, $\kappa=0.737$, turned on. The red entries denote solutions that
  differ from the $m=1$ ones, whereas the $^*$ denotes solutions that
  are different when the electromagnetic interactions are taken into
  account compared with when it is turned off.
  Skyrmions with Greek indices are new compared to the $m=1$,
  $\kappa=0$ case.
  Skyrmions that remain unaltered are not shown in this table.
}
\label{tab:B12change}
\end{table}
Let us turn to the $B=12$ Skyrmions.
We start with listing the changes to the solutions due to changing the
fit of the model from $m=1$, $\kappa=0$ to $m=0.65$, $\kappa=0.737$,
see tab.~\ref{tab:B12change}.
In the previous section, the EM interactions were not taken into
account.
Here we compute the new solutions with $m=0.65$ with and without the
EM interactions turned on; in both cases starting with the $m=1$,
$\kappa=0$ solution as the initial condition.
\begin{figure}[!htp]
  \centering
  ($\alpha$)\includegraphics[scale=\scaletwelf]{{B12c59_121k0m0.65s8_crop}}\quad
  ($\beta$)\includegraphics[scale=\scaletwelf]{{B12c13_121k0m0.65s8_crop}}\quad
  ($\gamma$)\includegraphics[scale=\scaletwelf]{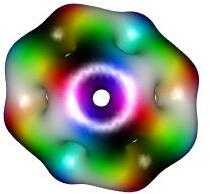}\quad
  ($\delta$)\includegraphics[scale=\scaletwelf]{{B12c73_121k0m0.65s8_crop}}
  \caption{$B=12$ Skyrmion solutions that are new compared with the
    solutions of the Sm\"org\aa sbord \cite{Gudnason:2022jkn}.
    Since these solutions are computed for the pion mass $m=0.65$,
    they may not exist for $m=1$.
  }
  \label{fig:B12new}
\end{figure}
The new solutions that appear in the process are shown in
fig.~\ref{fig:B12new}.
\begin{figure}[!htp]
  \begin{tabular}{c@{\raisebox{0.9cm}{$\quad\to\quad$}}cc}%
  (f)\includegraphics[scale=\scaletwelf]{{B12_121s8c61_crop}}&
  (c)\includegraphics[scale=\scaletwelf]{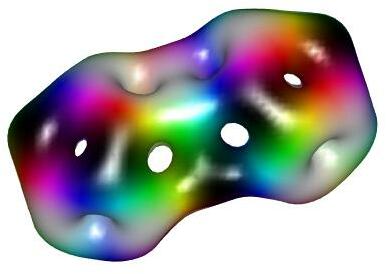}&
  (f)\includegraphics[scale=\scaletwelf]{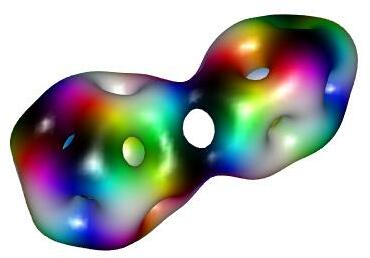}\\
  (s)\includegraphics[scale=\scaletwelf]{{B12_121s8c59_crop}}&
  ($\alpha$)\includegraphics[scale=\scaletwelf]{{B12c59_121k0m0.65s8_crop}}&
  (b)\includegraphics[scale=\scaletwelf]{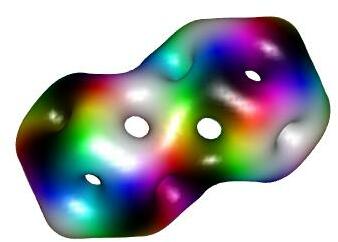}\\
  (u)\includegraphics[scale=\scaletwelf]{{B12_121s8c13_crop}}&
  ($\beta$)\includegraphics[scale=\scaletwelf]{{B12c13_121k0m0.65s8_crop}}&
  (d)\includegraphics[scale=\scaletwelf]{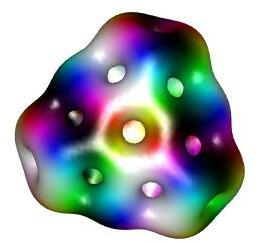}\\
  (aa)\includegraphics[scale=\scaletwelf]{{B12_121s8c12_crop}}&
  (k)\includegraphics[scale=\scaletwelf]{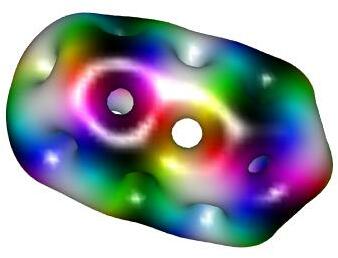}&
  (e)\includegraphics[scale=\scaletwelf]{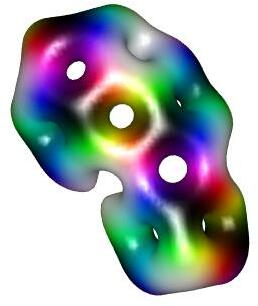}
  \end{tabular}
  \caption{$B=12$ Skyrmions that differ when the electromagnetic interactions
  are taken into account. The left-most column shows the initial
  conditions with $m=1$, $\kappa=0$, the middle column shows the solutions for
  $m=0.65$, $\kappa=0$ and the right-most one is for $m=0.65$,
  $\kappa=0.737$.
  }
  \label{fig:B12change}
\end{figure}
The four cases where the Skyrmions flow to two different solutions for
the Coulomb backreaction turned off ($\kappa=0$) and on
($\kappa=0.737$), are shown in fig.~\ref{fig:B12change}.
The $12_f$-Skyrmion remains the $12_f$-Skyrmion when the Coulomb force
is turned on, but changes to the $12_c$-Skyrmion with $\kappa=0$.
The $12_s$- and $12_u$-Skyrmions flow to two new Skyrmion solutions
when $\kappa=0$ and $m$ is lowered from $m=1$ to $m=0.65$, which are
denoted $12_\alpha$- and $12_\beta$-Skyrmions, respectively.
Finally, the $12_{aa}$-Skyrmion flows to two different existing
solutions, upon changing the pion mass and including the EM
interaction or not.
\begin{figure}[!htp]
  \begin{tabular}{c@{\raisebox{0.9cm}{$\quad\to\quad$}}c}%
    (v)\includegraphics[scale=\scaletwelf]{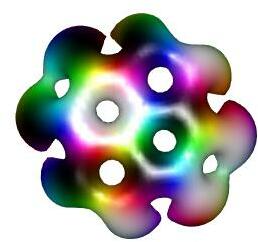}&
    ($\gamma$)\includegraphics[scale=\scaletwelf]{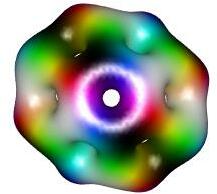}\\
    (ab)\includegraphics[scale=\scaletwelf]{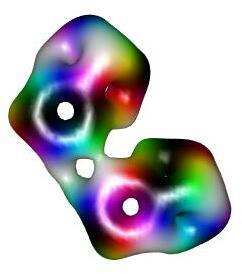}&
    ($\delta$)\includegraphics[scale=\scaletwelf]{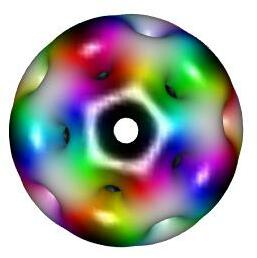}
  \end{tabular}
  \caption{$B=12$ Skyrmions that flow to the same \emph{new} Skyrmion
    solution due to the change of the pion mass $m=1\to0.65$,
    independently of whether the Coulomb force is turned on or not.
    The left-most column shows the initial conditions with $m=1$,
    $\kappa=0$, whereas the right-most column shows the $m=0.65$,
    $\kappa=0.737$ case. The $m=0.65$, $\kappa=0$ solutions are the same as
    the $\kappa=0.737$ ones.
  }
  \label{fig:B12change2}
\end{figure}
The remaining ten Skyrmions in tab.~\ref{tab:B12change} all flow to
the \emph{same} Skyrmion solution upon changing the pion mass from
$m=1$ to $m=0.65$, independently of whether the Coulomb force is
turned on or not.
In most cases, the Skyrmion flows to an existing lower-energy
solution.
However, in two cases, the Skyrmion flows to a new solution.
These two cases are shown in fig.~\ref{fig:B12change2}. 

\begin{figure}[!htp]
  \centering
  \renewcommand{\arraystretch}{0.1}
  \begin{tabular}{l@{\,}rccl@{\,}r}
  \raisebox{18pt}{1692.28} & \raisebox{18pt}{(b)}&
  \includegraphics[scale=0.15]{{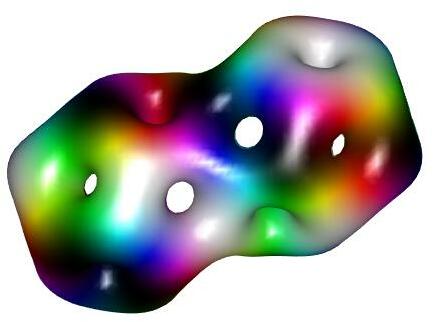}}&
  \includegraphics[scale=0.15]{{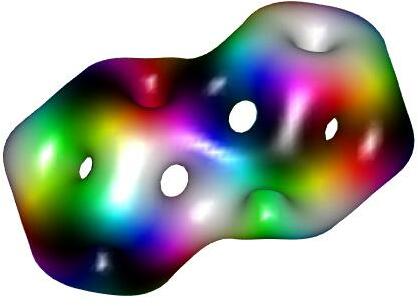}}&
  \raisebox{18pt}{(b)} & \raisebox{18pt}{1693.89}\\
  \raisebox{18pt}{1692.50} & \raisebox{18pt}{(c)}&
  \includegraphics[scale=0.2]{{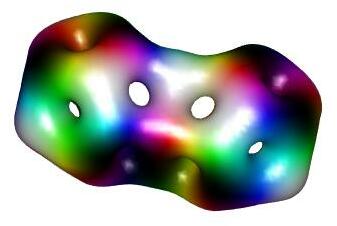}}&
  \includegraphics[scale=0.2]{{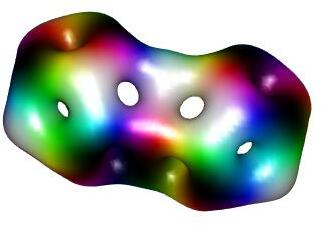}}&
  \raisebox{18pt}{(c)} & \raisebox{18pt}{1694.10}\\
  \raisebox{18pt}{1692.94} & \raisebox{18pt}{(d)}&
  \includegraphics[scale=0.21]{{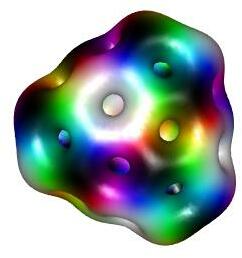}}&
  \includegraphics[scale=0.21]{{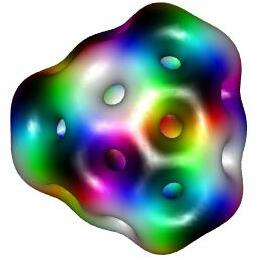}}&
  \raisebox{18pt}{(d)} & \raisebox{18pt}{1694.57}\\
  \raisebox{18pt}{1693.77} & \raisebox{18pt}{(e)}&
  \includegraphics[scale=0.18]{{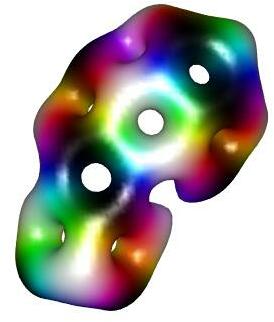}}&
  \includegraphics[scale=0.18]{{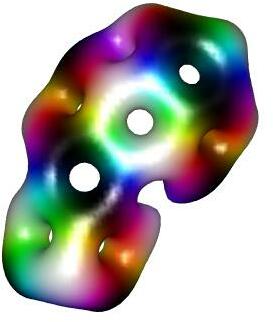}}&
  \raisebox{18pt}{(e)} & \raisebox{18pt}{1695.38}\\
  \raisebox{18pt}{1694.04} & \raisebox{18pt}{($\beta$)}&
  \includegraphics[scale=0.2]{{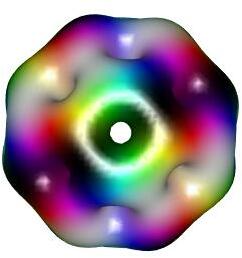}}&
  \includegraphics[scale=0.18]{{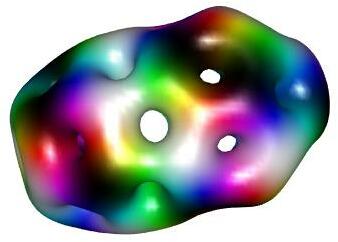}}&
  \raisebox{18pt}{(l)} & \raisebox{18pt}{1695.76}\\
  \raisebox{18pt}{1694.13} & \raisebox{18pt}{(l)}&
  \includegraphics[scale=0.18]{{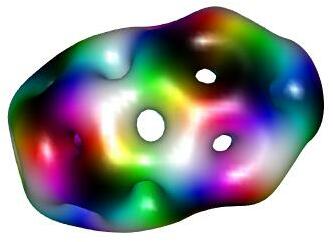}}&
  \includegraphics[scale=0.2]{{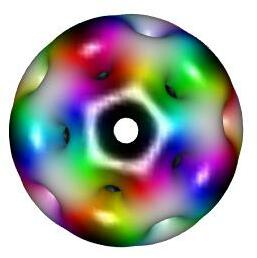}}&
  \raisebox{18pt}{($\delta$)} & \raisebox{18pt}{1695.94}\\
  \raisebox{18pt}{1694.14} & \raisebox{18pt}{($\alpha$)}&
  \includegraphics[scale=0.2]{{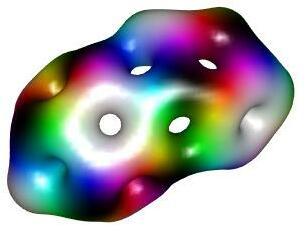}}&
  \includegraphics[scale=0.18]{{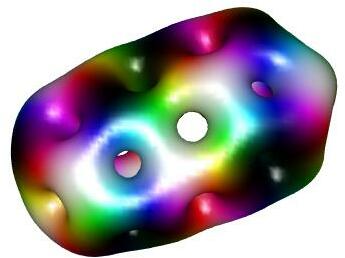}}&
  \raisebox{18pt}{(k)} & \raisebox{18pt}{1696.32}\\
  \raisebox{18pt}{1694.29} & \raisebox{18pt}{($\delta$)}&
  \includegraphics[scale=0.2]{{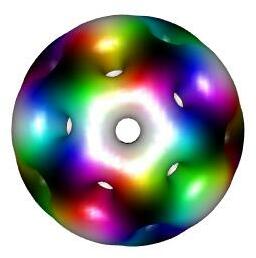}}&
  \includegraphics[scale=0.2]{{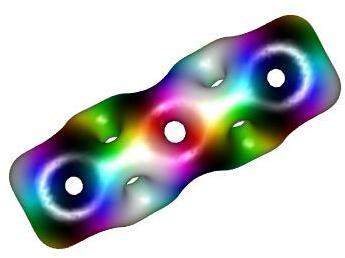}}&
  \raisebox{18pt}{(a)} & \raisebox{18pt}{1696.84}\\
  \raisebox{18pt}{1694.69} & \raisebox{18pt}{(k)}&
  \includegraphics[scale=0.18]{{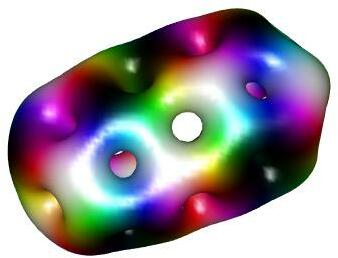}}&
  \includegraphics[scale=0.2]{{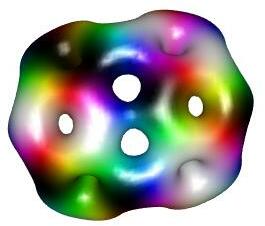}}&
  \raisebox{18pt}{(o)} & \raisebox{18pt}{1696.87}\\
  \raisebox{18pt}{1695.23} & \raisebox{18pt}{(o)}&
  \includegraphics[scale=0.2]{{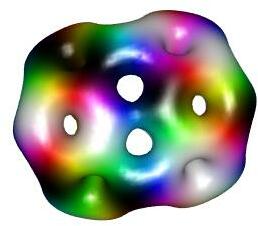}}&
  \includegraphics[scale=0.23]{{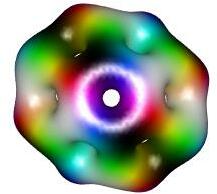}}&
  \raisebox{18pt}{($\gamma$)} & \raisebox{18pt}{1696.94}
  \end{tabular}
  \caption{The ten lightest $B=12$ Skyrmion solutions with $m=0.65$
    without Coulomb backreaction (left-hand side) and with the Coulomb
    backreaction taken into account (right-hand side). The solution
    labels refer to the solutions of figs.~\ref{fig:B12s} and
    \ref{fig:B12new}.
    Interestingly, the $12_a$ solution that is a chain of three cubes,
    is not the global energy minimizer in our calibration.
  }
  \label{fig:B12tenlightest}
\end{figure}
Finally, we show in fig.~\ref{fig:B12tenlightest} the ten lightest
$B=12$ Skyrmion solutions in our calibration (pion mass $m=0.65$)
without taking the Coulomb force into account (left-hand side) and
with taking it into account (right-hand side).
It is interesting to note that the expected global energy minimizer
(ground state), i.e.~the chain of three cubes (alpha particles), turns
out not to be the minimizer in our calibration.
Not only is it not the minimizer, it is not even in the top-ten
lightest states when the Coulomb backreaction is turned off
($\kappa=0$), whereas it figures as the 8th lightest state when it is
taken into account.
The four lightest states turn out to be the same, whether EM
interactions are backreacted or not.
In particular, the ground state for $B=12$ is the same solution:
namely the $12_b$-Skyrmion solution made up of two icosahedrally
symmetric Skyrmions that share a face.

We now turn to the rather large ensemble of Skyrmions, i.e.~the $B=16$
sector of the Sm\"org\aa sbord \cite{Gudnason:2022jkn}.
As one could expect from the results in the $B=12$ sector, the
alterations in this sector will be even more.
Due to the large number of $B=16$ solutions, we do not display all the
known solutions and refer to the labels of the solutions used in
ref.~\cite{Gudnason:2022jkn} using the Latin alphabet.
The new $B=16$ solutions are denoted using the Greek alphabet
(skipping omicron, since it looks like 'o').
We start by listing a table of the changes that happen when using the
$B=16$ Skyrmions as initial conditions for our calibration, i.e.~with
pion mass $m=0.65$ with and without taking into account the
backreaction of the Coulomb force, see tab.~\ref{tab:B16change}.

\begin{table}[!htp]
  \begin{minipage}{0.42\linewidth}
  {\small
\begin{tabular}{c||cc}
  config & $\kappa=0$  & $\kappa=0.737$ \\
  ($m=1$) & ($m=0.65$) & ($m=0.65$)\\
  \hline\hline
  $16_{i}$ & -- & --\\
  $16_{k}$ & \red{$16_{a}$} & \red{$16_{a}$}\\
  $16_{m}$ & \red{$16_{c}$} & \red{$16_{c}$}\\
  $16_{o}$ & -- & $^*16_{o}$\\
  $16_{t}$ & \red{$16_{c}$} & \red{$^*16_\alpha$}\\
  $16_{z}$ & -- & --\\
  $16_{ae}$ & -- & \red{$^*16_\beta$}\\
  $16_{ai}$ & -- & $\red{^*16_{aa}}$\\
  $16_{av}$ & -- & --\\
  $16_{ax}$ & \red{$16_\gamma$} & \red{$16_\gamma$}\\
  $16_{ay}$ & -- & --\\
  $16_{az}$ & \red{$16_{c}$} & \red{$16_{c}$}\\
  $16_{bb}$ & \red{$16_c$} & \red{$^*16_\delta$}\\
  $16_{bc}$ & \red{$16_{y}$} & \red{$16_{y}$}\\
  $16_{bd}$ & -- & --\\
  $16_{bf}$ & \red{$16_\varepsilon$} & \red{$16_\varepsilon$}\\
  $16_{bj}$ & \red{$16_c$} & \red{$16_c$}\\
  $16_{bl}$ & \red{$16_c$} & \red{$16_c$}\\
  $16_{bm}$ & -- & \red{$^*16_{p}$}\\
  $16_{bq}$ & -- & --\\
  $16_{bs}$ & \red{$16_{l}$} & \red{$^*16_{d}$}\\
  $16_{bu}$ & \red{$16_{s}$} & \red{$16_{s}$}\\
  $16_{bx}$ & \red{$16_{c}$} & \red{$16_{c}$}\\
  $16_{ce}$ & \red{$16_\zeta$} & \red{$16_\zeta$}\\
  $16_{ci}$ & \red{$16_{c}$} & $^*16_{ci}$\\
\end{tabular}}
\end{minipage}
  \begin{minipage}{0.42\linewidth}
    {\small
  \begin{tabular}{c||cc}
    config & $\kappa=0$  & $\kappa=0.737$ \\
  ($m=1$) & ($m=0.65$) & ($m=0.65$)\\
  \hline\hline
  $16_{cj}$ & \red{$16_\eta$} & \red{$16_\eta$}\\
  $16_{cl}$ & \red{$16_{p}$} & \red{$16_{p}$}\\
  $16_{cn}$ & \red{$16_c$} & \red{$16_c$}\\
  $16_{co}$ & \red{$16_a$} & \red{$16_a$}\\
  $16_{cr}$ & -- & \red{$^*16_{cp}$}\\
  $16_{cs}$ & -- & \red{$^*16_\theta$}\\
  $16_{cu}$ & -- & --\\
  $16_{cv}$ & \red{$16_{a}$} & $^{*}$--\\
  $16_{cy}$ & -- & $^*16_{cy}$\\
  $16_{cz}$ & \red{$16_\iota$} & \red{$16_\iota$}\\
  $16_{da}$ & \red{$16_\kappa$} & \red{$16_\kappa$}\\
  $16_{dd}$ & -- & $^*16_{dd}$\\
  $16_{de}$ & \red{$16_{aa}$} & \red{$16_{aa}$}\\
  $16_{dg}$ & \red{$16_\lambda$} & \red{$16_\lambda$}\\
  $16_{dn}$ & $16_{dn}$ & \red{$^*16_\mu$}\\
  $16_{do}$ & \red{$16_\nu$} & \red{$16_\nu$}\\
  $16_{dp}$ & -- & --\\
  $16_{du}$ & \red{$16_{c}$} & \red{$16_{c}$}\\
  $16_{dx}$ & \red{$16_{c}$} & \red{$16_{c}$}\\
  $16_{eb}$ & -- & --\\
  $16_{ec}$ & \red{$16_\xi$} & \red{$^*16_p$}\\
  $16_{eh}$ & \red{$16_\pi$} & \red{$16_\pi$}\\
  $16_{ej}$ & \red{$16_{dy}$} & \red{$16_{dy}$}\\
  $16_{em}$ & \red{$16_{bw}$} & \red{$16_{bw}$}\\
  $16_{eo}$ & \red{$16_\rho$} & \red{$16_\rho$}\\
    \end{tabular}}
  \end{minipage}
\caption{$B=16$ Skyrmions' change from $m=1$, $\kappa=0$ to
  the physical fit of $m=0.65$ with and without the physical EM
  coupling, $\kappa=0.737$, turned on. The -- denote
  solutions that do not exist, the red entries denote solutions that
  differ from the $m=1$ ones, whereas the $^*$ denotes solutions that
  are different when the electromagnetic interactions are taken into
  account compared with when it is turned off.
  Skyrmions with Greek indices are new compared to the $m=1$,
  $\kappa=0$ case.
  Skyrmions that remain unaltered are not shown in this table.
}
\label{tab:B16change}
\end{table}
There are 9 cases, where the Skyrmion solution disappears; this can
happen for example by the smaller pion mass not squeezing the Skyrmion
as much at $m=0.65$ as compared with at $m=1$ and hence several
Skyrmions inflate to the same more hollow solution.
Other possibilities include that the Skyrmion solution simply breaks
up into smaller chunks and hence does not count as a $B=16$ solution.

\def\scalesixteen{0.16}
\begin{figure}[!htp]
  ($\alpha$)\includegraphics[scale=\scalesixteen]{{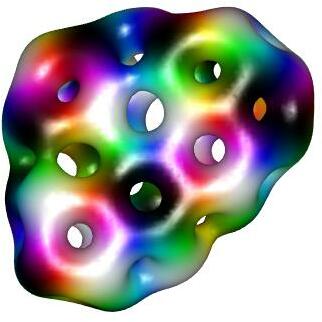}}
  ($\delta$)\includegraphics[scale=\scalesixteen]{{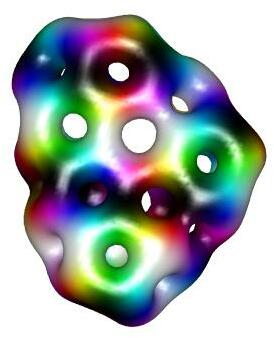}}
  ($\mu$)\includegraphics[scale=\scalesixteen]{{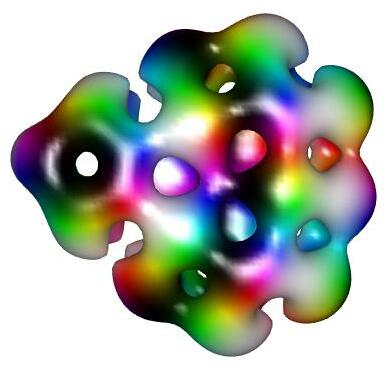}}
  \caption{New $B=16$ Skyrmion solutions that were found with Coulomb
    backreaction taken into account ($m=0.65$ and $\kappa=0.737$).
    In all these cases, the Skyrmion solution found without the
    Coulomb force turned on was a known one. In the case of the
    $16_\mu$-Skyrmion, the known solution was the same as the initial
    condition ($16_{dn}$).
  }
  \label{fig:B16_withEM_differentwithoutEM}
\end{figure}

\begin{figure}[!htp]
  ($\beta$)\includegraphics[scale=\scalesixteen]{{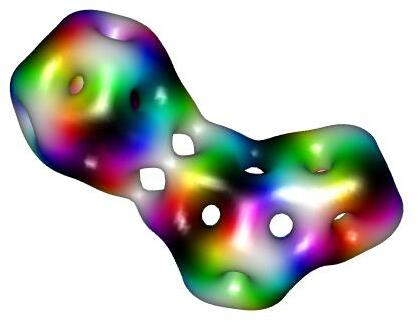}}
  ($\theta$)\includegraphics[scale=\scalesixteen]{{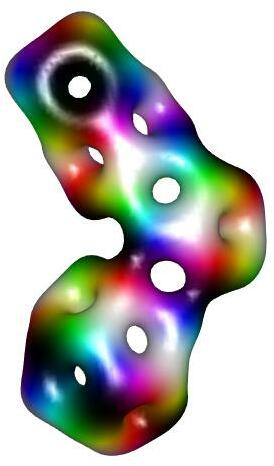}}
  \caption{New $B=16$ Skyrmion solutions that were found with Coulomb
    backreaction taken into account ($m=0.65$ and $\kappa=0.737$).
    In all these cases, no $B=16$ solution was found when the Coulomb
    force was not included.
  }
  \label{fig:B16_withEM_nowithoutEM}
\end{figure}

\begin{figure}[!htp]
  ($\xi$)\includegraphics[scale=\scalesixteen]{{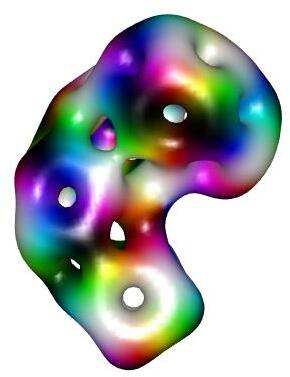}}
  \caption{New $B=16$ Skyrmion solution found by changing to our
    calibration, but without the Coulomb backreaction ($m=0.65$ and
    $\kappa=0$). Turning on the Coulomb backreaction gave a different,
    albeit known Skyrmion solution ($16_p$).
  }
  \label{fig:B16_onlywithoutEM_butdifferentwithEM}
\end{figure}

\begin{figure}[!htp]
  ($\gamma$)\includegraphics[scale=\scalesixteen]{{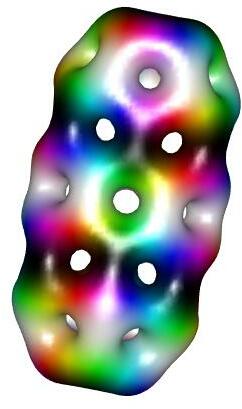}}
  ($\varepsilon$)\includegraphics[scale=\scalesixteen]{{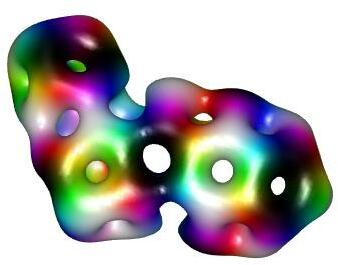}}
  ($\zeta$)\includegraphics[scale=\scalesixteen]{{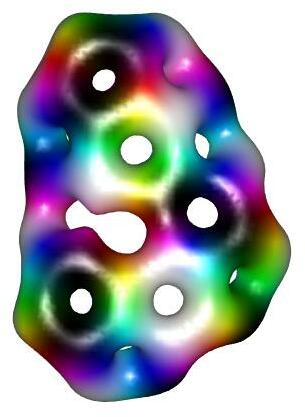}}
  ($\eta$)\includegraphics[scale=\scalesixteen]{{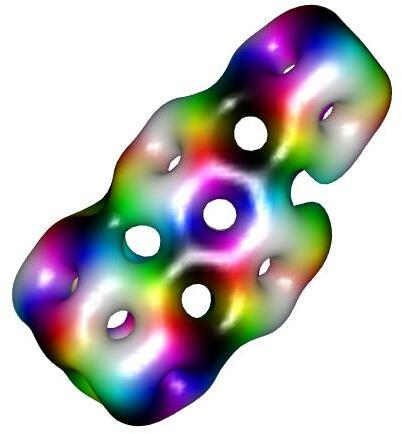}}
  ($\iota$)\includegraphics[scale=\scalesixteen]{{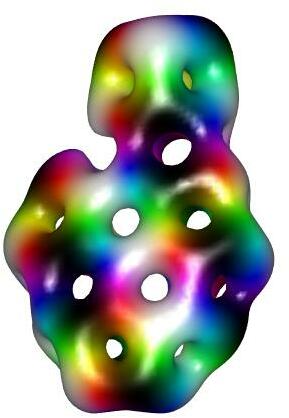}}
  ($\kappa$)\includegraphics[scale=\scalesixteen]{{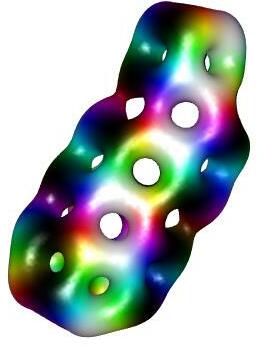}}
  ($\lambda$)\includegraphics[scale=\scalesixteen]{{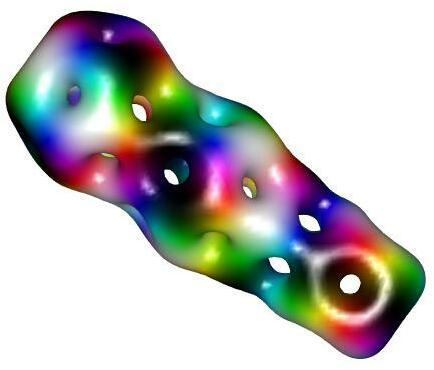}}
  ($\nu$)\includegraphics[scale=\scalesixteen]{{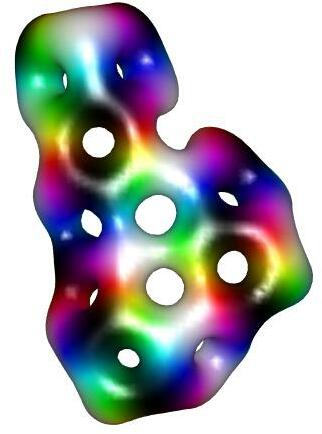}}
  ($\pi$)\includegraphics[scale=\scalesixteen]{{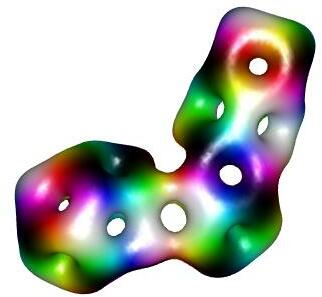}}
  ($\rho$)\includegraphics[scale=\scalesixteen]{{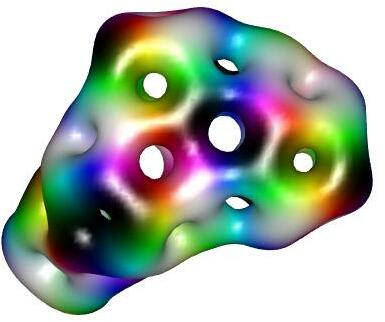}}
  \caption{New $B=16$ Skyrmion solutions found by changing to our
    calibration, but the solutions are qualitatively the same with and
    without the Coulomb backreaction taken into account.
    The figure shows the solutions for $m=0.65$ and $\kappa=0.737$. 
  }
  \label{fig:B16_withandwithoutEM}
\end{figure}

There are furthermore 15 cases, where the solutions are different when
the Coulomb backreaction (CBR) is taken into account from when it is not.
In 3 of the 15 cases, the Skyrmion remained the same with the Coulomb
backreaction turned on, but disappeared when it was switched off and
in 1 case, it remained with the backreaction ($16_{ci}$), but changed
to a low-energy solution when it was turned off ($16_c$).
Only in one case, did the Skyrmion solution disappear with CBR turned
on and in this case ($16_{cv}$) the solution without CBR decayed to
one of the lowest-energy solutions ($16_a$).
This may suggest that the Skyrmion solutions are more stable with the
CBR taken into account.
In 3 cases,  a new Skyrmion solution was found when CBR was taken into
account, whereas the solution flowed to a known solution without CBR; in
one of these cases, the solution without CBR remained the same, see
fig.~\ref{fig:B16_withEM_differentwithoutEM}. 
In 2 cases, the new Skyrmion solution was found with CBR taken into
account, whereas no solution exists without it, see
fig.~\ref{fig:B16_withEM_nowithoutEM}.
In further 3 cases, no solution exists without the CBR taken into
account, but with CBR the solution flowed to a different, albeit known
Skyrmion solution, see tab.~\ref{tab:B16change}.
In one case, the solutions flowed to 2 different known solutions.
Finally, in one case a new Skyrmion solution was found without the CBR
taken into account, whereas it flowed to a known solution ($16_p$)
with it, see fig.~\ref{fig:B16_onlywithoutEM_butdifferentwithEM}.

\begin{figure}[!htp]
  \centering
  \renewcommand{\arraystretch}{0.1}
  \begin{tabular}{l@{\,}rccl@{\,}r}
  \raisebox{18pt}{2248.38} & \raisebox{18pt}{(a)}&
  \includegraphics[scale=0.15]{{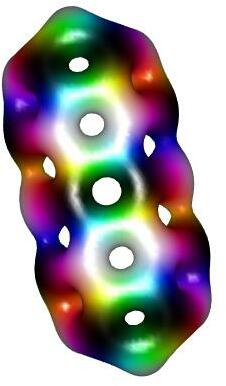}}&
  \includegraphics[scale=0.15]{{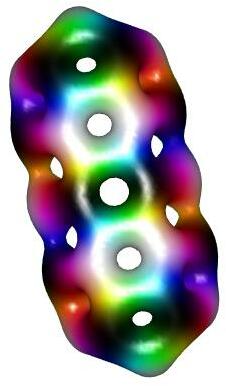}}&
  \raisebox{18pt}{(a)} & \raisebox{18pt}{2250.87}\\
  \raisebox{18pt}{2249.10} & \raisebox{18pt}{(d)}&
  \includegraphics[scale=0.15]{{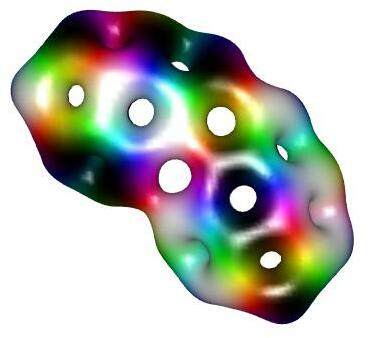}}&
  \includegraphics[scale=0.15]{{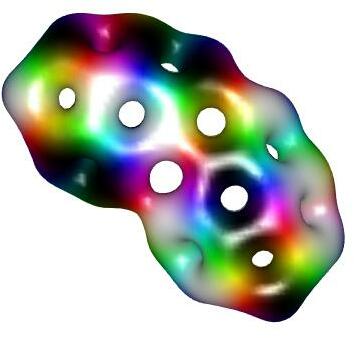}}&
  \raisebox{18pt}{(d)} & \raisebox{18pt}{2251.61}\\
  \raisebox{18pt}{2249.41} & \raisebox{18pt}{(e)}&
  \includegraphics[scale=0.13]{{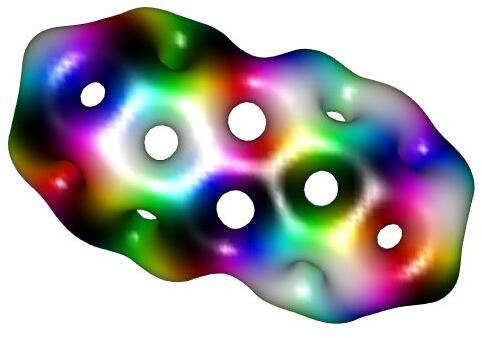}}&
  \includegraphics[scale=0.13]{{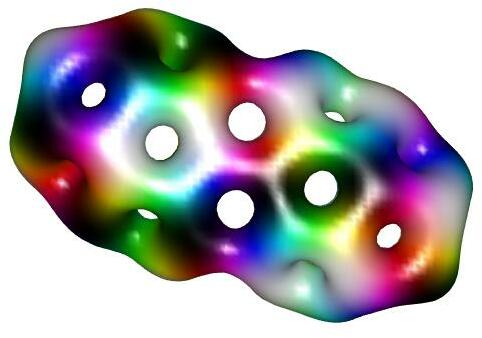}}&
  \raisebox{18pt}{(e)} & \raisebox{18pt}{2251.93}\\
  \raisebox{18pt}{2249.98} & \raisebox{18pt}{(c)}&
  \includegraphics[scale=0.16]{{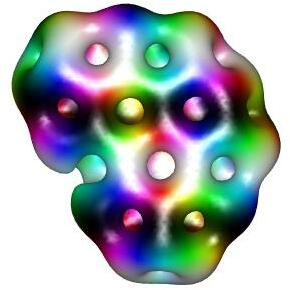}}&
  \includegraphics[scale=0.16]{{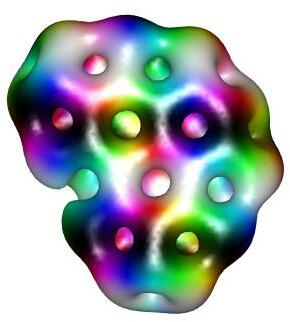}}&
  \raisebox{18pt}{(c)} & \raisebox{18pt}{2252.54}\\
  \raisebox{18pt}{2250.35} & \raisebox{18pt}{(b)}&
  \includegraphics[scale=0.13]{{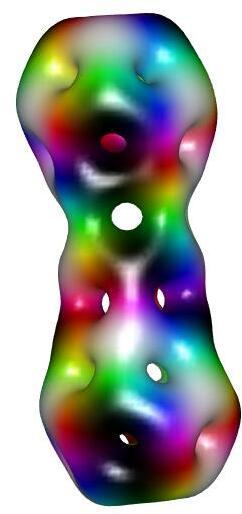}}&
  \includegraphics[scale=0.13]{{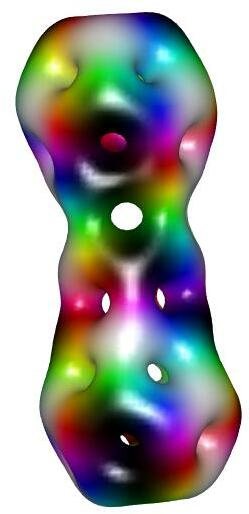}}&
  \raisebox{18pt}{(b)} & \raisebox{18pt}{2252.72}\\
  \raisebox{18pt}{2250.50} & \raisebox{18pt}{($\gamma$)}&
  \includegraphics[scale=0.15]{{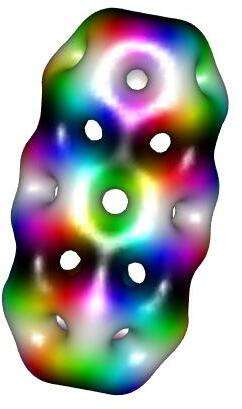}}&
  \includegraphics[scale=0.15]{{B16c68_121k0.737m0.65s8_crop.jpg}}&
  \raisebox{18pt}{($\gamma$)} & \raisebox{18pt}{2253.02}\\
  \raisebox{18pt}{2250.98} & \raisebox{18pt}{(h)}&
  \includegraphics[scale=0.15]{{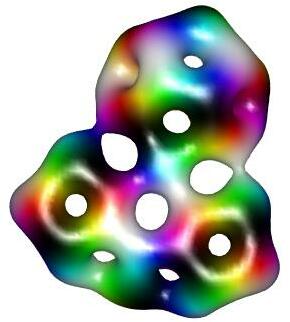}}&
  \includegraphics[scale=0.15]{{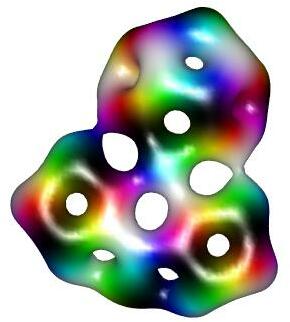}}&
  \raisebox{18pt}{(h)} & \raisebox{18pt}{2253.51}\\
  \raisebox{18pt}{2251.44} & \raisebox{18pt}{(l)}&
  \includegraphics[scale=0.15]{{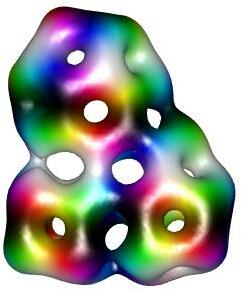}}&
  \includegraphics[scale=0.15]{{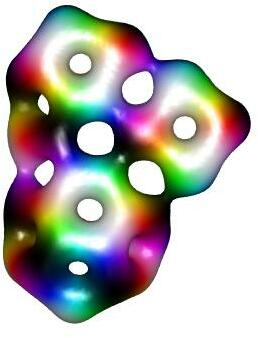}}&
  \raisebox{18pt}{(l)} & \raisebox{18pt}{2253.96}\\
  \raisebox{18pt}{2250.03} & \raisebox{18pt}{(o)}&
  \includegraphics[scale=0.15]{{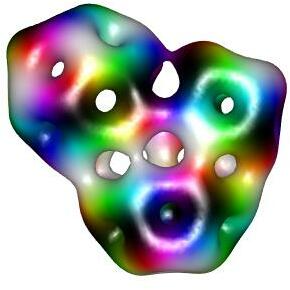}}&
  \includegraphics[scale=0.15]{{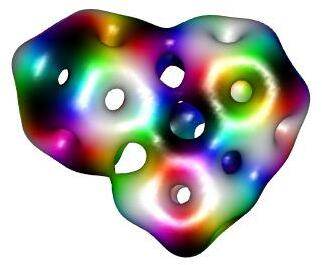}}&
  \raisebox{18pt}{(o)} & \raisebox{18pt}{2254.23}\\
  \raisebox{18pt}{2250.05} & \raisebox{18pt}{(n)}&
  \includegraphics[scale=0.13]{{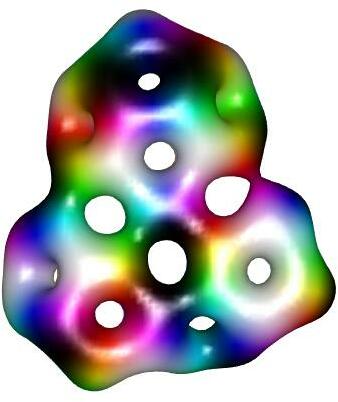}}&
  \includegraphics[scale=0.13]{{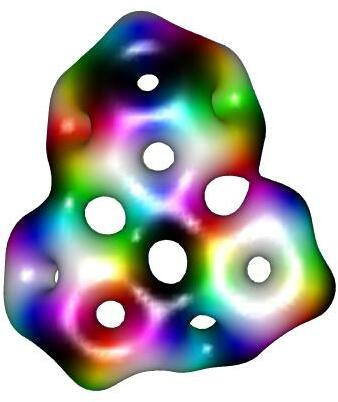}}&
  \raisebox{18pt}{(n)} & \raisebox{18pt}{2254.25}
  \end{tabular}
  \caption{The ten lightest $B=16$ Skyrmion solutions with $m=0.65$
    without Coulomb backreaction (left-hand side) and with the Coulomb
    backreaction taken into account (right-hand side). The solution
    labels refer to the solutions of ref.~\cite{Gudnason:2022jkn} and
    $\gamma$ refers to the new Skyrmion solution of
    fig.~\ref{fig:B16_withandwithoutEM}.
    In this sector, all the ten lightest Skyrmions are the same with
    and without the Coulomb backreaction taken into account.
  }
  \label{fig:B16tenlightest}
\end{figure}
We now conclude the $B=16$ sector by showing the ten lightest Skyrmion
solutions in our calibration (pion mass $m=0.65$) in
fig.~\ref{fig:B16tenlightest}, without taking the CBR into account
(left-hand side) and with it (right-hand side).
Unlike the case of the $B=12$ Skyrmions, the ten lightest $B=16$
Skyrmions are the same whether CBR is taken into account or not.
Interestingly though, there appears a new Skyrmion as $16_\gamma$ as
the 6th lightest Skyrmion in our calibration.
The ground state is however the same $16_a$ solution as in
ref.~\cite{Gudnason:2022jkn}.

For the $B=40$ Skyrmions, we seed the computations with 60 random
configurations using the algorithm of ref.~\cite{Gudnason:2022jkn} and
run the arrested Newton flow to a final solution with and without the
Coulomb energy backreacted to the Skyrme fields.
First some statistics.
We find that 15 of them come out equal.
5 of them do not give rise to solutions with $B=40$ without taking the
Coulomb effect into account, but do flow to solutions when its
backreaction is taken into account.
Of these 5 solutions, using the final solutions as input, in 1 case
the solution is equal when the Coulomb effect is turned off and in
another the solution changes; finally, in 3 cases the solution ceases
to exist (with baryon number 40).
The remaining 40 solution come out different; that is, starting with
the same random initial configuration of 1-Skyrmions placed randomly,
the arrested Newton flow algorithm find two \emph{different} solutions
when taking into account the backreaction of the Coulomb force and
when not taking it into account.
However, in 37 of these 40 cases, even though they flow to different
solutions, taking each of these final solutions as input, they both
exist when respectively turning off or switching on the Coulomb force.
In 2 cases, the solution ceases to exist (with baryon number 40) when
the Coulomb force is switched off.
Finally, in 1 single case, interestingly, the solution without the
Coulomb effect taken into account flows to the same solution from
that without it taken into account once the Coulomb force is turned
on.

This rather limited statistical sample shows that, although the
Coulomb effect is somewhat small, it does have physical importance for
the existence of certain Skyrmion states, but it appears that it is
more important for the \emph{dynamics} of nuclei.
We note, however, that the dynamics here is not quite the physical
dynamics, as the arrested Newton flow artificially removes the kinetic
energy to speed up the process of finding a minimum of the static
energy (although this removal of kinetic energy somehow crudely mimics
energy being carried away by radiation).

\begin{figure}[!htp]
  \centering\renewcommand{\arraystretch}{0.1}
  \begin{minipage}[t]{0.49\linewidth}
    \vspace{0pt}
\begin{tabular}{cc}
(b) 5573.86 & 5585.62 (a)\\[5pt]
\includegraphics[scale=0.2]{{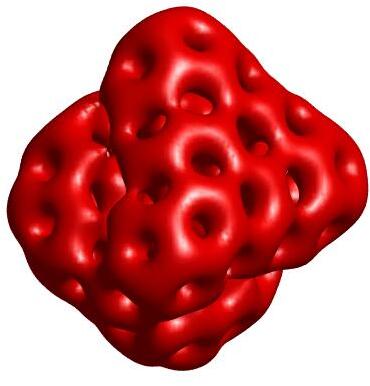}}&
\includegraphics[scale=0.23]{{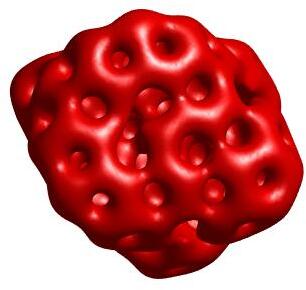}}\\
\includegraphics[scale=0.2]{{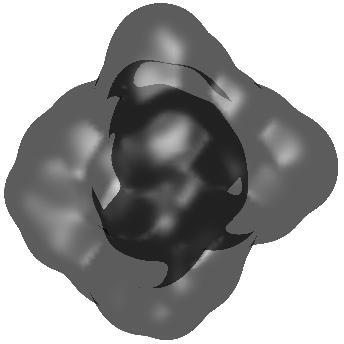}}&
\includegraphics[scale=0.23]{{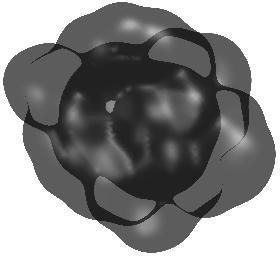}}\\
\includegraphics[scale=0.2]{{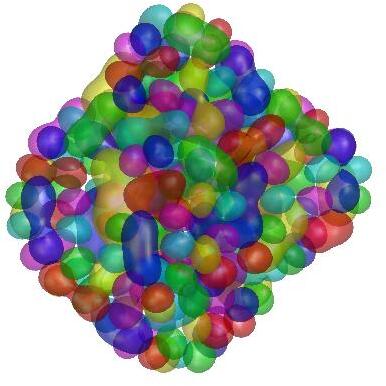}}&
\includegraphics[scale=0.23]{{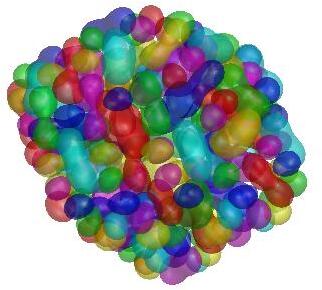}}\\[5pt]
$\kappa=0$ & $\kappa=0.737$
\end{tabular}
  \end{minipage}\hfill\vrule\hfill  
  \begin{minipage}[t]{0.49\linewidth}
    \vspace{0pt}
\begin{tabular}{cc}
(a) 5573.87 & 5585.65 (b)\\[5pt]
\includegraphics[scale=0.23]{{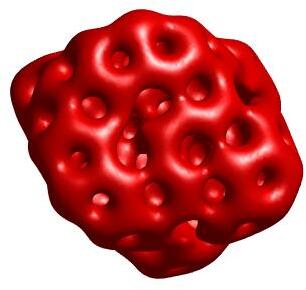}}&
\includegraphics[scale=0.2]{{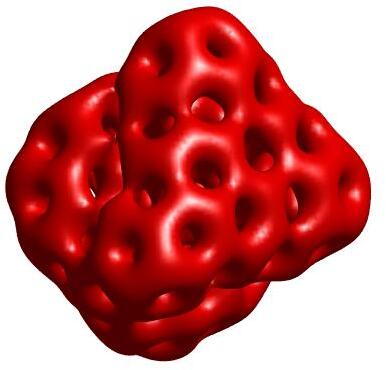}}\\
\includegraphics[scale=0.23]{{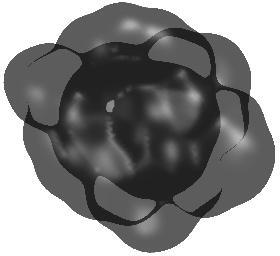}}&
\includegraphics[scale=0.2]{{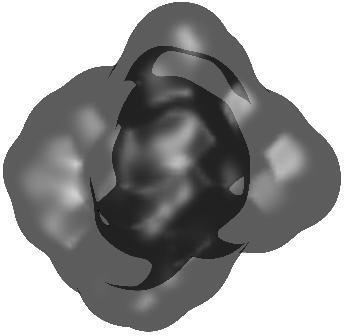}}\\
\includegraphics[scale=0.23]{{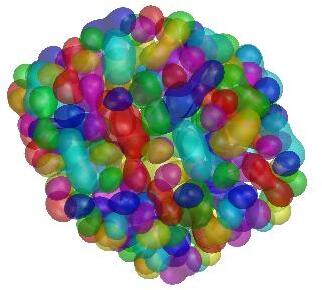}}&
\includegraphics[scale=0.2]{{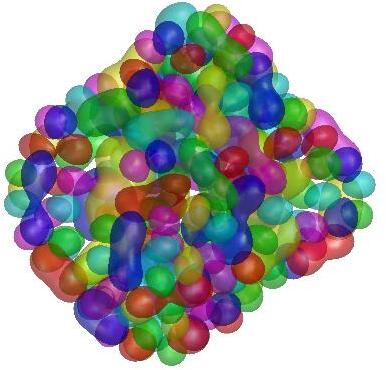}}\\[5pt]
$\kappa=0$ & $\kappa=0.737$
\end{tabular}
  \end{minipage}
\caption{The lightest (left) and next-to-lightest (right) $B=40$
  Skyrmions.
  The three rows displays the baryon charge
  isosurface at 1/4 of the maximum density, the $\sigma=\phi_0$
  isosurface at $\phi_0=0$ and the pion isosurfaces with
  $\phi_{1,2,3}=3/4$ ($\phi_{1,2,3}=-3/4$) corresponding to red,
  green, blue (cyan, magenta, yellow).
  The labels of $B=40$ Skyrmion solutions are ordered according to
  the energies of the Coulomb backreacted solutions (i.e.~those with
  $\kappa=0.737$).
}
\label{fig:1-2lightest}
\end{figure}

\begin{figure}[!htp]
  \centering\renewcommand{\arraystretch}{0.1}
  \begin{minipage}[t]{0.49\linewidth}
    \vspace{0pt}
\begin{tabular}{cc}
(c) 5575.73 & 5587.50 (c)\\[5pt]
\includegraphics[scale=0.2]{{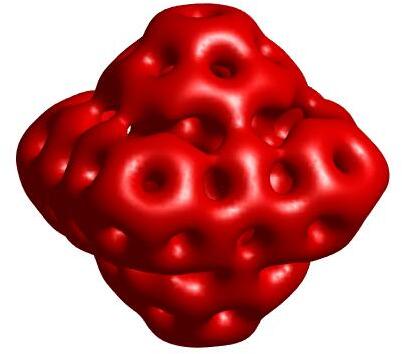}}&
\includegraphics[scale=0.2]{{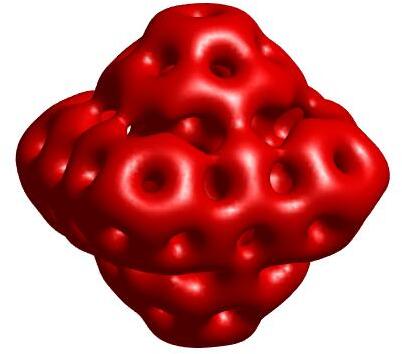}}\\
\includegraphics[scale=0.2]{{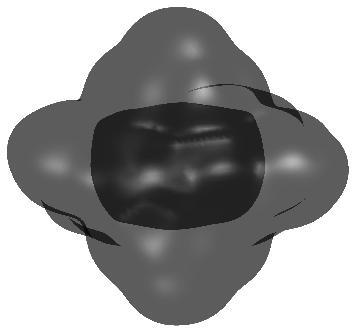}}&
\includegraphics[scale=0.2]{{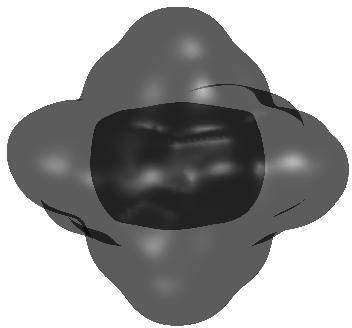}}\\
\includegraphics[scale=0.2]{{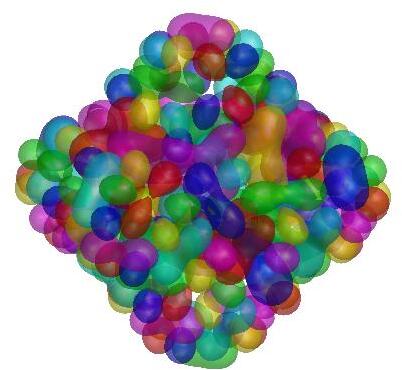}}&
\includegraphics[scale=0.2]{{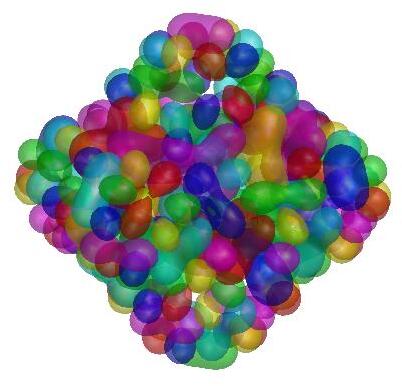}}\\[5pt]
$\kappa=0$ & $\kappa=0.737$
\end{tabular}
  \end{minipage}\hfill\vrule\hfill  
  \begin{minipage}[t]{0.49\linewidth}
    \vspace{0pt}
\begin{tabular}{cc}
(e) 5576.91 & 5588.68 (d)\\[5pt]
\includegraphics[scale=0.2]{{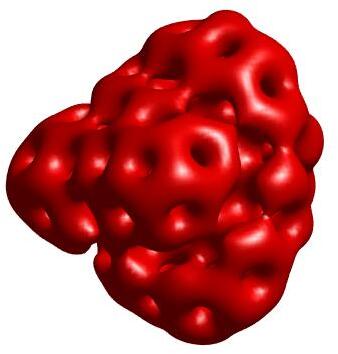}}&
\includegraphics[scale=0.2]{{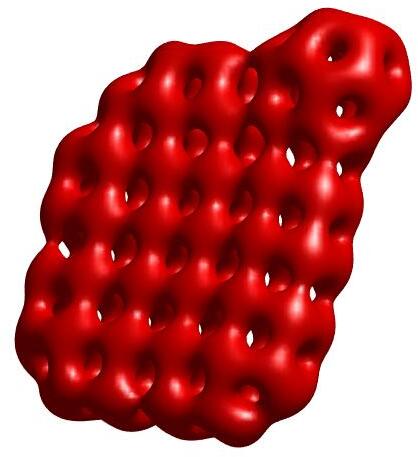}}\\
\includegraphics[scale=0.2]{{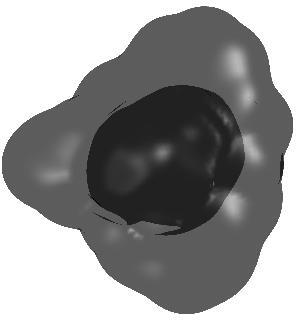}}&
\includegraphics[scale=0.2]{{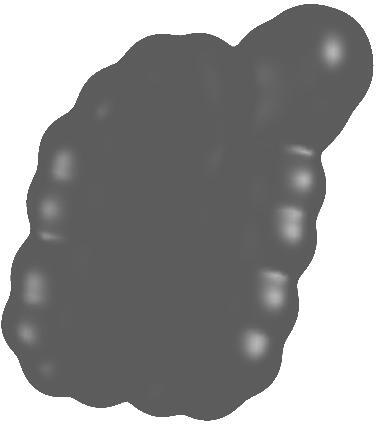}}\\
\includegraphics[scale=0.2]{{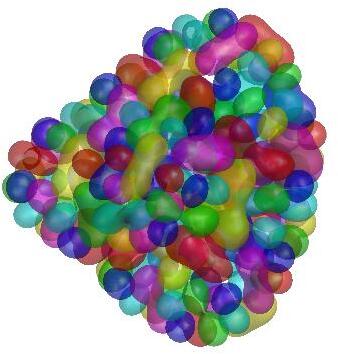}}&
\includegraphics[scale=0.2]{{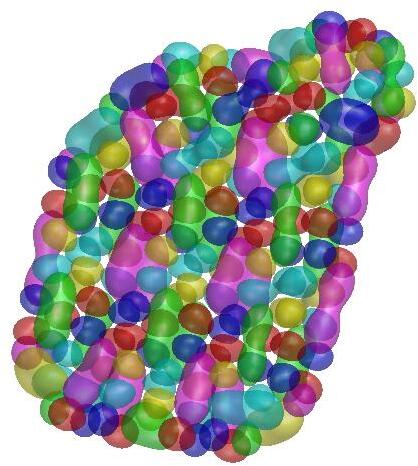}}\\[5pt]
$\kappa=0$ & $\kappa=0.737$
\end{tabular}
  \end{minipage}
\caption{The 3rd (left) and 4th (right) lightest $B=40$ Skyrmions.
  For details, see the caption of fig.~\ref{fig:1-2lightest}.}
\label{fig:3-4lightest}
\end{figure}

\begin{figure}[!htp]
  \centering\renewcommand{\arraystretch}{0.1}
  \begin{minipage}[t]{0.49\linewidth}
    \vspace{0pt}
\begin{tabular}{cc}
(f) 5576.99 & 5588.72 (e)\\[5pt]
\includegraphics[scale=0.22]{{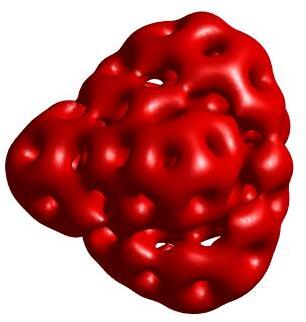}}&
\includegraphics[scale=0.2]{{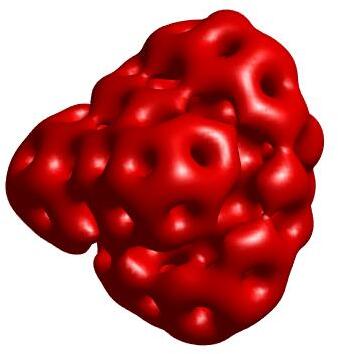}}\\
\includegraphics[scale=0.22]{{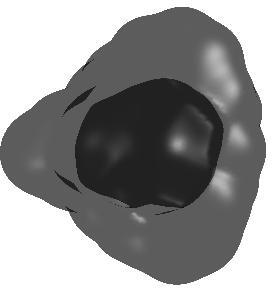}}&
\includegraphics[scale=0.2]{{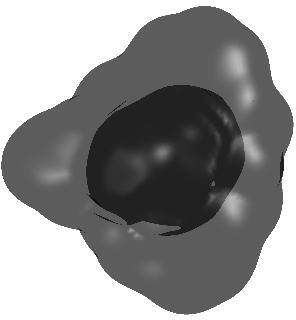}}\\
\includegraphics[scale=0.22]{{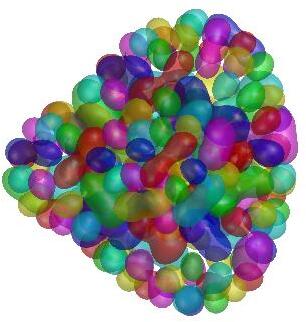}}&
\includegraphics[scale=0.2]{{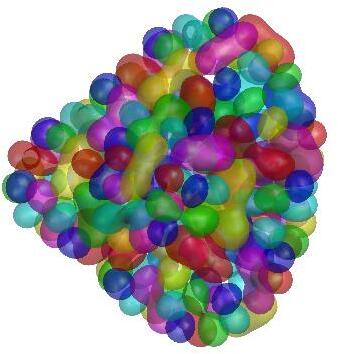}}\\[5pt]
$\kappa=0$ & $\kappa=0.737$
\end{tabular}
  \end{minipage}\hfill\vrule\hfill  
  \begin{minipage}[t]{0.49\linewidth}
    \vspace{0pt}
\begin{tabular}{cc}
(h) 5577.77 & 5588.80 (f)\\[5pt]
\includegraphics[scale=0.2]{{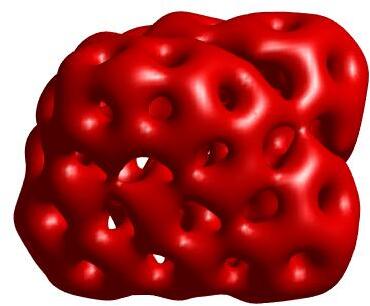}}&
\includegraphics[scale=0.22]{{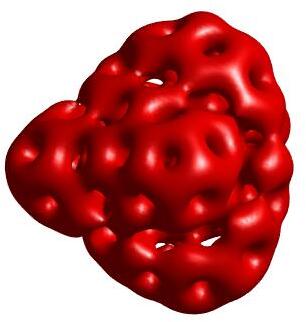}}\\
\includegraphics[scale=0.2]{{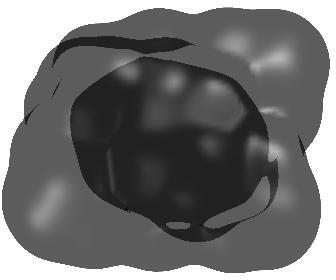}}&
\includegraphics[scale=0.22]{{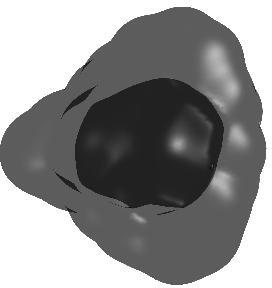}}\\
\includegraphics[scale=0.2]{{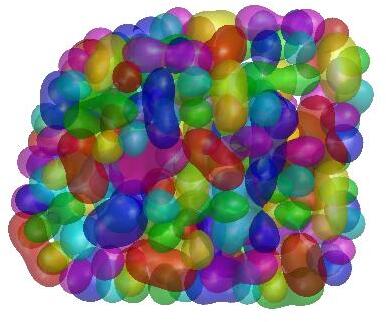}}&
\includegraphics[scale=0.22]{{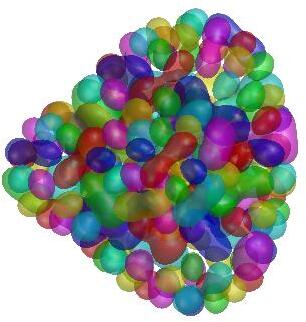}}\\[5pt]
$\kappa=0$ & $\kappa=0.737$
\end{tabular}
  \end{minipage}
\caption{The 5th (left) and 6th (right) lightest $B=40$ Skyrmions.
  For details, see the caption of fig.~\ref{fig:1-2lightest}.}
\label{fig:5-6lightest}
\end{figure}

\begin{figure}[!htp]
  \centering\renewcommand{\arraystretch}{0.1}
    \begin{minipage}[t]{0.49\linewidth}
    \vspace{0pt}
\begin{tabular}{cc}
(g) 5577.86 & 5589.56 (g)\\[5pt]
\includegraphics[scale=0.2]{{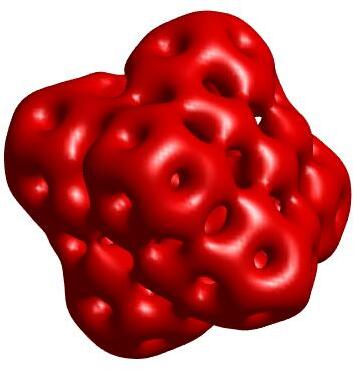}}&
\includegraphics[scale=0.2]{{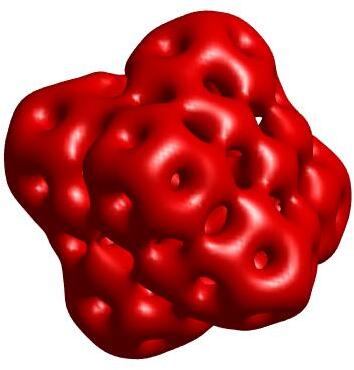}}\\
\includegraphics[scale=0.2]{{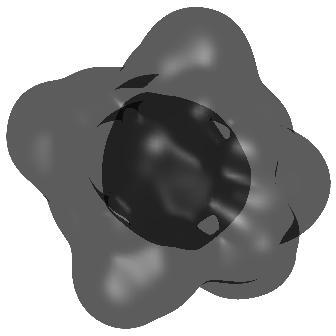}}&
\includegraphics[scale=0.2]{{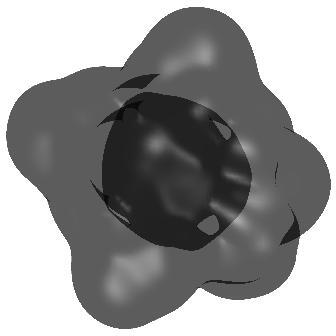}}\\
\includegraphics[scale=0.2]{{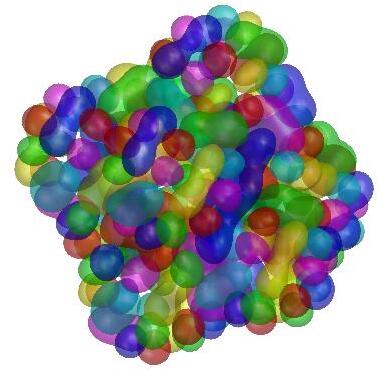}}&
\includegraphics[scale=0.2]{{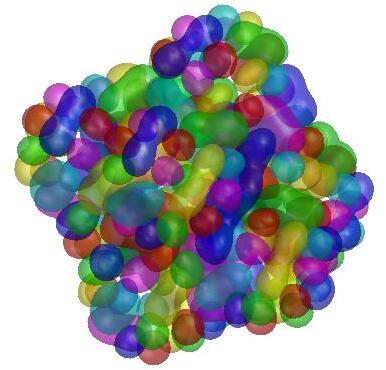}}\\[5pt]
$\kappa=0$ & $\kappa=0.737$
\end{tabular}
  \end{minipage}\hfill\vrule\hfill  
  \begin{minipage}[t]{0.49\linewidth}
    \vspace{0pt}
\begin{tabular}{cc}
(i) 5577.92 & 5589.57 (h)\\[5pt]
\includegraphics[scale=0.2]{{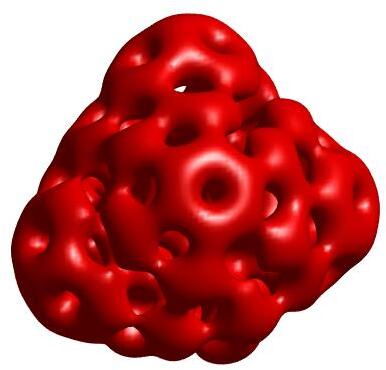}}&
\includegraphics[scale=0.2]{{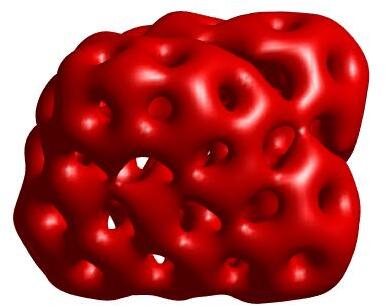}}\\
\includegraphics[scale=0.2]{{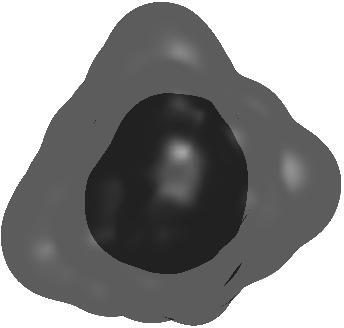}}&
\includegraphics[scale=0.2]{{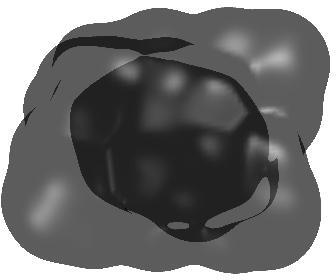}}\\
\includegraphics[scale=0.2]{{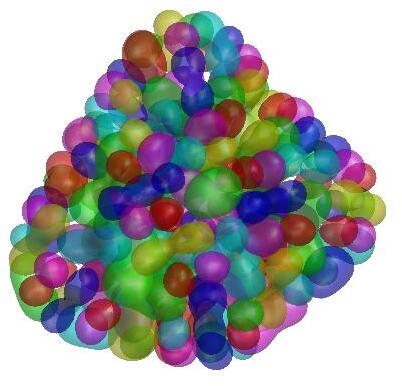}}&
\includegraphics[scale=0.2]{{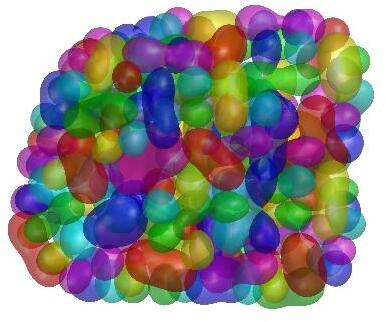}}\\[5pt]
$\kappa=0$ & $\kappa=0.737$
\end{tabular}
  \end{minipage}
\caption{The 7th (left) and 8th (right) lightest $B=40$ Skyrmions.
  For details, see the caption of fig.~\ref{fig:1-2lightest}.}
\label{fig:7-8lightest}
\end{figure}

\begin{figure}[!htp]
  \centering\renewcommand{\arraystretch}{0.1}
    \begin{minipage}[t]{0.49\linewidth}
    \vspace{0pt}
\begin{tabular}{cc}
(j) 5577.96 & 5589.75 (i)\\[5pt]
\includegraphics[scale=0.23]{{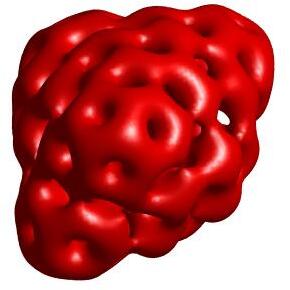}}&
\includegraphics[scale=0.2]{{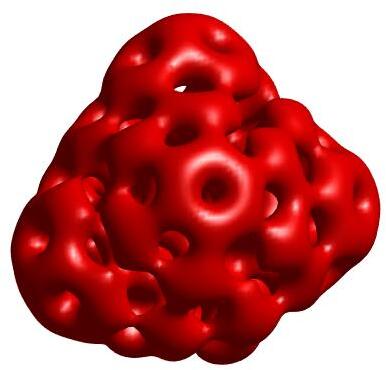}}\\
\includegraphics[scale=0.23]{{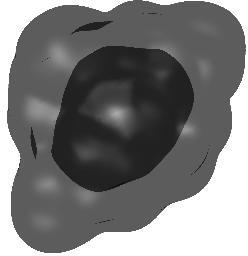}}&
\includegraphics[scale=0.2]{{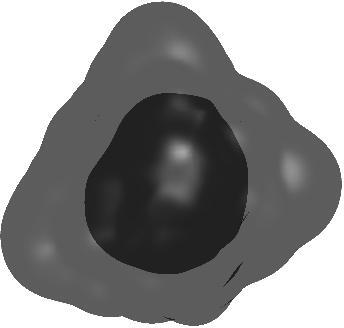}}\\
\includegraphics[scale=0.23]{{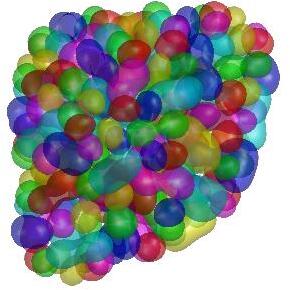}}&
\includegraphics[scale=0.2]{{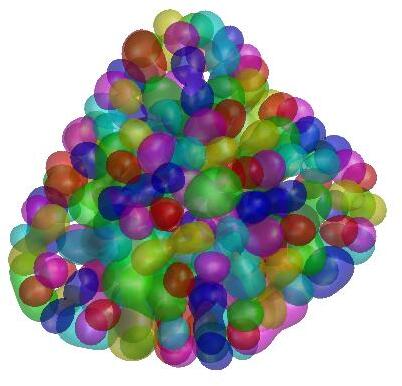}}\\[5pt]
$\kappa=0$ & $\kappa=0.737$
\end{tabular}
  \end{minipage}\hfill\vrule\hfill  
  \begin{minipage}[t]{0.49\linewidth}
    \vspace{0pt}
\begin{tabular}{cc}
(k) 5578.03 & 5589.81 (j)\\[5pt]
\includegraphics[scale=0.22]{{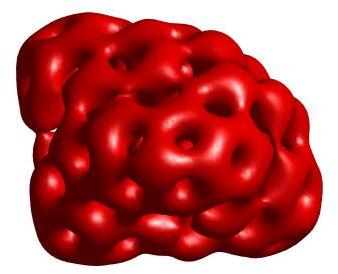}}&
\includegraphics[scale=0.22]{{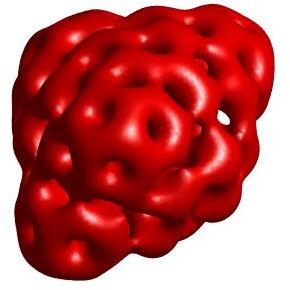}}\\
\includegraphics[scale=0.22]{{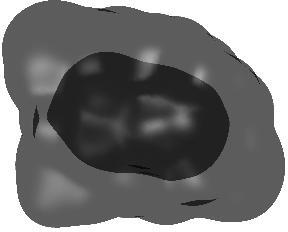}}&
\includegraphics[scale=0.22]{{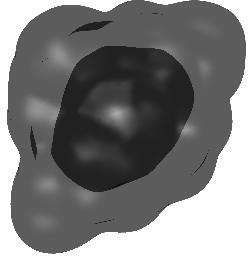}}\\
\includegraphics[scale=0.22]{{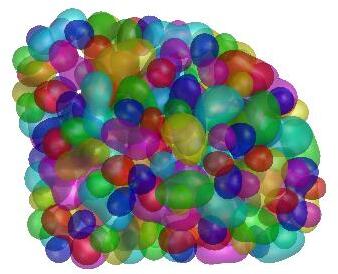}}&
\includegraphics[scale=0.22]{{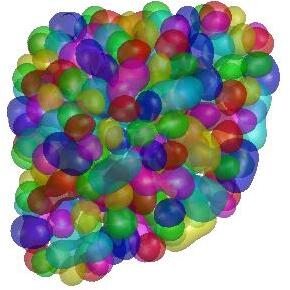}}\\[5pt]
$\kappa=0$ & $\kappa=0.737$
\end{tabular}
  \end{minipage}
\caption{The 9th (left) and 10th (right) lightest $B=40$ Skyrmions.
  For details, see the caption of fig.~\ref{fig:1-2lightest}.}    
\label{fig:9-10lightest}
\end{figure}

We start by comparing the ten lightest $B=40$ Skyrmion solutions in
our calibration, with and without the CBR taken into account.
The figs.~\ref{fig:1-2lightest}-\ref{fig:9-10lightest} show the $n$-th
lightest solution ($n=1,\ldots,10$) with the CBR turned off on the
left-hand side and taken into account on the right-hand side.
The three rows correspond to the baryon-charge isosurface, the
$\phi_0=0$ or $\sigma=0$ isosurface which is the midpoint between the
vacuum and the anti-vacuum, and finally the pion clouds at
$\phi_{1,2,3}=3/4$ shown by red, green and blue and
$\phi_{1,2,3}=-3/4$ shown by the corresponding anti-colours.
Although all found solutions have been tested to exist both with and
without CBR taken into account, it is a quite nice result that for
such large Skyrmions as $B=40$ like the calcium nucleus, the ordering
of the lowest-energy states differ when CBR is taken into account or
not.
We expected this on physical grounds, since it is the largest stable
isospin-0 ground state with $B=4n$ and hence for baryon numbers higher
than 40, we expect that the amount of protons is too large to give
rise to a stable ground state and hence the isospin must be
nonvanishing (yielding more neutrons than protons).
The only reason for the number of neutrons and the number of protons
being the same, is due to a fact of the strong interactions, i.e.~the
symmetry energy favours the isospin-0 states, whereas the Coulomb force
obviously prefers more neutrons.
Clearly a lot of the same solutions appear among the ten lightest
$B=40$ Skyrmions.
The two lightest states switch their order when the CBR is turned on,
see fig.~\ref{fig:1-2lightest}.
The 3rd lightest state is the same with and without CBR, see
fig.~\ref{fig:3-4lightest}(left).
The 4th lightest state only appears in top-10 with Coulomb
interactions turned on, see fig.~\ref{fig:3-4lightest}.
Apart from that, the order is the same -- with eliminating the flat
$40_d$-Skyrmion from the CBR-off list, until the 7th-lightest state:
This happens to be the same with and without CBR, see
fig.~\ref{fig:7-8lightest}.
Essentially the $40_g$-Skyrmion and the $40_h$-Skyrmion switch order
when the CBR switched off.
We also note that with the exception of the $40_d$ (the 4th lightest
solution), all the lightest Skyrmion solutions have a hollow $\phi_0$
structure: They basically have a ``hole'' or a less dense region inside
the nucleus that is covered by a shell.
It is completely different from the fullerene-type solutions of
refs.~\cite{Battye:1997qq,Houghton:1997kg,Battye:2000se,Battye:2001qn}
since the ``hole'' is not the anti-vacuum ($\phi_0=-1$),
but approximately the true vacuum ($\phi_0=1$), see
figs.~\ref{fig:1-2lightest}-\ref{fig:9-10lightest}.
The fourth-lightest $B=40$ Skyrmion ($40_d$) is different, as it is
planar in the sense of being of the ``graphene''-type solution, see
ref.~\cite{Gudnason:2022jkn}.

\begin{figure}[!htp]
  \centering\renewcommand{\arraystretch}{0.1}
  \begin{minipage}[t]{0.49\linewidth}
    \vspace{0pt}
\begin{tabular}{cc}
  (bd) $\begin{array}{r@{.}l}5584&71\\[2pt](0&)\end{array}$ &
  $\begin{array}{r@{.}l}5593&08\\[2pt](5596&28)\end{array}$ (aj)\\[5pt]
\includegraphics[scale=0.23]{{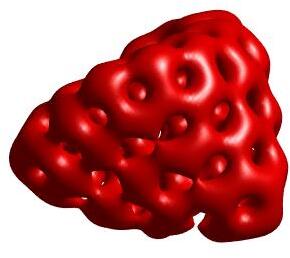}}&
\includegraphics[scale=0.23]{{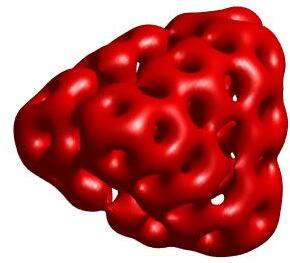}}\\
\includegraphics[scale=0.23]{{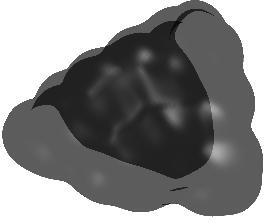}}&
\includegraphics[scale=0.23]{{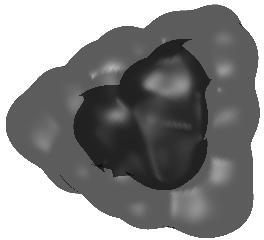}}\\
\includegraphics[scale=0.23]{{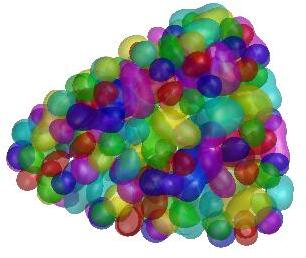}}&
\includegraphics[scale=0.23]{{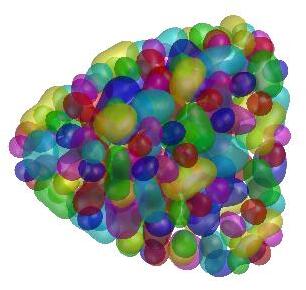}}\\[5pt]
$\kappa=0$ & $\kappa=0.737$
\end{tabular}
  \end{minipage}\hfill\vrule\hfill
  \begin{minipage}[t]{0.49\linewidth}
    \vspace{0pt}
\begin{tabular}{cc}
  (cq) $\begin{array}{r@{.}l}5598&93\\[2pt](0&)\end{array}$ &
  $\begin{array}{r@{.}l}5602&49\\(5610&34)\end{array}$ (ca)\\[5pt]
  \includegraphics[scale=0.2]{{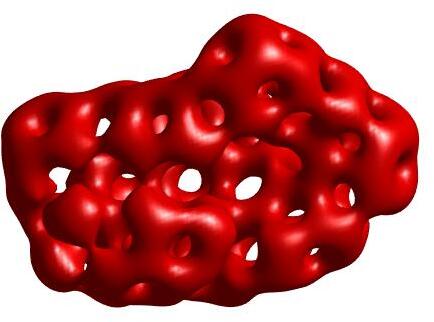}}&
\includegraphics[scale=0.2]{{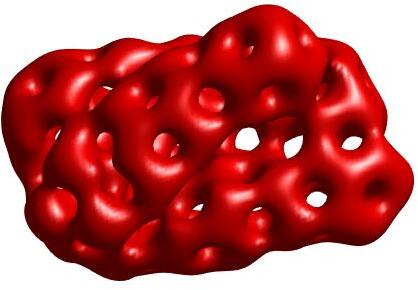}}\\
\includegraphics[scale=0.2]{{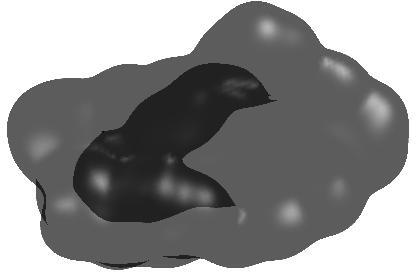}}&
\includegraphics[scale=0.2]{{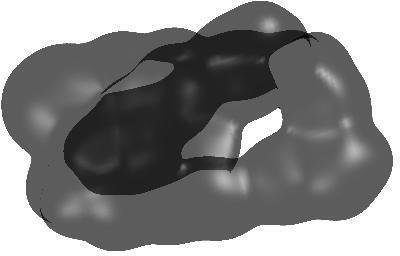}}\\
\includegraphics[scale=0.2]{{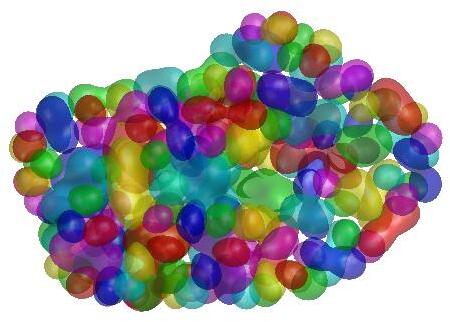}}&
\includegraphics[scale=0.2]{{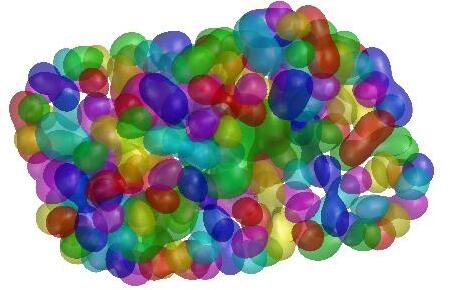}}\\[5pt]
$\kappa=0$ & $\kappa=0.737$
\end{tabular}
  \end{minipage}
\caption{Two cases of a $B=40$ Skyrmion that flows from an initial
  condition to two different solutions with and without CBR: 
  Turning on the Coulomb effect on the $\kappa=0$ solution changes it
  only slightly (it looks like the $\kappa=0$ figure and its energy is
  written in the $\kappa=0.737$ column in parentheses).
  Turning off the CBR on the $\kappa=0.737$ solution, however yields
  no stable $B=40$ Skyrmion, in both cases.
  For details, see the caption of fig.~\ref{fig:1-2lightest}.
}
\label{fig:conf20+27}
\end{figure}
We will now discuss a few selected Skyrmion configurations that
differ from each other when CBR is taken into account or not.
First in fig.~\ref{fig:conf20+27} are shown two cases where the
initial configuration (not shown) flows to two different solutions.
Taking the two end results, the $\kappa=0$ solution only changes
slightly by taking into account CBR, but the $\kappa=0.737$ solution
disappears when switching off Coulomb interactions.

\begin{figure}[!htp]
  \centering\renewcommand{\arraystretch}{0.1}
  \begin{floatrow}
    \ffigbox{
\begin{tabular}{cc}
  (bv) $\begin{array}{r@{.}l}5593&16\\(5590&28)\end{array}$ &
  $\begin{array}{r@{.}l}5601&46\\(5601&46)\end{array}$ (bw)\\[5pt]
  \includegraphics[scale=0.2]{{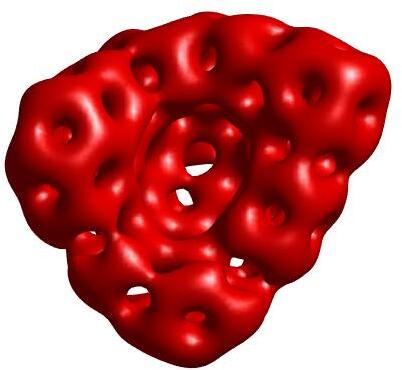}}&
  \includegraphics[scale=0.2]{{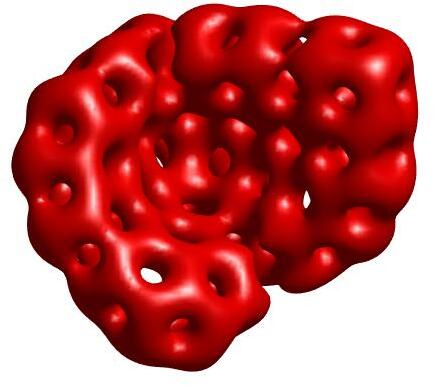}}\\
  \includegraphics[scale=0.2]{{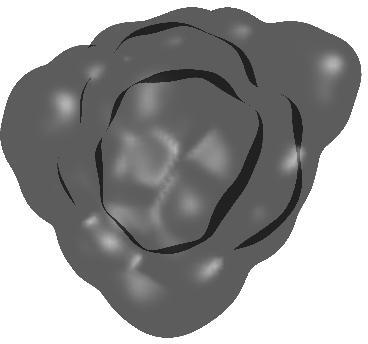}}&
  \includegraphics[scale=0.2]{{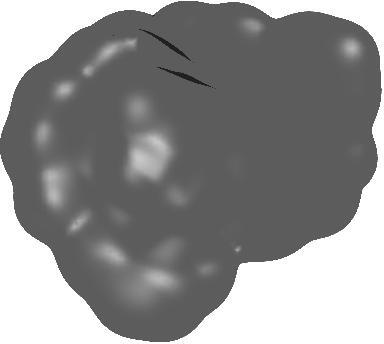}}\\
  \includegraphics[scale=0.2]{{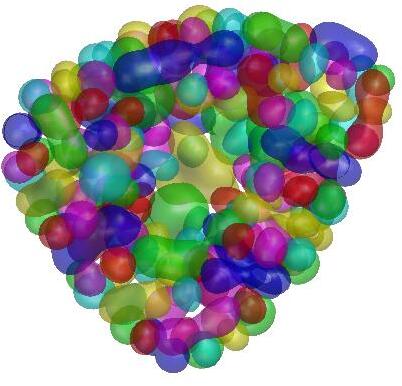}}&
  \includegraphics[scale=0.2]{{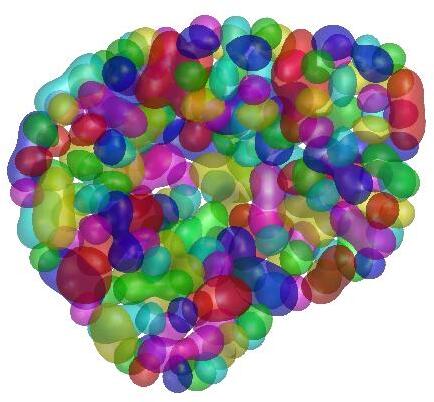}}\\[5pt]
  $\kappa=0$ & $\kappa=0.737$
\end{tabular}}{
\caption{A $B=40$ Skyrmion that flows from an initial condition to two
  different solutions with and without CBR: 
  Interestingly, turning on the Coulomb effect on the $\kappa=0$
  solution flows it to the same solution as the initial solution
  flowed to.
  Turning off the Coulomb force also does not change the
  $\kappa=0.737$ solution.
}
\label{fig:conf4}}
    \ffigbox{
\begin{tabular}{cc}
  $\begin{array}{r@{.}l}0&\\(5579&19)\end{array}$ &
  $\begin{array}{r@{.}l}5590&19\\(0&)\end{array}$ (m)\\[5pt]
  \includegraphics[scale=0.19]{{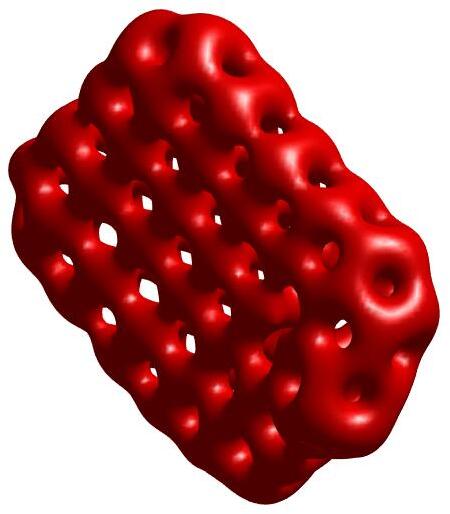}}&
  \includegraphics[scale=0.19]{{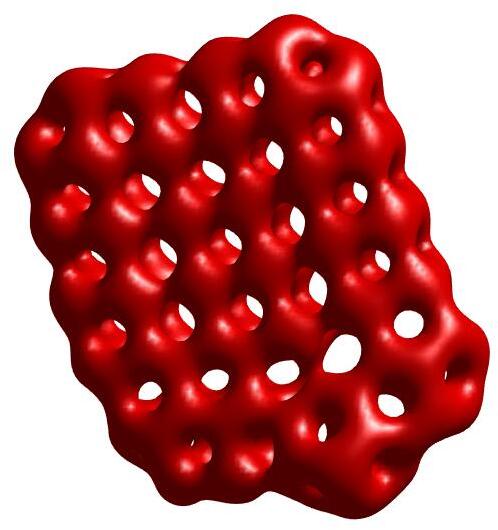}}\\
  \includegraphics[scale=0.19]{{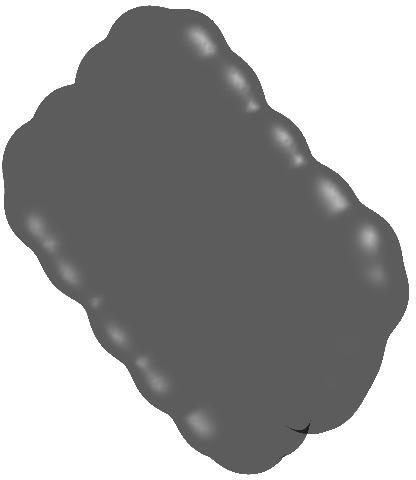}}&
  \includegraphics[scale=0.19]{{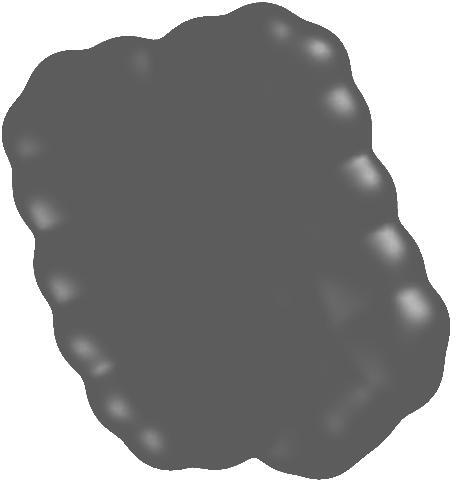}}\\
  \includegraphics[scale=0.19]{{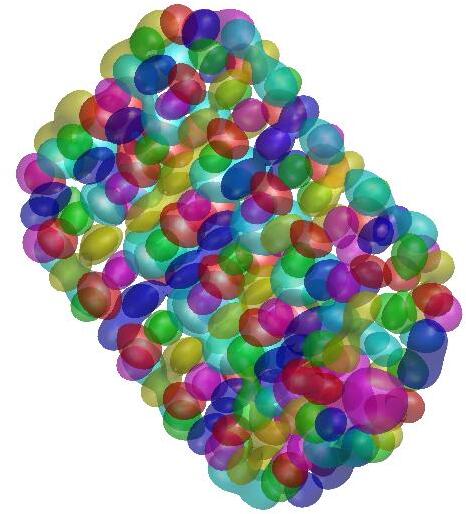}}&
  \includegraphics[scale=0.19]{{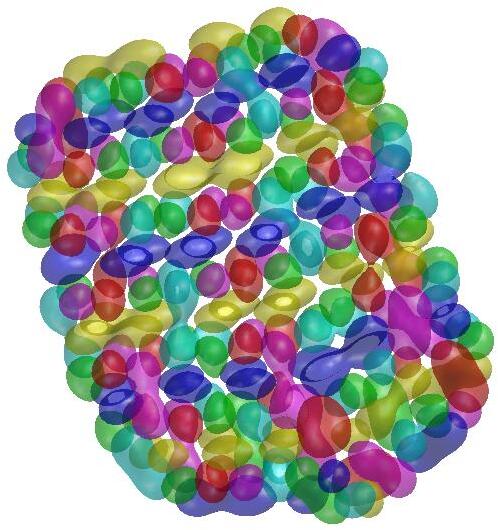}}\\[5pt]
  $\kappa=0$ & $\kappa=0.737$
\end{tabular}}{
\caption{A $B=40$ Skyrmion that only flows to a $B=40$ solution with
  the Coulomb effect turned on. However, switching it off flows this
  solution to a slightly different solution (shown on the left).
  For details, see the caption of fig.~\ref{fig:1-2lightest}.
}
\label{fig:conf12}}
    \end{floatrow}
\end{figure}
Another interesting case is shown in fig.~\ref{fig:conf4}, where the
configuration flows to two different solution depending on whether CBR
is taken into account or not. Interestingly, taking the $\kappa=0$
solution and turning on the Coulomb interaction, the solution flows to
the \emph{same} $\kappa=0.737$ solution as it flowed to from the
initial condition.
Turning off the Coulomb interaction, however, just yields a slightly
deformed version of the $\kappa=0.737$ solution.

Another odd case is shown in fig.~\ref{fig:conf12}, where the solution
does not exist when CBR is turned off during the flow from the initial
configuration.
Turning off the Coulomb interaction, however, yields a somewhat
deformed version of the $\kappa=0.737$ solution.

\begin{figure}[!htp]
\centering\renewcommand{\arraystretch}{0.1}
\begin{tabular}{lcc}
\includegraphics[scale=0.2]{{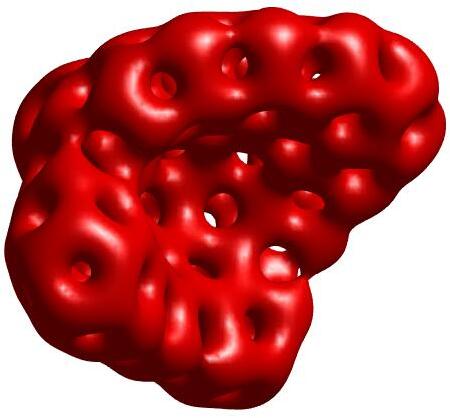}}&
\includegraphics[scale=0.2]{{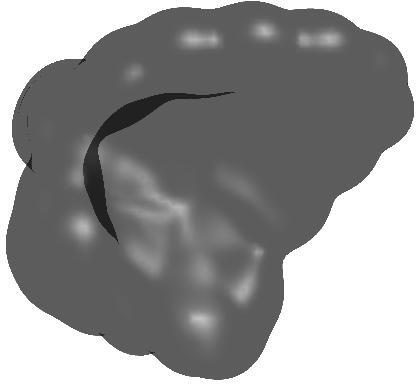}}&
\includegraphics[scale=0.2]{{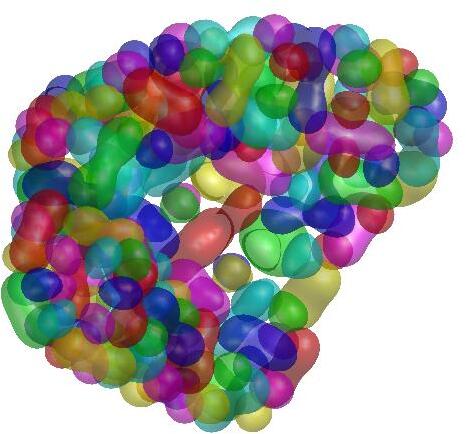}}\\
(bh) 5596.62 (5585.44)
\end{tabular}
\caption{A $B=40$ Skyrmion that only flows to a solution with the
  Coulomb effect turned on.
  Switching off the Coulomb effect, however, yields the same solution
  (energy shown in parentheses).
  For details, see the caption of fig.~\ref{fig:1-2lightest}.
} 
\label{fig:conf13}
\end{figure}
Again a situation that only yields a $B=40$ solution from a certain
initial condition when CBR is taken into account, is shown in
fig.~\ref{fig:conf13}.
In this case, the solution does not change when the Coulomb
interaction is turned off (but the total energy decreases slightly,
of course, which is shown in parenthesis in the figure). 

\begin{figure}[!htp]
\centering\renewcommand{\arraystretch}{0.1}
\begin{tabular}{lcc}
\includegraphics[scale=0.2]{{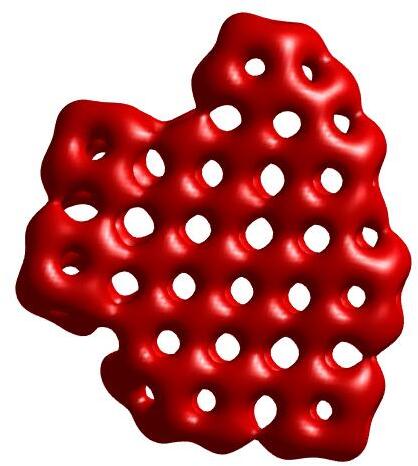}}&
\includegraphics[scale=0.2]{{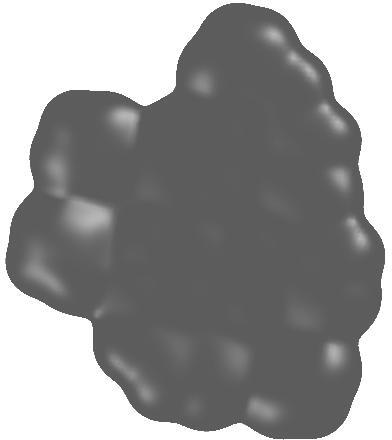}}&
\includegraphics[scale=0.2]{{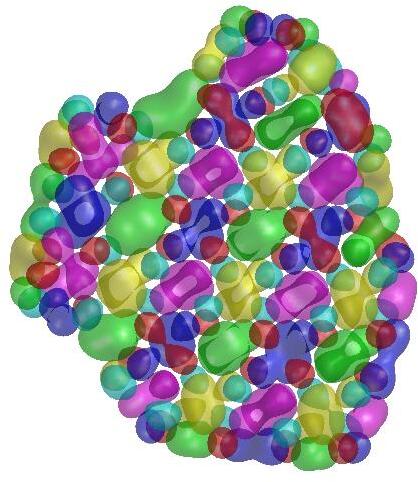}}\\
(at) 5594.57\\
\includegraphics[scale=0.2]{{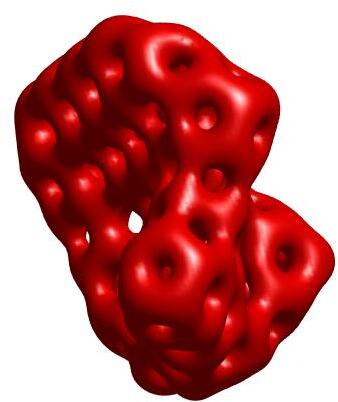}}&
\includegraphics[scale=0.2]{{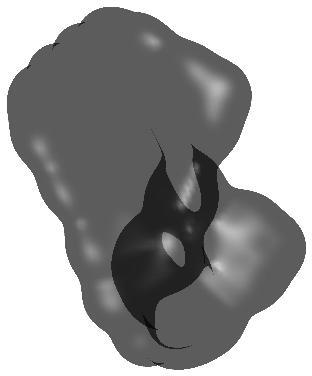}}&
\includegraphics[scale=0.2]{{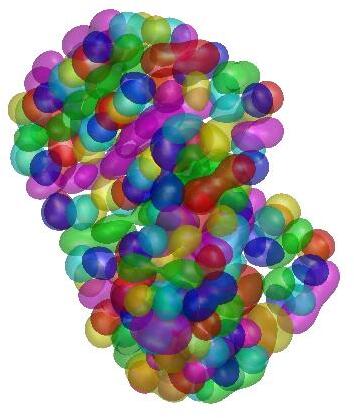}}\\
(ck) 5605.88\\
\includegraphics[scale=0.2]{{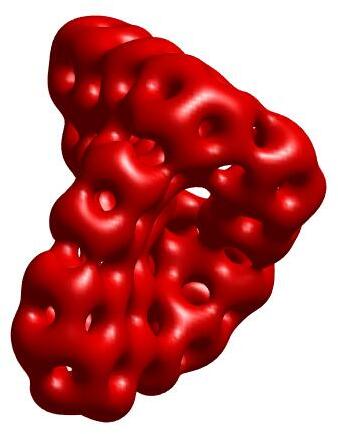}}&
\includegraphics[scale=0.2]{{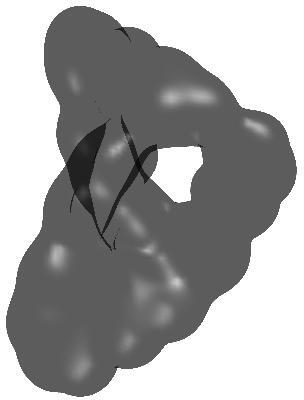}}&
\includegraphics[scale=0.2]{{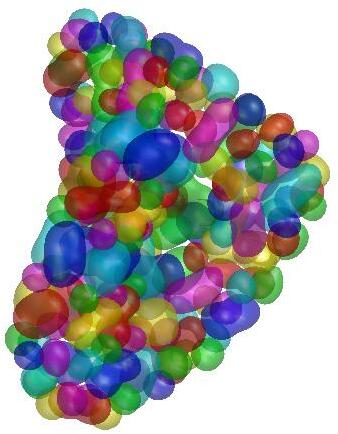}}\\
(co) 5609.94
\end{tabular}
\caption{$B=40$ Skyrmions that only flow to a solution with the
  Coulomb effect turned on.
  Switching off the Coulomb effect also yields no solution with
  $B=40$.
  For details, see the caption of fig.~\ref{fig:1-2lightest}.
} 
\label{fig:conf_kappaonly_noflowdown}
\end{figure}

The final three odd cases are shown in
fig.~\ref{fig:conf_kappaonly_noflowdown}; again the solution only
exists when CBR is taken into account and more interestingly, taking
these final solutions as inputs and switching off the Coulomb
interaction gives no stable solution with baryon number $B=40$.

For completeness, we show the remaining 79 $B=40$ solutions with
higher energies in appendix \ref{app:moreB40s}.

\subsection{The Coulomb energy and the effect of its backreaction}\label{sec:backreaction}

We are finally in a position to summarize the Coulomb energy for a
broad range of Skyrmions with baryon numbers 4, 8, 12, 16 and 40 as
well as study the detailed effects of the backreaction of the Coulomb
force onto the Skyrmions.
We begin with the latter.

\begin{figure}[!htp]
  \centering
  \mbox{
  \subfloat[]{\includegraphics[scale=0.3]{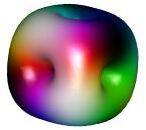}}
  \subfloat[]{\includegraphics[scale=0.3]{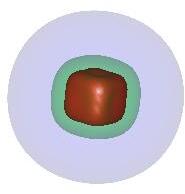}}
  \subfloat[]{\includegraphics[scale=0.3]{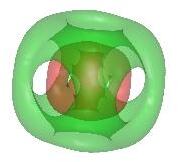}}
  \subfloat[]{\includegraphics[scale=0.3]{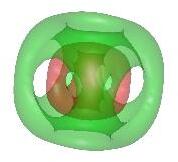}}}
  \caption{The only $B=4$ Skyrmion, corresponding to Helium-4:
    (a) displays the baryon charge isosurface at 1/4 of its
    maximum density and (b) the Coulomb potential $V$ at 98\% (red), 90\%
    (green) and 50\% (purple) of its maximum density. The last two
    panels show the Skyrmion (c) baryon charge density and (d) energy
    density for the backreacted solution with the non-backreacted
    solution subtracted off.
    The green isosurface is showing the positive region at
    half-maximum density and the red shows the negative region at
    half-maximum (negative) density.
    The fact that these differences are positive at larger radii
    negative at smaller radii means that the backreacted Skyrmion
    solution is \emph{larger} than the non-backreacted one.
    The levelsets are at (a) $0.074$, (c) green $4.8\times10^{-5}$,
    (c) red $-2.0\times10^{-4}$, (d) green $6.4\times10^{-3}$, (d) red
    $-0.034$ and the maximum energy density is for
    comparison $43.9$.
  }
  \label{fig:B4diff}
\end{figure}

\begin{figure}[!htp]
  \centering
  \mbox{
  \subfloat[]{\includegraphics[scale=0.3]{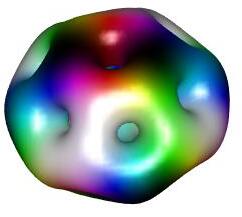}}
  \subfloat[]{\includegraphics[scale=0.3]{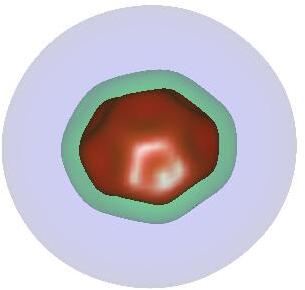}}
  \subfloat[]{\includegraphics[scale=0.3]{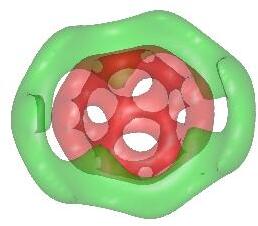}}
  \subfloat[]{\includegraphics[scale=0.3]{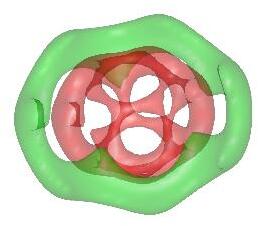}}}
  \caption{The lightest $B=8$ Skyrmion in our calibration ($m=0.65$
    and $\kappa=0.737$), corresponding to Beryllium-8.
    For details, see the caption of fig.~\ref{fig:B4diff}.
    The levelsets are at (a) $0.072$, (c) green $1.4\times10^{-4}$,
    (c) red $-3.6\times10^{-4}$, (d) green $0.021$, (d) red
    $-0.061$ and the maximum energy density is for
    comparison $43.7$. 
  }
  \label{fig:B8diff}
\end{figure}

\begin{figure}[!htp]
  \centering
  \mbox{
  \subfloat[]{\includegraphics[scale=0.25]{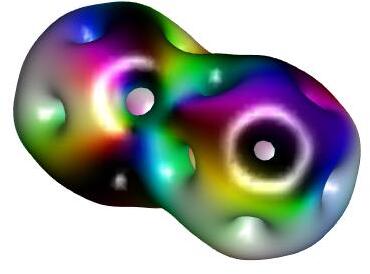}}
  \subfloat[]{\includegraphics[scale=0.25]{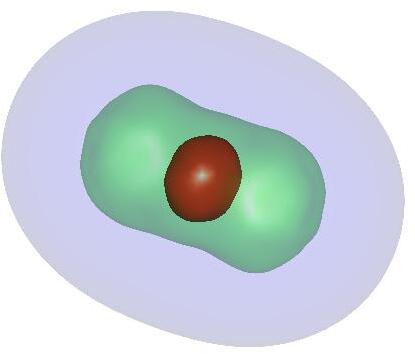}}
  \subfloat[]{\includegraphics[scale=0.25]{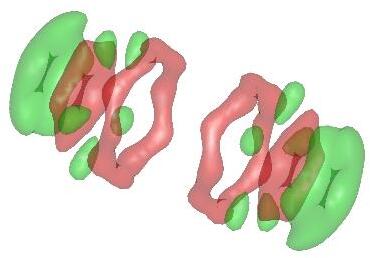}}
  \subfloat[]{\includegraphics[scale=0.25]{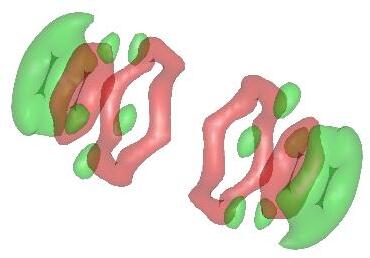}}}
  \caption{The lightest $B=12$ Skyrmion in our calibration ($m=0.65$
    and $\kappa=0.737$), corresponding to Carbon-12.
    For details, see the caption of fig.~\ref{fig:B4diff}.
    The levelsets are at (a) $0.072$, (c) green $8.4\times10^{-4}$,
    (c) red $-1.0\times10^{-3}$, (d) green $0.12$, (d) red
    $-0.16$ and the maximum energy density is for
    comparison $42.6$. 
  }
  \label{fig:B12diff}
\end{figure}

\begin{figure}[!htp]
  \centering
  \mbox{
  \subfloat[]{\includegraphics[scale=0.25]{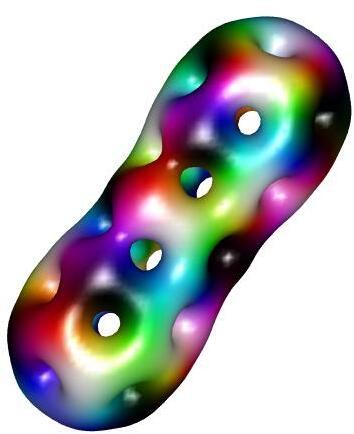}}
  \subfloat[]{\includegraphics[scale=0.25]{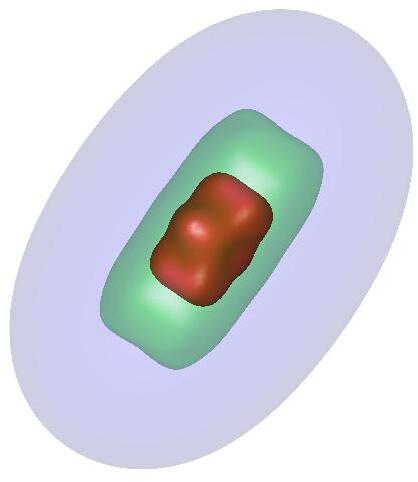}}
  \subfloat[]{\includegraphics[scale=0.25]{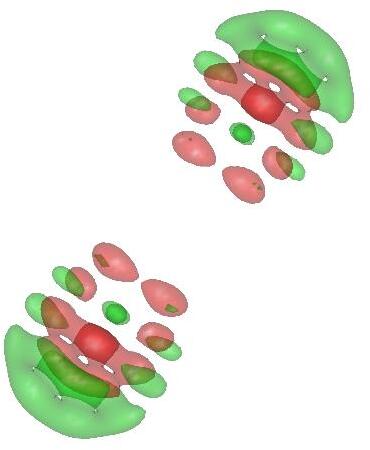}}
  \subfloat[]{\includegraphics[scale=0.25]{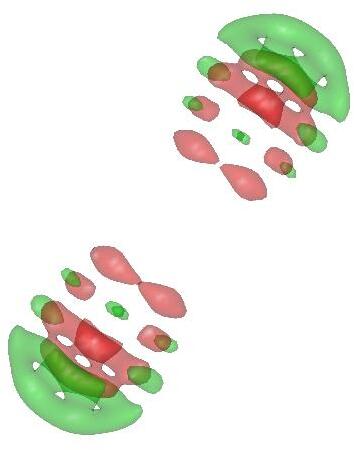}}}
  \caption{The lightest $B=16$ Skyrmion in our calibration ($m=0.65$
    and $\kappa=0.737$), corresponding to Oxygen-16.
    For details, see the caption of fig.~\ref{fig:B4diff}.
    The levelsets are at (a) $0.084$, (c) green $6.8\times10^{-4}$,
    (c) red $-1.0\times10^{-3}$, (d) green $0.10$, (d) red
    $-0.16$ and the maximum energy density is for
    comparison $50.1$. 
  }
  \label{fig:B16diff}
\end{figure}

\begin{figure}[!htp]
  \centering
  \mbox{
  \subfloat[]{\includegraphics[scale=0.3]{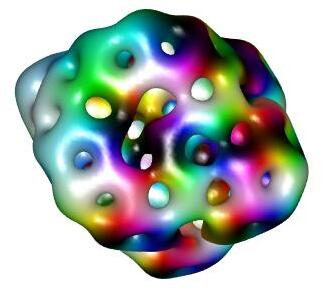}}
  \subfloat[]{\includegraphics[scale=0.3]{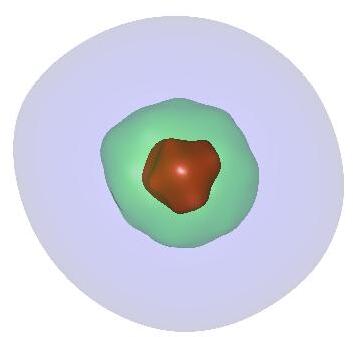}}
  \subfloat[]{\includegraphics[scale=0.3]{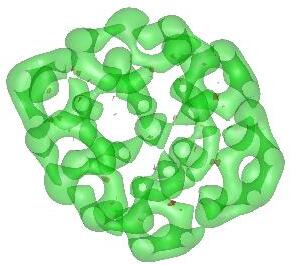}}
  \subfloat[]{\includegraphics[scale=0.3]{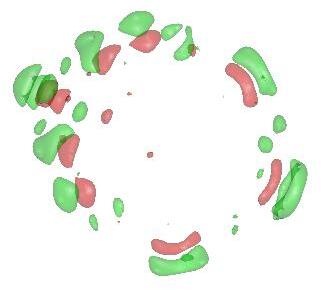}}}
  \caption{The lightest $B=40$ Skyrmion in our calibration ($m=0.65$
    and $\kappa=0.737$), corresponding to Calcium-40.
    For details, see the caption of fig.~\ref{fig:B4diff}.
    The levelsets are at (a) $0.084$, (c) green $0.34$,
    (c) red $-2.5\times10^{-3}$, (d) green $0.31$, (d) red
    $-0.42$ and the maximum energy density is for
    comparison $50.4$. 
  }
  \label{fig:B40diff}
\end{figure}
Figs.~\ref{fig:B4diff} through \ref{fig:B40diff} show the (a) baryon
charge density, (b) Coulomb (electric) potential, (c) [(d)] difference
between the baryon charge density [energy density] with backreaction
of the Coulomb energy and without it.
Notice that the red surface (in panel (b)), corresponding to nearly the maximum
density of the electric potential ($V$) has a distinct shape, whereas
the half-maximum density (purple) isosurface interpolates between the
red surface's shape and a sphere.
It is interesting to see in what way the Skyrmions are altered by the
backreaction of the Coulomb energy.
In the $B=4$ Skyrmion, the backreaction amounts to increasing the cube
isotropically, see fig.~\ref{fig:B4diff}.
This is consistent with the fact that the Coulomb interaction acts
similarly to the Skyrme term, but with a much smaller effect due to
the smallness of the electromagnetic coupling (charge).
For the $B=8$ Skyrmion, the solution increases not isotropically but
more in the direction of the rim, see fig.~\ref{fig:B8diff}.
In the larger $B=12$ and $B=16$ Skyrmions, the increase of the
Skyrmions' sizes takes place along the axis aligned with the
Skyrmion's longest length.
Finally, the $B=40$ Skyrmion also grows, but in this case it seems
more complicated than with the simpler Skyrmions.

\begin{table}[!htp]
  \centering
  {\small
    \begin{tabular}{l|llll}
      \hline\hline
&$B=4$&Helium-4&abs. diff.&rel. diff.\\\hline
$E$ &	575.41\\
$E$ &	$3.797\GeV$ &	$3.727\GeV$ &	$69.5\MeV$ &	1.86\%\\
$E_C$ &	$1.919\MeV$ &	$1.575\MeV$ &	$0.3\MeV$ &	21.87\%\\
$R$ &	$1.422\fm$ &	$1.678\fm$ &	$-0.256\fm$ &	-15.24\%\\
\hline\hline
&$B=8$&Beryllium-8&abs. diff.&rel. diff.\\\hline
$E$ &	1133.65\\
$E$ &	$7.481\GeV$ &	$7.455\GeV$ &	$25.7\MeV$ &	0.34\%\\
$E_C$ &	$5.713\MeV$ &	$5.000\MeV$ &	$0.7\MeV$ &	14.26\%\\
$R$ &	$1.908\fm$ &	-- &	-- &	--\\
\hline\hline
&$B=12$&Carbon-12&abs. diff.&rel. diff.\\\hline
$E$ &	1693.85\\
$E$ &	$11.177\GeV$ &	$11.177\GeV$ &	$0.0\MeV$ &	0.00\%\\
$E_C$ &	$10.591\MeV$ &	$9.828\MeV$ &	$0.8\MeV$ &	7.77\%\\
$R$ &	$2.471\fm$ &	$2.470\fm$ &	$0.001\fm$ &	0.04\%\\
\hline\hline
&$B=16$&Oxygen-16&abs. diff.&rel. diff.\\\hline
$E$ &	2250.87\\
$E$ &	$14.853\GeV$ &	$14.895\GeV$ &	$-42.5\MeV$ &	-0.29\%\\
$E_C$ &	$16.389\MeV$ &	$15.874\MeV$ &	$0.5\MeV$ &	3.25\%\\
$R$ &	$3.002\fm$ &	$2.699\fm$ &	$0.303\fm$ &	11.23\%\\
\hline\hline
&$B=40$&Calcium-40&abs. diff.&rel. diff.\\\hline
$E$ &	5585.62\\
$E$ &	$36.858\GeV$ &	$37.215\GeV$ &	$-357.4\MeV$ &	-0.96\%\\
$E_C$ &	$77.388\MeV$ &	$73.100\MeV$ &	$4.3\MeV$ &	5.86\%\\
$R$ &	$3.525\fm$ &	$3.478\fm$ &	$0.047\fm$ &	1.35\%\\
\hline\hline
  \end{tabular}}
  \caption{Total energy in Skyrme units, total energy in GeV compared
    with experimental data
    \cite{helium4,beryllium8,carbon12,oxygen16,calcium40}, Coulomb
    energy in MeV compared with the fit $0.156B^{\frac53}$ of
    ref.~\cite{Tian:2014uka}, and charge radii compared with
    experimental data
    \cite{helium4,beryllium8,carbon12,oxygen16,calcium40}.
    The comparison is made both in absolute values and in
    percentages. 
    }
  \label{tab:groundstate_details}
\end{table}
Table \ref{tab:groundstate_details} shows the total energies, Coulomb
energies and charge radii of the lightest Skyrmion solutions with
baryon numbers 4, 8, 12, 16 and 40.
Additionally, experimental data is used for comparison from the NuDat3
database \cite{helium4,beryllium8,carbon12,oxygen16,calcium40}.
The total energies are surprisingly well described by the model in our
calibration, with the largest deviation of 1.86\% (excess) for
Helium-4 and only $-0.96\%$ deviation for Calcium-40.
The fact that there is an excess in the energies for $B<12$ and a lack
for $B>12$ illustrates that the Skyrme model is too tightly bound --
i.e.~the good old story of the too large binding energies.
Nevertheless, the problem is much milder than one would anticipate,
when looking only at large Skyrmions and ignoring the energy for the
single nucleon ($B=1$).
Of course, these energies are classical and no quantum corrections
from vibrational modes etc.~have been taken into account here (recall
that for spin-0 and isospin-0 states, there are no contribution from
the zeromode quantization of the Skyrmions).

Even though the total energies are predicted very precisely in our
calibration, the Coulomb energies are more imprecise and of the usual
order of magnitude of errors in Skyrme-type models.
Specifically, the Coulomb energies are consistently overestimated
by between $3.3\%$ and $21.9\%$, with Oxygen-16 being the most precise
and Helium-4 the most imprecise.
The charge radii are slightly better with the Helium-4 being 15.2\%
too small and Oxygen-16 being 11.2\% too large, as the two extremes.

\begin{figure}[!htp]
  \centering
  \mbox{
    \subfloat[]{\includegraphics[width=0.49\linewidth]{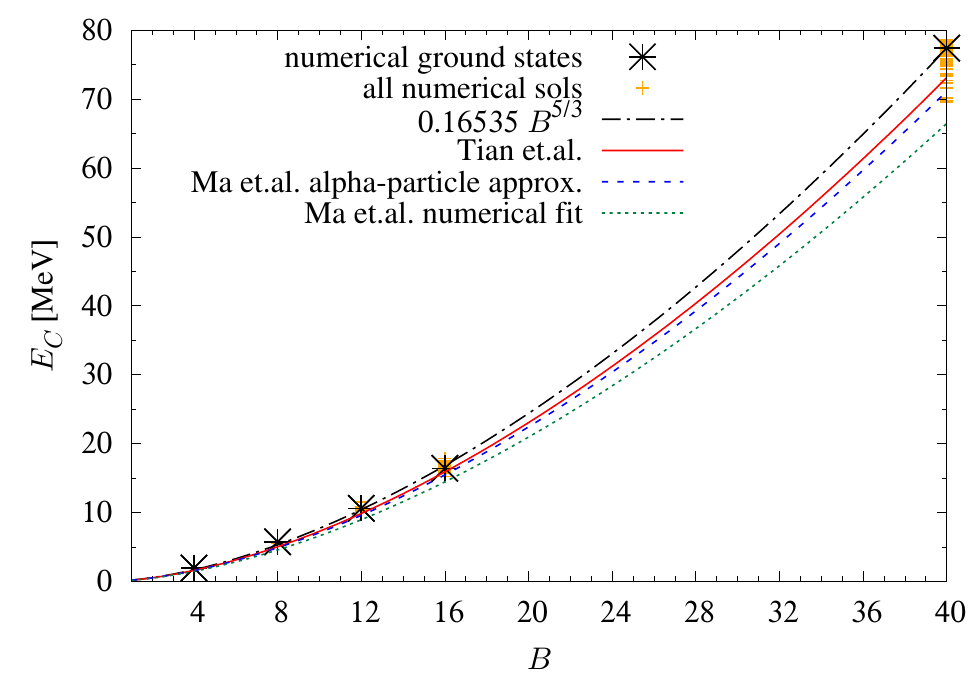}}
    \subfloat[]{\includegraphics[width=0.49\linewidth]{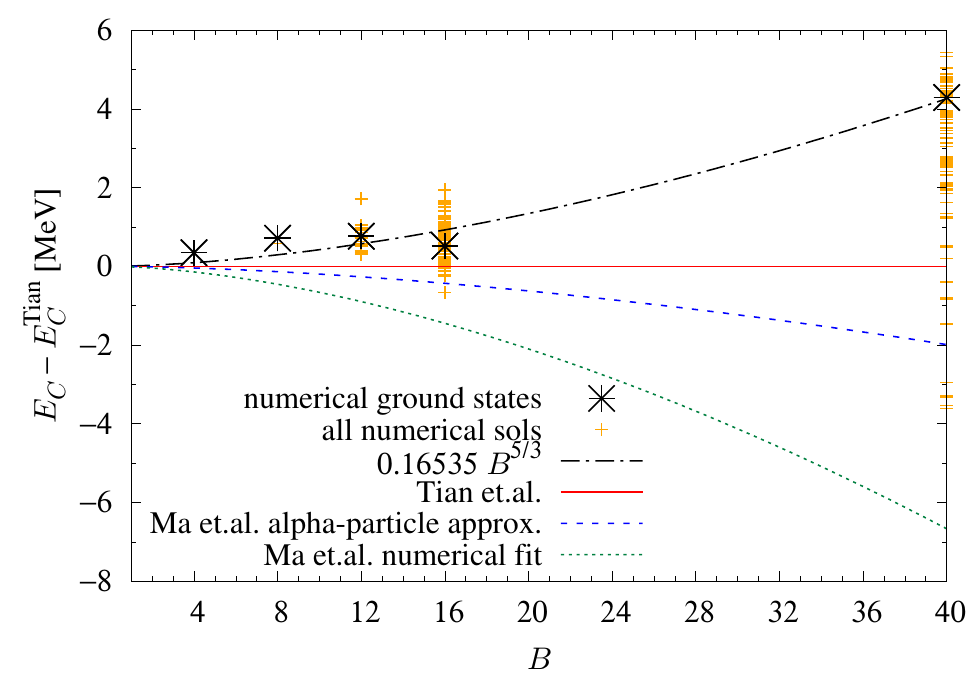}}}
  \caption{Coulomb energies of the lightest Skyrmions (black crosses)
    as well as all remaining solutions (orange pluses). The lightest
    Skyrmions are fitted to the power-law shown with a black
    dot-dashed line.
    In comparison, the experimental fit of Tian
    et.al.~\cite{Tian:2014uka} and two fits of Ma
    et.al.~\cite{Ma:2019fvk} are shown.
  Panel (b) shows the same information, but relative to the fit of
  Tian et.al.~\cite{Tian:2014uka}.
  }
  \label{fig:Acoulomb}
\end{figure}
Compiling the Coulomb energies into a figure, we show in
fig.~\ref{fig:Acoulomb} the Coulomb energies computed for the ground
states (the global energy minimizers for each baryon number) with a
large black cross and all the remaining solutions with smaller orange
pluses.
Taking as the ``experimental'' fit, the fitted result of Tian
et.al.~\cite{Tian:2014uka}, our ground states are all slightly above
the nuclear ``experimental'' fit, but only about 4 MeV for Calcium-40
and less or about 1 MeV for the other 4 nuclei.
We summarize the four different Coulomb fits that are shown in
fig.~\ref{fig:Acoulomb}:
\begin{align}
E_C^{\rm Tian} &= 0.156 B^{\frac53}\MeV, \non
E_C^{\rm Ma,\,APA} &= 0.152 B^{\frac53}\MeV, \non
E_C^{\rm Ma,\,num} &= 0.142 B^{\frac53}\MeV, \non
E_C^{\rm fit} &= 0.165 B^{\frac53}\MeV,
\end{align}
with $E_C^{\rm Tian}$ being the simplest fit of
ref.~\cite{Tian:2014uka}, $E_C^{\rm Ma,\,APA}$ being the fit to the
alpha-particle approximation of ref.~\cite{Ma:2019fvk},
$E_C^{\rm Ma,\,num}$ being a fit to their numerical alpha-particle
solutions, and finally $E_C^{\rm fit}$ being a fit to our ground
states in our calibration.
It is worth mentioning that the fit of ref.~\cite{Tian:2014uka} is
based on nuclei in the range $B\sim11$-$75$ and hence is not fitted on
our two lightest nuclei ($B=4$ and $B=8$).
It is perhaps also important to notice that their fit is made on
nuclei, not only having vanishing isospin, so a fit in the isospin-0
sector may be more precise than the fit of ref.~\cite{Tian:2014uka}.

\section{Conclusion and discussion}

In this paper, we have considered the standard Skyrme model with the
addition of the Maxwell term and a source term for the electric field
that matches with the Gell-Mann-Nishijima relation.
Formally, the equations are identical to those of the $\omega$-Skyrme
model \cite{Gudnason:2020arj}, with the identification of the electric
potential $A_0$ and $\omega_0$ and with the exception that $A_0$ is
massless whereas $\omega_0$ is massive.
This gives rise to a technical complication, because $A_0$ with a
power-like falloff requires in principle very large domains for
obtaining its solution.
We solve this problem with the approximation of assuming that the
electric charge is approximately interchangeable with a point charge
at the centre-of-mass of the Skyrmion, as seen at the distance of the
boundary of the box in which the numerical equations are solved.
This would be a poor approximation, if not used in conjunction with
the trick of integrating-by-parts, so that the Coulomb energy is
evaluated as the product of the electric charge density (Skyrmion
baryon charge density) and the electric potential, integrated over space.
Since the Skyrmion baryon charge density is exponentially spatially localized in the
massive Skyrme model, so is this product.
We calibrate the model to the size and energy of Carbon-12, which is a
relatively large nucleus, thus giving a larger range of
``good behaviour'' of our fit in baryon numbers.
The fit, however, requires a pion mass parameter smaller than unity
(often used in the literature) and this fact changes the solutions and
ground states (lightest Skyrmion solutions) with respect to the
Sm\"org\aa sbord of ref.~\cite{Gudnason:2022jkn}.
We find that the ``dynamics'' of the Skyrmions is more sensitive to
the inclusion of the Coulomb backreaction (CBR) than the final state
outcome.
That is, starting with a particular initial condition, turning CBR on
or off may very likely result in different (local) solutions, whereas
the global minimizer of the energy is often the same.
It turns out that for the largest stable nucleus with
baryon number $B=4n$ and isospin zero in the ground state, Calcium-40,
the lightest Skyrmion solutions differ with the CBR turned on or
off.
We fit the Coulomb energies of our Skyrmions and find that the
coefficient of the power-law is larger than the phenomenological one
of ref.~\cite{Tian:2014uka}, which in turn is larger than the result
of ref.~\cite{Ma:2019fvk}.
Nevertheless, the Coulomb energies are within about 15\% of the
phenomenological fits, which is quite reasonable for a simple
over-bound Skyrme model without quantization taken into account.
We notice that the total energies are remarkably well described by our
solutions in our calibration, with the largest deviation of 1.86\%.
Finally, we should point out that the 100 $B=40$ solutions are new and
due to the complexity of such large Skyrmions, we cannot for sure say
that we have found the global minimizer of the energy.

Clearly, the Coulomb effect is important for large nuclei and taking
it into account with a source term is just the simplest approach.
As explained in detail in the introduction, there is a more
elaborate approach of gauging the Skyrme model and including
Wess-Zumino-like or Callan-Witten terms that reproduce the QCD
anomalies.
Such an approach would complicate the model significantly and the fact
that the Coulomb effect is most pronounced for large nuclei, requires
one to study large Skyrmions, like the $B=40$ Skyrmions studied in
this paper.

\subsection*{Acknowledgements}

S.~B.~G.~thanks the Outstanding Talent Program of Henan University and
the Ministry of Education of Henan Province for partial support.
The work of S.~B.~G.~is supported by the National Natural Science
Foundation of China (Grant No.~12071111) and by the Ministry of
Science and Technology of China (Grant No.~G2022026021L).

\bibliographystyle{JHEP}
\bibliography{{references}}

\providecommand{\href}[2]{#2}\begingroup\raggedright\begin{thebibliography}{10}

\bibitem{Scherer:2002tk}
S.~Scherer, {\it {Introduction to chiral perturbation theory}},  {\em Adv.
  Nucl. Phys.} {\bf 27} (2003) 277,
  [\href{http://arxiv.org/abs/hep-ph/0210398}{{\tt hep-ph/0210398}}].

\bibitem{Skyrme:1961vq}
T.~H.~R. Skyrme, {\it {A Nonlinear field theory}},  {\em Proc. Roy. Soc. Lond.
  A} {\bf 260} (1961) 127--138.

\bibitem{Skyrme:1962vh}
T.~H.~R. Skyrme, {\it {A Unified Field Theory of Mesons and Baryons}},  {\em
  Nucl. Phys.} {\bf 31} (1962) 556--569.

\bibitem{Krasnov:2011up}
K.~Krasnov, {\it {Gravity as a diffeomorphism invariant gauge theory}},  {\em
  Phys. Rev. D} {\bf 84} (2011) 024034,
  [\href{http://arxiv.org/abs/1101.4788}{{\tt arXiv:1101.4788}}].

\bibitem{Pais:1973mi}
A.~Pais, {\it {Remark on baryon conservation}},  {\em Phys. Rev. D} {\bf 8}
  (1973) 1844--1846.

\bibitem{Callan:1983nx}
C.~G. Callan, Jr. and E.~Witten, {\it {Monopole Catalysis of Skyrmion Decay}},
  {\em Nucl. Phys. B} {\bf 239} (1984) 161--176.

\bibitem{Piette:1997ny}
B.~M. A.~G. Piette and D.~H. Tchrakian, {\it {Static solutions in the U(1)
  gauged Skyrme model}},  {\em Phys. Rev. D} {\bf 62} (2000) 025020,
  [\href{http://arxiv.org/abs/hep-th/9709189}{{\tt hep-th/9709189}}].

\bibitem{Navarro-Lerida:2023hbv}
F.~Navarro-Lerida, E.~Radu, and D.~H. Tchrakian, {\it {The role of the
  Callan\textendash{}Witten anomaly density as a Chern\textendash{}Simons term
  in Skyrme model $^{*}$}},  {\em J. Phys. A} {\bf 56} (2023), no.~46 465401,
  [\href{http://arxiv.org/abs/2304.12648}{{\tt arXiv:2304.12648}}].

\bibitem{Radu:2005jp}
E.~Radu and D.~H. Tchrakian, {\it {Spinning U(1) gauged skyrmions}},  {\em
  Phys. Lett. B} {\bf 632} (2006) 109--113,
  [\href{http://arxiv.org/abs/hep-th/0509014}{{\tt hep-th/0509014}}].

\bibitem{Livramento:2023keg}
L.~R. Livramento, E.~Radu, and Y.~Shnir, {\it {Solitons in the Gauged
  Skyrme-Maxwell Model}},  {\em SIGMA} {\bf 19} (2023) 042,
  [\href{http://arxiv.org/abs/2301.12848}{{\tt arXiv:2301.12848}}].

\bibitem{Livramento:2023tmm}
L.~R. Livramento and Y.~Shnir, {\it {Multisolitons in a gauged Skyrme-Maxwell
  model}},  {\em Phys. Rev. D} {\bf 108} (2023), no.~6 065010,
  [\href{http://arxiv.org/abs/2307.05756}{{\tt arXiv:2307.05756}}].

\bibitem{Kirichenkov:2023omy}
R.~Kirichenkov, J.~Kunz, N.~Sawado, and Y.~Shnir, {\it {Skyrmions and pion
  stars in the gauged U(1) Einstein-Skyrme model}},  {\em Phys. Rev. D} {\bf
  109} (2024), no.~4 045002, [\href{http://arxiv.org/abs/2311.12432}{{\tt
  arXiv:2311.12432}}].

\bibitem{Cork:2021ylu}
J.~Cork, D.~Harland, and T.~Winyard, {\it {A model for gauged skyrmions with
  low binding energies}},  {\em J. Phys. A} {\bf 55} (2022), no.~1 015204,
  [\href{http://arxiv.org/abs/2109.06886}{{\tt arXiv:2109.06886}}].

\bibitem{Cork:2023pft}
J.~Cork and D.~Harland, {\it {Geometry of Gauged Skyrmions}},  {\em SIGMA} {\bf
  19} (2023) 071, [\href{http://arxiv.org/abs/2303.02623}{{\tt
  arXiv:2303.02623}}].

\bibitem{Gudnason:2023jpq}
S.~B. Gudnason and C.~Halcrow, {\it {Quantum binding energies in the Skyrme
  model}},  {\em Phys. Lett. B} {\bf 850} (2024) 138526,
  [\href{http://arxiv.org/abs/2307.09272}{{\tt arXiv:2307.09272}}].

\bibitem{Ohtani:2004aw}
M.~Ohtani and K.~Ohta, {\it {Skyrmions coupled with the electromagnetic field
  via the gauged Wess-Zumino term}},  {\em Phys. Rev. D} {\bf 70} (2004)
  096014, [\href{http://arxiv.org/abs/hep-ph/0406173}{{\tt hep-ph/0406173}}].

\bibitem{Ohtani:2005xv}
M.~Ohtani and K.~Ohta, {\it {Spin polarized Skyrmions in the U(EM)(1) gauged
  Wess-Zumino action}},  {\em Nucl. Phys. A} {\bf 755} (2005) 661--664.

\bibitem{He:2016oqk}
B.-R. He, {\it {Skyrme model study of proton and neutron properties in a strong
  magnetic field}},  {\em Phys. Lett. B} {\bf 765} (2017) 109--112,
  [\href{http://arxiv.org/abs/1609.09055}{{\tt arXiv:1609.09055}}].

\bibitem{Aviles:2017hro}
L.~Avil\'es, F.~Canfora, N.~Dimakis, and D.~Hidalgo, {\it {Analytic
  topologically nontrivial solutions of the (3+1)-dimensional $U(1)$ gauged
  Skyrme model and extended duality}},  {\em Phys. Rev. D} {\bf 96} (2017),
  no.~12 125005, [\href{http://arxiv.org/abs/1711.07408}{{\tt
  arXiv:1711.07408}}].

\bibitem{Canfora:2018clt}
F.~Canfora, M.~Lagos, S.~H. Oh, J.~Oliva, and A.~Vera, {\it {Analytic
  (3+1)-dimensional gauged Skyrmions, Heun, and Whittaker-Hill equations and
  resurgence}},  {\em Phys. Rev. D} {\bf 98} (2018), no.~8 085003,
  [\href{http://arxiv.org/abs/1809.10386}{{\tt arXiv:1809.10386}}].

\bibitem{Canfora:2019asc}
F.~Canfora, N.~Dimakis, and A.~Paliathanasis, {\it {Analytic Studies of Static
  and Transport Properties of (Gauged) Skyrmions}},  {\em Eur. Phys. J. C} {\bf
  79} (2019), no.~2 139, [\href{http://arxiv.org/abs/1902.01563}{{\tt
  arXiv:1902.01563}}].

\bibitem{Sakai:2004cn}
T.~Sakai and S.~Sugimoto, {\it {Low energy hadron physics in holographic QCD}},
   {\em Prog. Theor. Phys.} {\bf 113} (2005) 843--882,
  [\href{http://arxiv.org/abs/hep-th/0412141}{{\tt hep-th/0412141}}].

\bibitem{Rebhan:2014rxa}
A.~Rebhan, {\it {The Witten-Sakai-Sugimoto model: A brief review and some
  recent results}},  {\em EPJ Web Conf.} {\bf 95} (2015) 02005,
  [\href{http://arxiv.org/abs/1410.8858}{{\tt arXiv:1410.8858}}].

\bibitem{Atiyah:1989dq}
M.~F. Atiyah and N.~S. Manton, {\it {Skyrmions From Instantons}},  {\em Phys.
  Lett. B} {\bf 222} (1989) 438--442.

\bibitem{Bolognesi:2013nja}
S.~Bolognesi and P.~Sutcliffe, {\it {The Sakai-Sugimoto soliton}},  {\em JHEP}
  {\bf 01} (2014) 078, [\href{http://arxiv.org/abs/1309.1396}{{\tt
  arXiv:1309.1396}}].

\bibitem{Witten:1998qj}
E.~Witten, {\it {Anti-de Sitter space and holography}},  {\em Adv. Theor. Math.
  Phys.} {\bf 2} (1998) 253--291,
  [\href{http://arxiv.org/abs/hep-th/9802150}{{\tt hep-th/9802150}}].

\bibitem{Johnson:2008vna}
C.~V. Johnson and A.~Kundu, {\it {External Fields and Chiral Symmetry Breaking
  in the Sakai-Sugimoto Model}},  {\em JHEP} {\bf 12} (2008) 053,
  [\href{http://arxiv.org/abs/0803.0038}{{\tt arXiv:0803.0038}}].

\bibitem{Bergman:2008sg}
O.~Bergman, G.~Lifschytz, and M.~Lippert, {\it {Response of Holographic QCD to
  Electric and Magnetic Fields}},  {\em JHEP} {\bf 05} (2008) 007,
  [\href{http://arxiv.org/abs/0802.3720}{{\tt arXiv:0802.3720}}].

\bibitem{Bonenfant:2012kt}
E.~Bonenfant, L.~Harbour, and L.~Marleau, {\it {Near-BPS Skyrmions: Non-shell
  configurations and Coulomb effects}},  {\em Phys. Rev. D} {\bf 85} (2012)
  114045, [\href{http://arxiv.org/abs/1205.1414}{{\tt arXiv:1205.1414}}].

\bibitem{Adam:2013tda}
C.~Adam, C.~Naya, J.~Sanchez-Guillen, and A.~Wereszczynski, {\it {Nuclear
  binding energies from a Bogomol'nyi-Prasad-Sommerfield Skyrme model}},  {\em
  Phys. Rev. C} {\bf 88} (2013), no.~5 054313,
  [\href{http://arxiv.org/abs/1309.0820}{{\tt arXiv:1309.0820}}].

\bibitem{Adam:2013wya}
C.~Adam, C.~Naya, J.~Sanchez-Guillen, and A.~Wereszczynski, {\it
  {Bogomol\textquoteright{}nyi-Prasad-Sommerfield Skyrme Model and Nuclear
  Binding Energies}},  {\em Phys. Rev. Lett.} {\bf 111} (2013), no.~23 232501,
  [\href{http://arxiv.org/abs/1312.2960}{{\tt arXiv:1312.2960}}].

\bibitem{Ma:2019fvk}
N.~Ma, C.~J. Halcrow, and H.~Zhang, {\it {Effect of the Coulomb energy on
  Skyrmions}},  {\em Phys. Rev. C} {\bf 99} (2019), no.~4 044312,
  [\href{http://arxiv.org/abs/1901.06025}{{\tt arXiv:1901.06025}}].

\bibitem{10.1119/1.1969367}
B.~C. Carlson and G.~L. Morley, {\it {Multipole Expansion of Coulomb Energy}},
  {\em American Journal of Physics} {\bf 31} (03, 1963) 209--211,
  [\href{http://arxiv.org/abs/https://pubs.aip.org/aapt/ajp/article-pdf/31/3/209/10110064/209\_1\_online.pdf}{{\tt
  https://pubs.aip.org/aapt/ajp/article-pdf/31/3/209/10110064/209\_1\_online.pdf}}].

\bibitem{Gudnason:2022jkn}
S.~B. Gudnason and C.~Halcrow, {\it {A Sm\"org\r{a}sbord of Skyrmions}},  {\em
  JHEP} {\bf 08} (2022) 117, [\href{http://arxiv.org/abs/2202.01792}{{\tt
  arXiv:2202.01792}}].

\bibitem{Gudnason:2020arj}
S.~B. Gudnason and J.~M. Speight, {\it {Realistic classical binding energies in
  the $\omega$-Skyrme model}},  {\em JHEP} {\bf 07} (2020) 184,
  [\href{http://arxiv.org/abs/2004.12862}{{\tt arXiv:2004.12862}}].

\bibitem{Manton:2004tk}
N.~S. Manton and P.~Sutcliffe, {\em {Topological solitons}}.
\newblock Cambridge Monographs on Mathematical Physics. Cambridge University
  Press, 2004.

\bibitem{Derrick:1964ww}
G.~H. Derrick, {\it {Comments on nonlinear wave equations as models for
  elementary particles}},  {\em J. Math. Phys.} {\bf 5} (1964) 1252--1254.

\bibitem{Harland:2013rxa}
D.~Harland, {\it {Topological energy bounds for the Skyrme and Faddeev models
  with massive pions}},  {\em Phys. Lett. B} {\bf 728} (2014) 518--523,
  [\href{http://arxiv.org/abs/1311.2403}{{\tt arXiv:1311.2403}}].

\bibitem{Adam:2013tga}
C.~Adam and A.~Wereszczynski, {\it {Topological energy bounds in generalized
  Skyrme models}},  {\em Phys. Rev. D} {\bf 89} (2014), no.~6 065010,
  [\href{http://arxiv.org/abs/1311.2939}{{\tt arXiv:1311.2939}}].

\bibitem{Zyla:2020zbs}
{\bf Particle Data Group} Collaboration, P.~A. Zyla et~al., {\it {Review of
  Particle Physics}},  {\em PTEP} {\bf 2020} (2020), no.~8 083C01.

\bibitem{Angeli:2013epw}
I.~Angeli and K.~P. Marinova, {\it {Table of experimental nuclear ground state
  charge radii: An update}},  {\em Atom. Data Nucl. Data Tabl.} {\bf 99}
  (2013), no.~1 69--95.

\bibitem{Adkins:1983ya}
G.~S. Adkins, C.~R. Nappi, and E.~Witten, {\it {Static Properties of Nucleons
  in the Skyrme Model}},  {\em Nucl. Phys. B} {\bf 228} (1983) 552.

\bibitem{Manton:2006tq}
N.~S. Manton and S.~W. Wood, {\it {Reparametrising the Skyrme model using the
  lithium-6 nucleus}},  {\em Phys. Rev. D} {\bf 74} (2006) 125017,
  [\href{http://arxiv.org/abs/hep-th/0609185}{{\tt hep-th/0609185}}].

\bibitem{Battye:1997qq}
R.~A. Battye and P.~M. Sutcliffe, {\it {Symmetric skyrmions}},  {\em Phys. Rev.
  Lett.} {\bf 79} (1997) 363--366,
  [\href{http://arxiv.org/abs/hep-th/9702089}{{\tt hep-th/9702089}}].

\bibitem{Houghton:1997kg}
C.~J. Houghton, N.~S. Manton, and P.~M. Sutcliffe, {\it {Rational maps,
  monopoles and Skyrmions}},  {\em Nucl. Phys. B} {\bf 510} (1998) 507--537,
  [\href{http://arxiv.org/abs/hep-th/9705151}{{\tt hep-th/9705151}}].

\bibitem{Battye:2000se}
R.~M. Battye and P.~M. Sutcliffe, {\it {Solitonic fullerenes}},  {\em Phys.
  Rev. Lett.} {\bf 86} (2001) 3989--3992,
  [\href{http://arxiv.org/abs/hep-th/0012215}{{\tt hep-th/0012215}}].

\bibitem{Battye:2001qn}
R.~A. Battye and P.~M. Sutcliffe, {\it {Skyrmions, fullerenes and rational
  maps}},  {\em Rev. Math. Phys.} {\bf 14} (2002) 29--86,
  [\href{http://arxiv.org/abs/hep-th/0103026}{{\tt hep-th/0103026}}].

\bibitem{helium4}
``Nudat3 database -- helium-4.''
\newblock Accessed on September 26, 2024.

\bibitem{beryllium8}
``Nudat3 database -- beryllium-8.''
\newblock Accessed on September 26, 2024.

\bibitem{carbon12}
``Nudat3 database -- carbon-12.''
\newblock Accessed on September 26, 2024.

\bibitem{oxygen16}
``Nudat3 database -- oxygen-16.''
\newblock Accessed on September 26, 2024.

\bibitem{calcium40}
``Nudat3 database -- calcium-40.''
\newblock Accessed on September 26, 2024.

\bibitem{Tian:2014uka}
J.~Tian, H.~Cui, N.~Wang, and K.~Zheng, {\it {Effect of Coulomb energy on the
  symmetry energy coefficients of finite nuclei}},  {\em Phys. Rev. C} {\bf 90}
  (2014), no.~2 024313, [\href{http://arxiv.org/abs/1403.6560}{{\tt
  arXiv:1403.6560}}].

\end{thebibliography}\endgroup

\appendix
\section*{Appendices}
\section{Proof of Lemma \ref{loclem}}\label{app:loclem_proof}

Let $\rho:\R^3\ra\R$ satisfy $|\rho(x)|\leq Ce^{-|x|/C}$ and $V:\R^3\ra\R$ be the electrostatic potential it induces, that is, the unique solution of \eqref{gauss} decaying at infinity. Then
\beq\label{Vgreen}
V(x)=\frac{1}{4\pi\eps_0}\int_{\R^3}\frac{\rho(x')}{|x-x'|}\d^3x',
\eeq
and it follows immediately that $|V(x)|\leq C^3V_*(x/C)$, where $V_*$ is the potential induced by $\rho_*(x)=e^{-|x|}$. The electric field $E_*=-\nabla V_*$ induced by $\rho_*$ is radial, so we may compute $V_*$ explicitly by an application of the divergence theorem, obtaining
\beq
V_*(x)=\frac{1}{\eps_0}\left[\frac{2}{r}-e^{-r}\left(1+\frac{2}{r}\right)\right],
\eeq
and the claimed localization of $V$ immediately follows. 

Let $E=-\nabla V$ be the electric field induced by $\rho$. We seek an upper bound on $|n\cdot E|$ where $n=x/r$ is the unit radial vector. Now \eqref{Vgreen} implies that
\beq
E=\frac{1}{4\pi\eps_0}\int_{\R^3}\frac{x-x'}{|x-x'|^3}\rho(x')\d^3x',
\eeq
and hence that
\bea
|n\cdot E|&\leq&\frac{1}{4\pi\eps_0}\int_{\R^3}\frac{|\rho(x')|}{|x-x'|^2}\d^3x'\\
&\leq&\frac{C}{4\pi\eps_0}\int_{\R^3}\frac{e^{-|x'|/C}}{|x-x'|^2}\d^3x'
=\frac{C}{4\pi\eps_0}f(r)
\eea
where
\bea
f(r)&=&2\pi\int_{-1}^1\d z\int_0^\infty\d s\frac{s^2e^{-s/C}}{s^2+r^2-2zsr}\nonumber \\
&=&2\pi\int_0^\infty\d s \, s e^{-s/C}\log\left|\frac{s+r}{s-r}\right|\nonumber \\
&=&2\pi r\int_0^\infty\d s\,  s e^{-(r/C)s}\log\left|\frac{s+1}{s-1}\right|
\eea
Choose and fix $\eps\in(0,1/2)$. Then there exists $K>0$ such that
\beq
\log\left|\frac{s+1}{s-1}\right|\leq\left\{
\begin{array}{cl}
Ks, & s\in[0,1-\eps],\\
K\log\left(\frac{1}{1-s}\right), & s\in[1-\eps,1), \\
K\log\left(\frac{1}{s-1}\right), & s\in (1,1+\eps], \\
K/s, & s\in[1+\eps,\infty).
\end{array}
\right.
\eeq
Hence
\bea
f(r)&\leq&2\pi rK\left\{2\left(\frac{C}{r}\right)^3+\frac{e^{-r/C}}{r/C}
+\eps(1-\log\eps)e^{-(r/C)(1-\eps)}+\eps(1-\log\eps)(1+\eps)e^{-r/C}\right\}
\nonumber \\
&\leq&\frac{K'}{r^2}
\eea
for some $K'>0$. The claimed localization of $|\cd V/\cd r|=|n\cdot E|$ now follows.

\section{The remaining \texorpdfstring{$B=40$}{B=40} solutions}\label{app:moreB40s}

\begin{figure}[!htp]
\centering
\begin{tabular}{lcc}
\includegraphics[scale=0.22]{{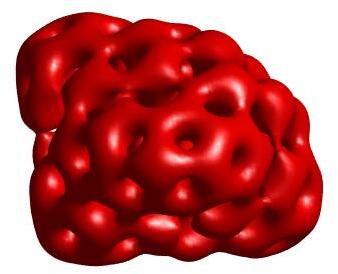}}&
\includegraphics[scale=0.22]{{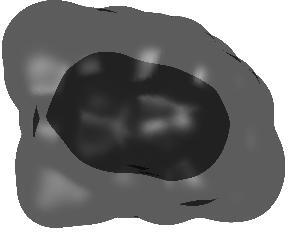}}&
\includegraphics[scale=0.22]{{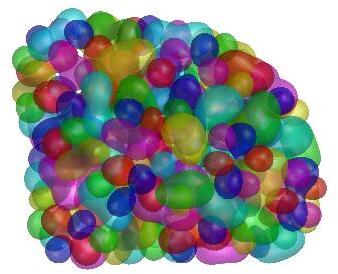}}\\
(k) 5589.89\\
\includegraphics[scale=0.22]{{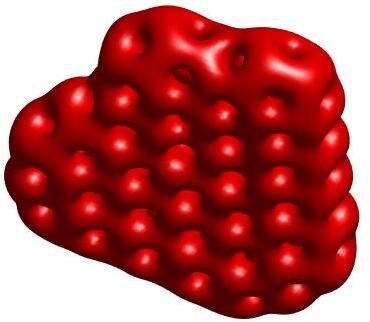}}&
\includegraphics[scale=0.22]{{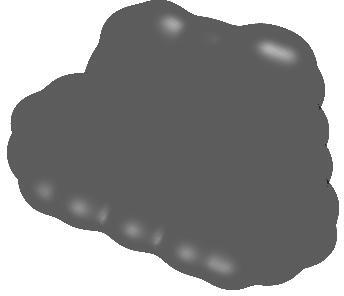}}&
\includegraphics[scale=0.22]{{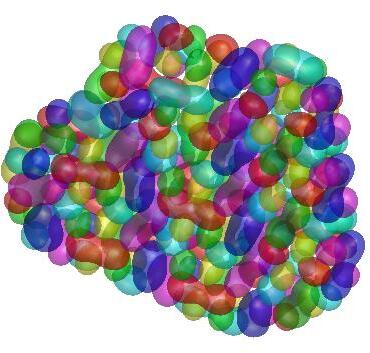}}\\
(l) 5590.00\\
\includegraphics[scale=0.22]{{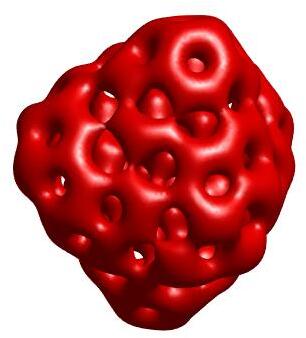}}&
\includegraphics[scale=0.22]{{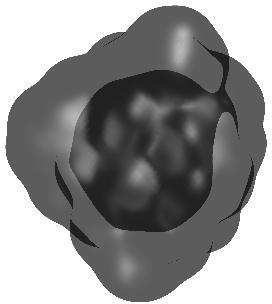}}&
\includegraphics[scale=0.22]{{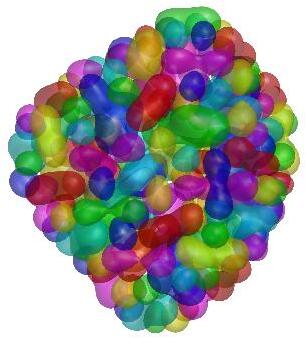}}\\
(n) 5590.49\\
\includegraphics[scale=0.16]{{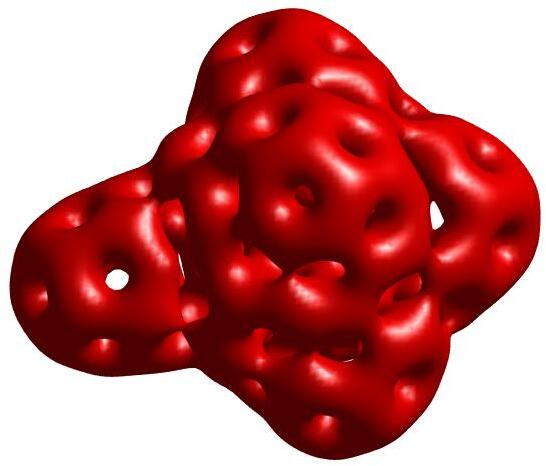}}&
\includegraphics[scale=0.16]{{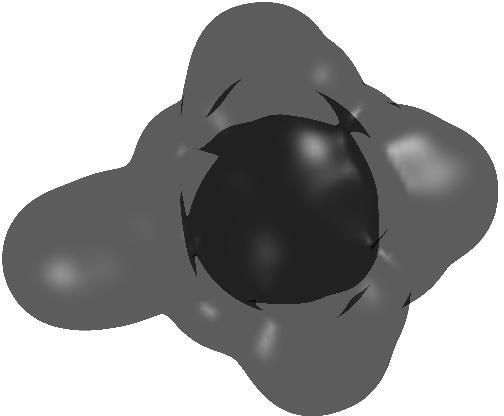}}&
\includegraphics[scale=0.16]{{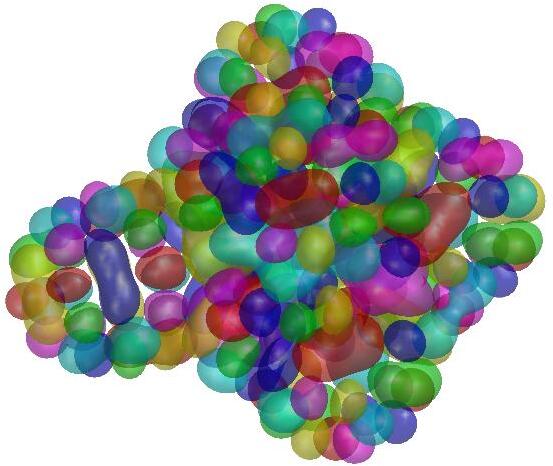}}\\
(o) 5590.50\\
\includegraphics[scale=0.23]{{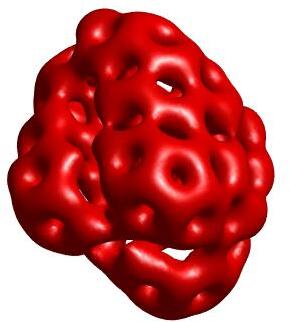}}&
\includegraphics[scale=0.23]{{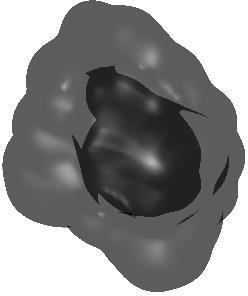}}&
\includegraphics[scale=0.23]{{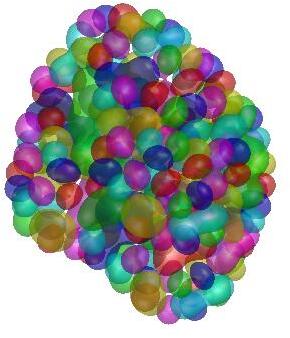}}\\
(p) 5590.64\\
\end{tabular}
\caption{$B=40$ solutions with $\kappa=0.737$ ordered by increasing
  static energy, excluding the 10 lightest solutions as well as that
  of fig.~\ref{fig:conf12}. }
\label{fig:B40rest1}
\end{figure}

\begin{figure}[!htp]
\centering
\begin{tabular}{lcc}
\includegraphics[scale=0.2]{{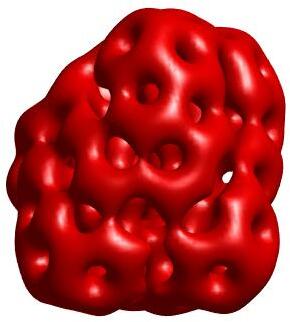}}&
\includegraphics[scale=0.2]{{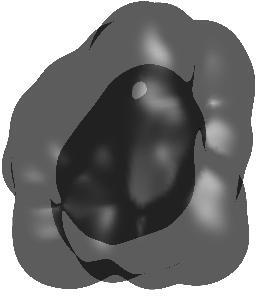}}&
\includegraphics[scale=0.2]{{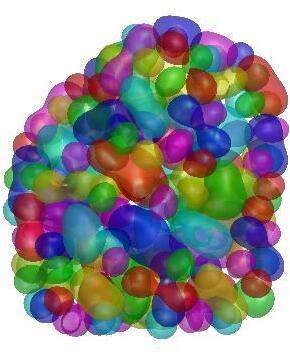}}\\
(q) 5590.73\\
\includegraphics[scale=0.2]{{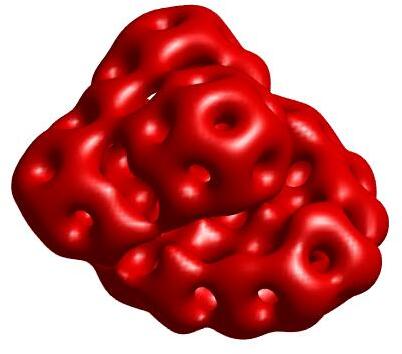}}&
\includegraphics[scale=0.2]{{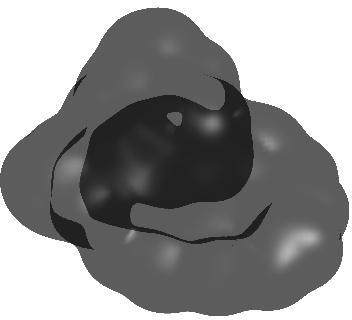}}&
\includegraphics[scale=0.2]{{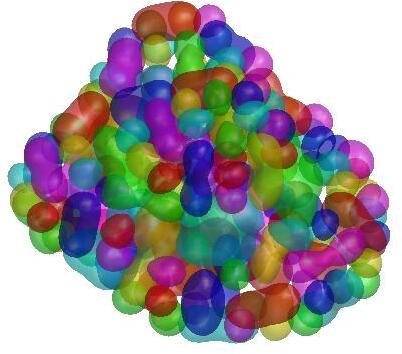}}\\
(r) 5590.75\\
\includegraphics[scale=0.22]{{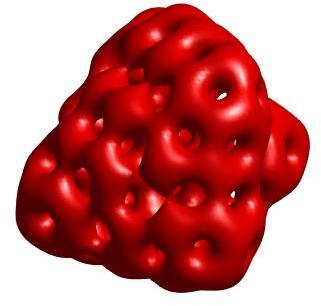}}&
\includegraphics[scale=0.22]{{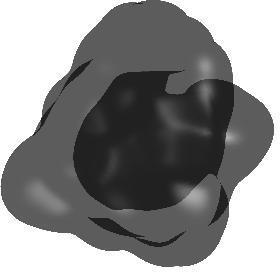}}&
\includegraphics[scale=0.22]{{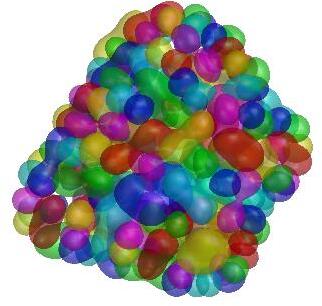}}\\
(s) 5590.76\\
\includegraphics[scale=0.2]{{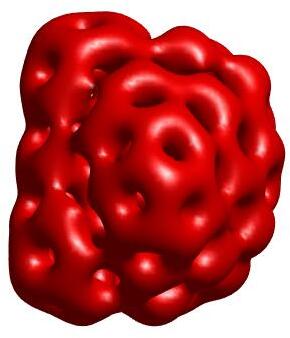}}&
\includegraphics[scale=0.2]{{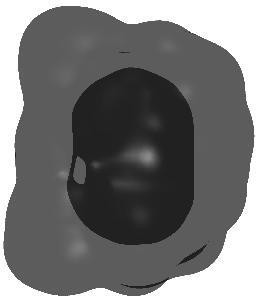}}&
\includegraphics[scale=0.2]{{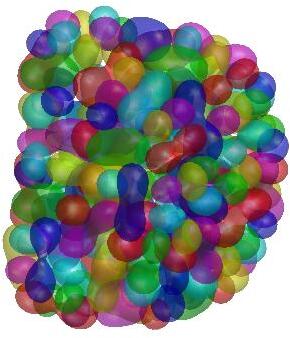}}\\
(t) 5590.91\\
\includegraphics[scale=0.16]{{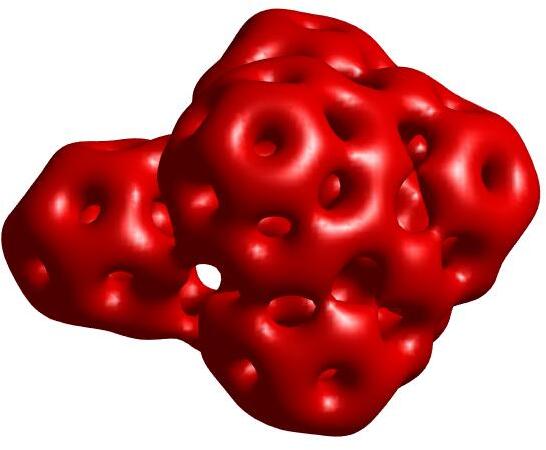}}&
\includegraphics[scale=0.16]{{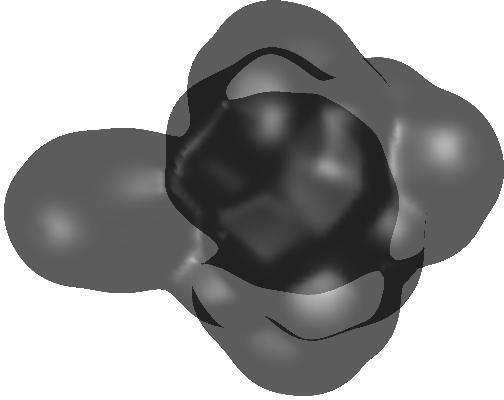}}&
\includegraphics[scale=0.16]{{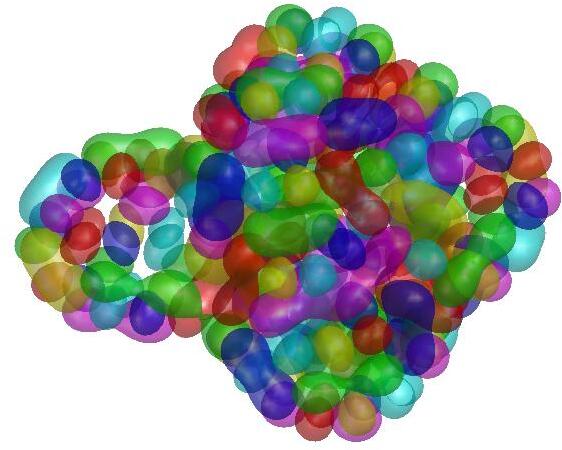}}\\
(u) 5591.09\\
\end{tabular}
\caption{$B=40$ solutions with $\kappa=0.737$ ordered by increasing
  static energy, continued from fig.~\ref{fig:B40rest1}.
}
\label{fig:B40rest2}
\end{figure}

\begin{figure}[!htp]
\centering
\begin{tabular}{lcc}
\includegraphics[scale=0.22]{{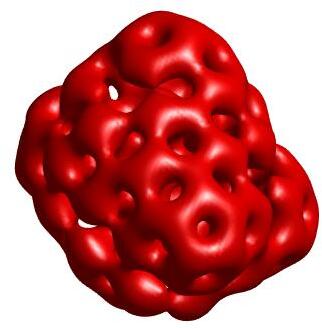}}&
\includegraphics[scale=0.22]{{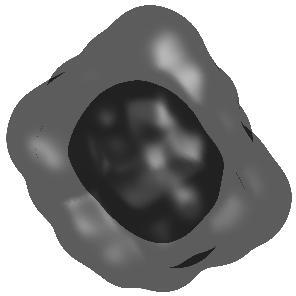}}&
\includegraphics[scale=0.22]{{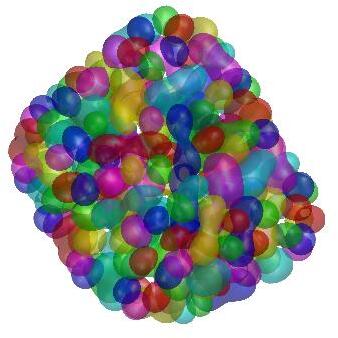}}\\
(v) 5591.09\\
\includegraphics[scale=0.22]{{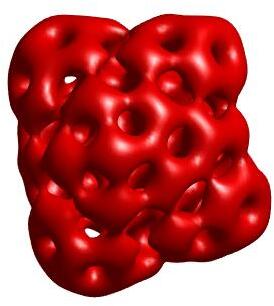}}&
\includegraphics[scale=0.22]{{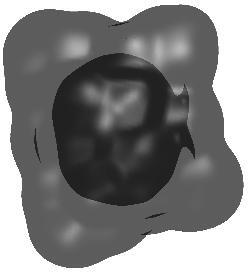}}&
\includegraphics[scale=0.22]{{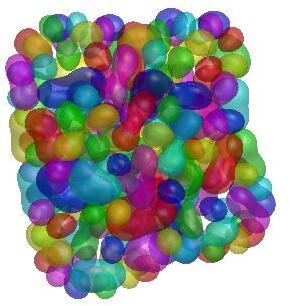}}\\
(w) 5591.45\\
\includegraphics[scale=0.22]{{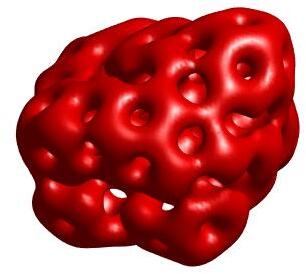}}&
\includegraphics[scale=0.22]{{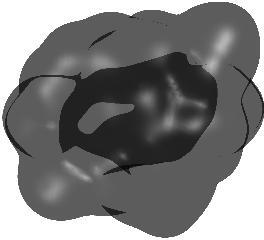}}&
\includegraphics[scale=0.22]{{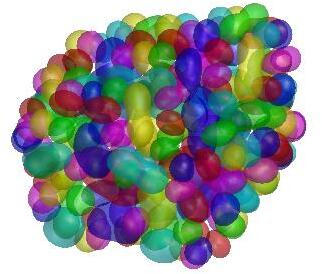}}\\
(x) 5591.48\\
\includegraphics[scale=0.21]{{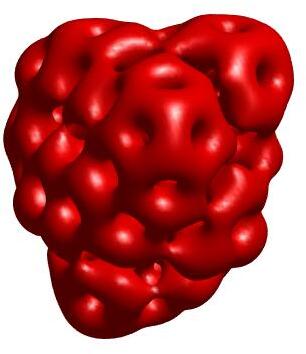}}&
\includegraphics[scale=0.21]{{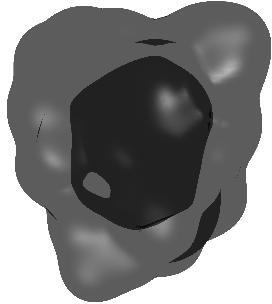}}&
\includegraphics[scale=0.21]{{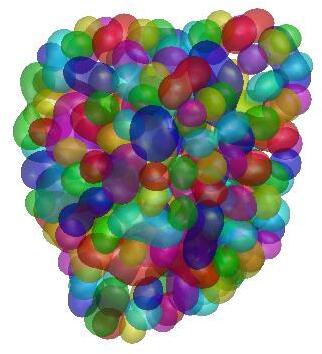}}\\
(y) 5591.70\\
\includegraphics[scale=0.2]{{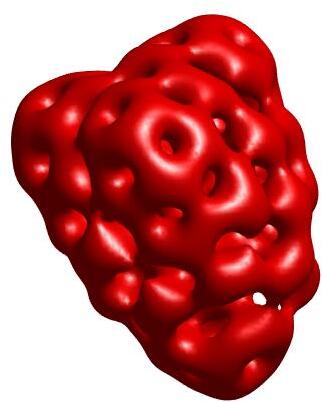}}&
\includegraphics[scale=0.2]{{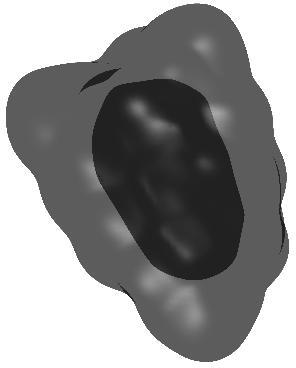}}&
\includegraphics[scale=0.2]{{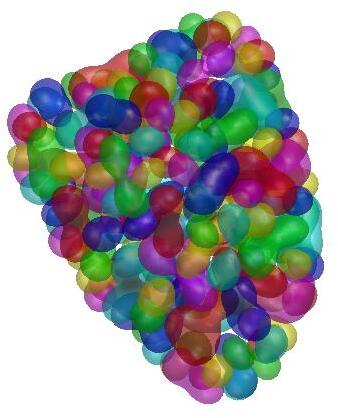}}\\
(z) 5591.81\\
\end{tabular}
\caption{$B=40$ solutions with $\kappa=0.737$ ordered by increasing
  static energy, continued from fig.~\ref{fig:B40rest2}.
}
\label{fig:B40rest3}
\end{figure}

\begin{figure}[!htp]
\centering
\begin{tabular}{lcc}
\includegraphics[scale=0.23]{{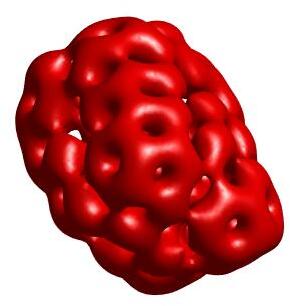}}&
\includegraphics[scale=0.23]{{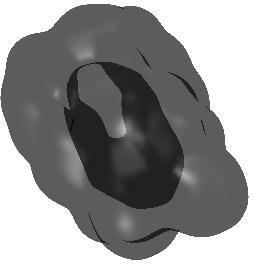}}&
\includegraphics[scale=0.23]{{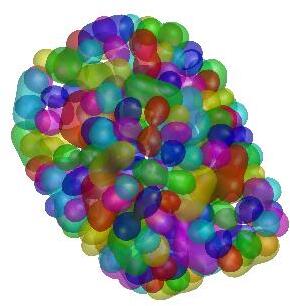}}\\
(aa) 5591.83\\
\includegraphics[scale=0.2]{{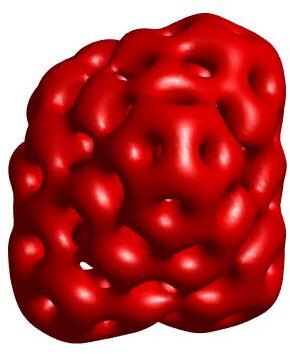}}&
\includegraphics[scale=0.2]{{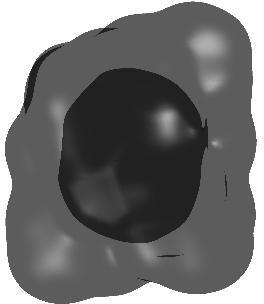}}&
\includegraphics[scale=0.2]{{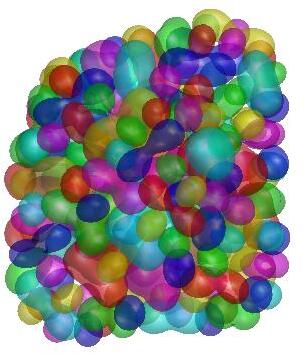}}\\
(ab) 5591.98\\
\includegraphics[scale=0.2]{{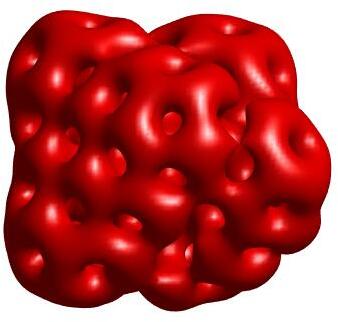}}&
\includegraphics[scale=0.2]{{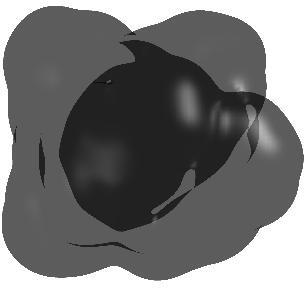}}&
\includegraphics[scale=0.2]{{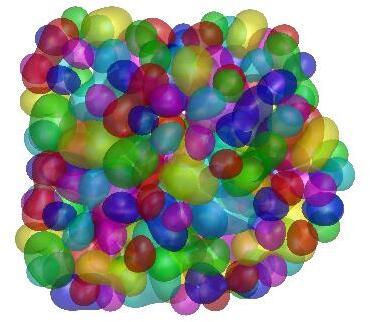}}\\
(ac) 5591.99\\
\includegraphics[scale=0.2]{{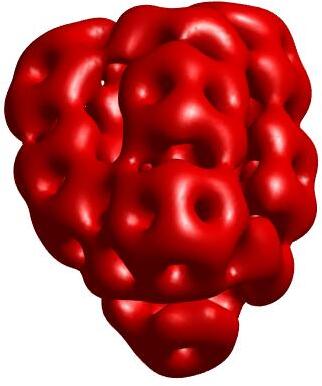}}&
\includegraphics[scale=0.2]{{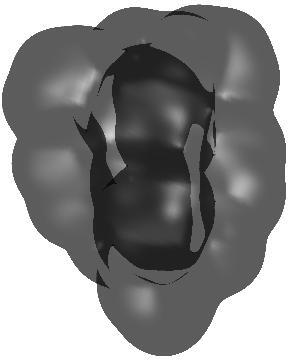}}&
\includegraphics[scale=0.2]{{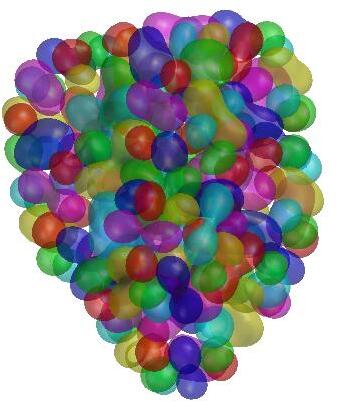}}\\
(ad) 5592.16\\
\includegraphics[scale=0.18]{{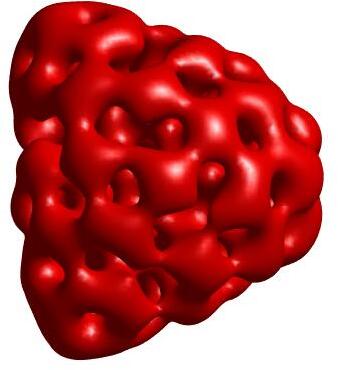}}&
\includegraphics[scale=0.18]{{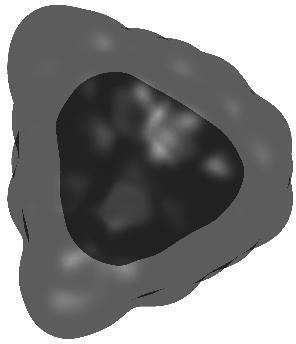}}&
\includegraphics[scale=0.18]{{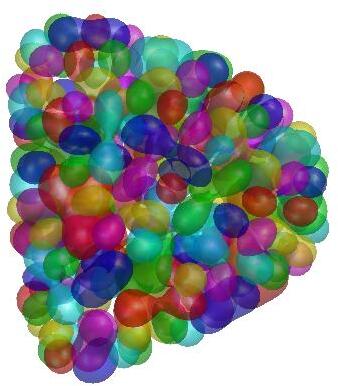}}\\
(ae) 5592.23\\
\end{tabular}
\caption{$B=40$ solutions with $\kappa=0.737$ ordered by increasing
  static energy, continued from fig.~\ref{fig:B40rest3}.
}
\label{fig:B40rest4}
\end{figure}

\begin{figure}[!htp]
\centering
\begin{tabular}{lcc}
\includegraphics[scale=0.23]{{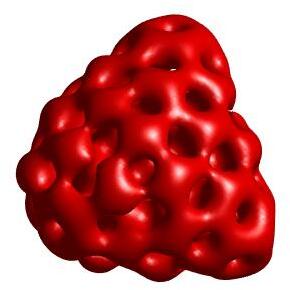}}&
\includegraphics[scale=0.23]{{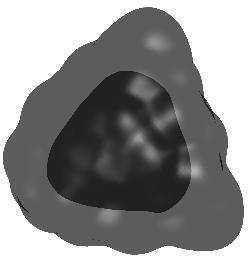}}&
\includegraphics[scale=0.23]{{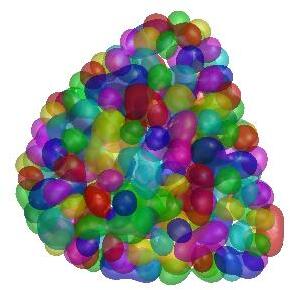}}\\
(af) 5592.26\\
\includegraphics[scale=0.21]{{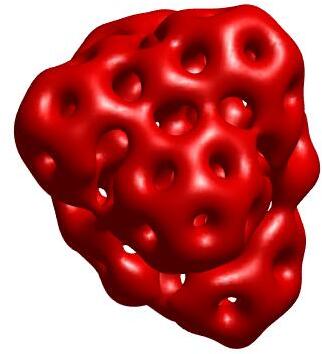}}&
\includegraphics[scale=0.21]{{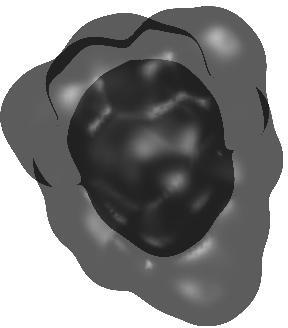}}&
\includegraphics[scale=0.21]{{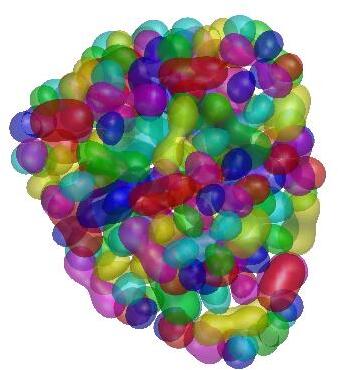}}\\
(ag) 5592.26\\
\includegraphics[scale=0.19]{{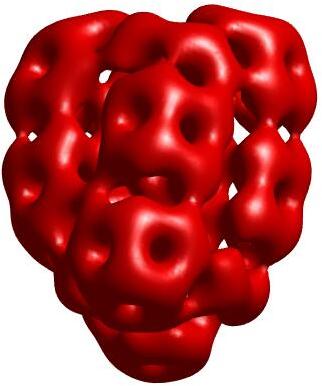}}&
\includegraphics[scale=0.19]{{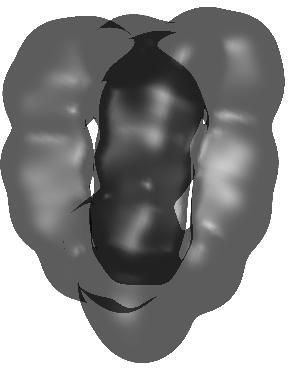}}&
\includegraphics[scale=0.19]{{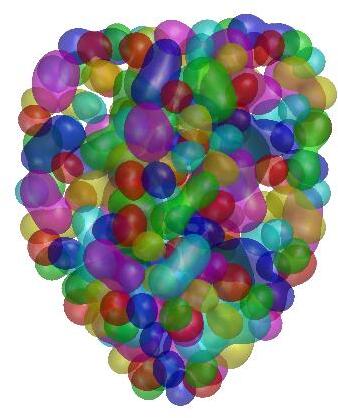}}\\
(ah) 5592.29\\
\includegraphics[scale=0.21]{{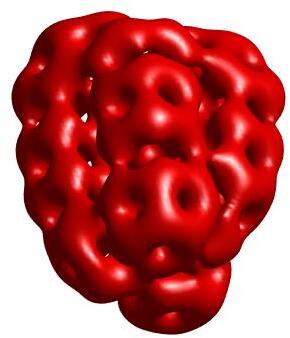}}&
\includegraphics[scale=0.21]{{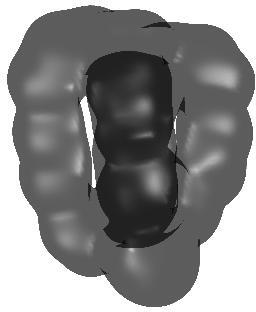}}&
\includegraphics[scale=0.21]{{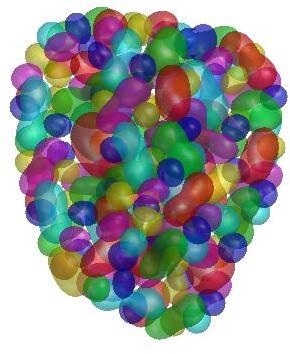}}\\
(ai) 5592.32\\
\includegraphics[scale=0.2]{{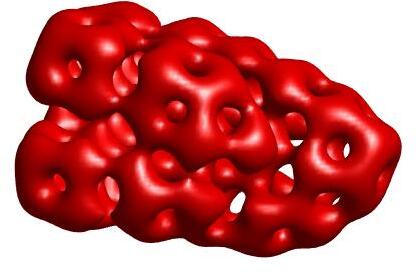}}&
\includegraphics[scale=0.2]{{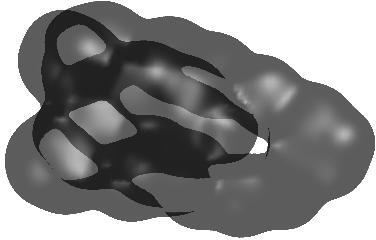}}&
\includegraphics[scale=0.2]{{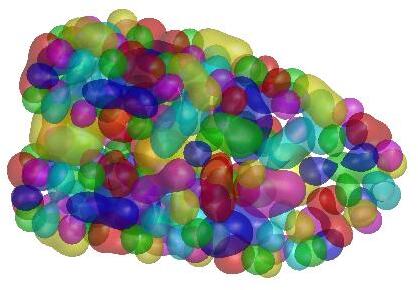}}\\
(ak) 5593.23\\
\end{tabular}
\caption{$B=40$ solutions with $\kappa=0.737$ ordered by increasing
  static energy, continued from fig.~\ref{fig:B40rest4} and excluding
  that of fig.~\ref{fig:conf20+27}(left).
}
\label{fig:B40rest5}
\end{figure}

\begin{figure}[!htp]
\centering
\begin{tabular}{lcc}
\includegraphics[scale=0.21]{{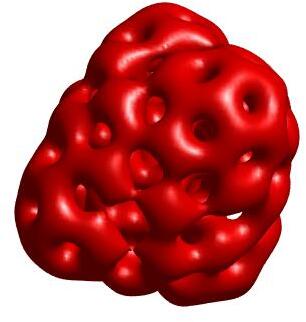}}&
\includegraphics[scale=0.21]{{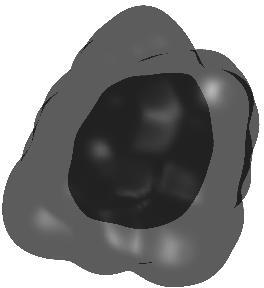}}&
\includegraphics[scale=0.21]{{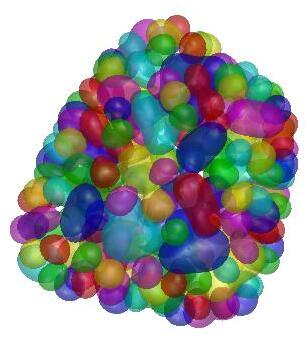}}\\
(al) 5593.72\\
\includegraphics[scale=0.21]{{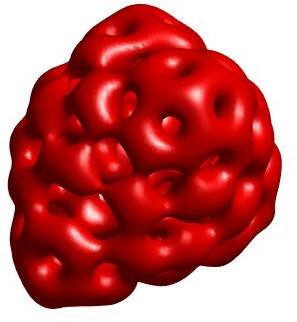}}&
\includegraphics[scale=0.21]{{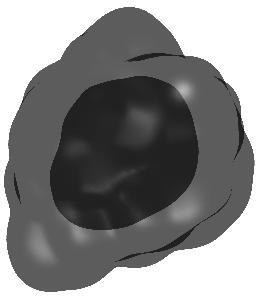}}&
\includegraphics[scale=0.21]{{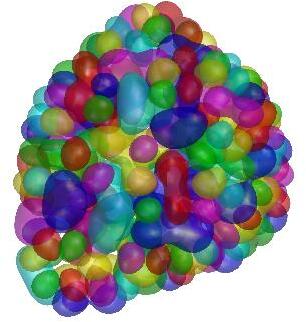}}\\
(am) 5593.78\\
\includegraphics[scale=0.21]{{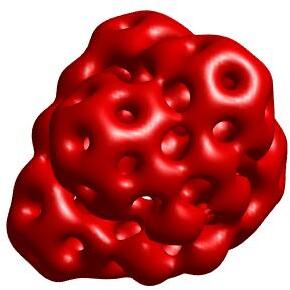}}&
\includegraphics[scale=0.21]{{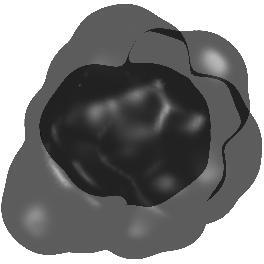}}&
\includegraphics[scale=0.21]{{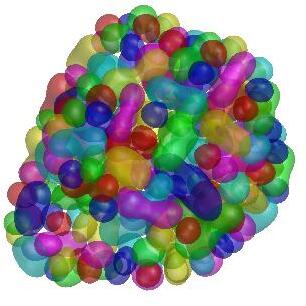}}\\
(an) 5593.81\\
\includegraphics[scale=0.21]{{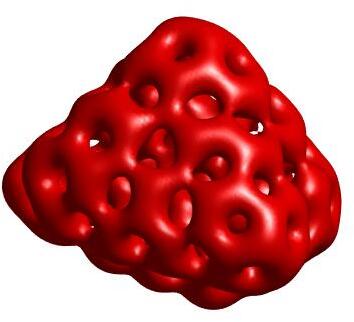}}&
\includegraphics[scale=0.21]{{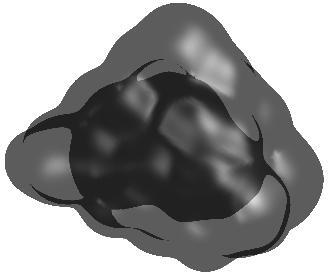}}&
\includegraphics[scale=0.21]{{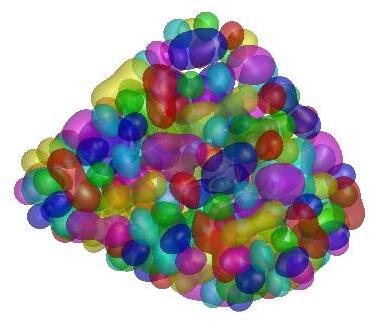}}\\
(ao) 5593.90\\
\includegraphics[scale=0.2]{{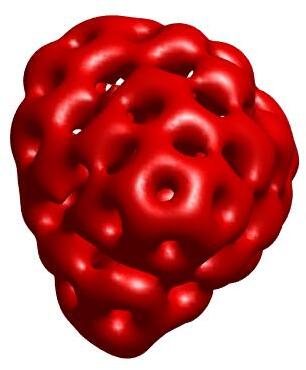}}&
\includegraphics[scale=0.2]{{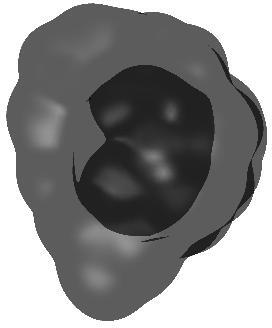}}&
\includegraphics[scale=0.2]{{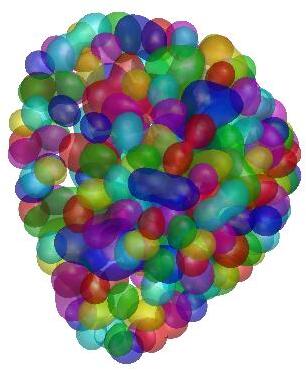}}\\
(ap) 5593.91\\
\end{tabular}
\caption{$B=40$ solutions with $\kappa=0.737$ ordered by increasing
  static energy, continued from fig.~\ref{fig:B40rest5}.
}
\label{fig:B40rest6}
\end{figure}

\begin{figure}[!htp]
\centering
\begin{tabular}{lcc}
\includegraphics[scale=0.21]{{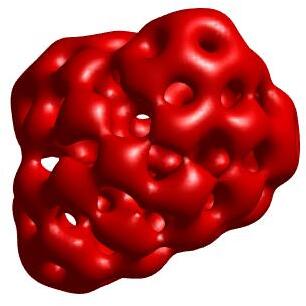}}&
\includegraphics[scale=0.21]{{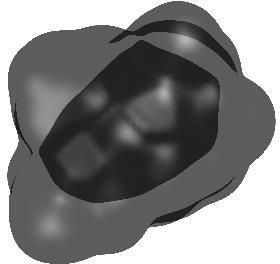}}&
\includegraphics[scale=0.21]{{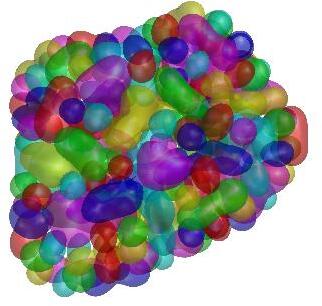}}\\
(aq) 5594.12\\
\includegraphics[scale=0.21]{{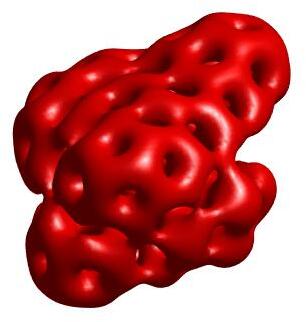}}&
\includegraphics[scale=0.21]{{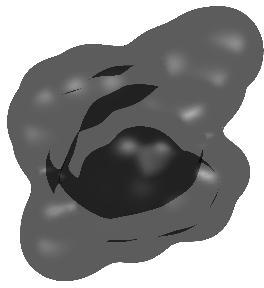}}&
\includegraphics[scale=0.21]{{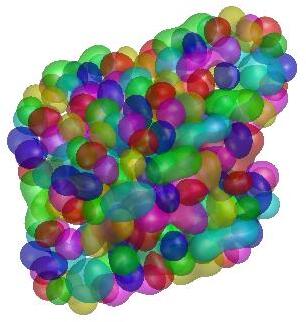}}\\
(ar) 5594.14\\
\includegraphics[scale=0.2]{{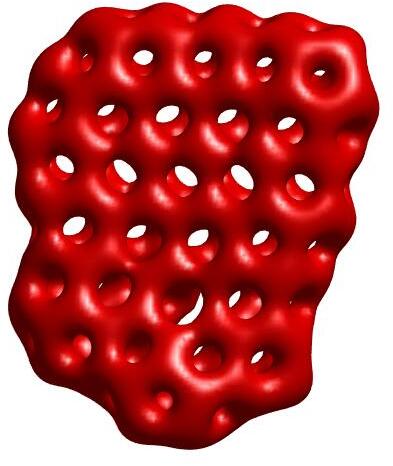}}&
\includegraphics[scale=0.2]{{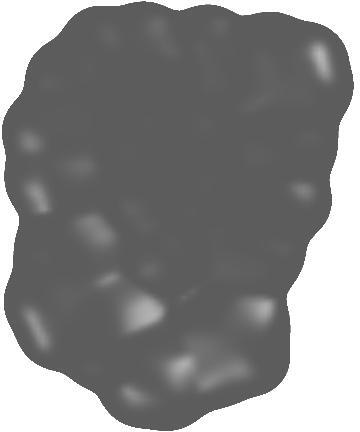}}&
\includegraphics[scale=0.2]{{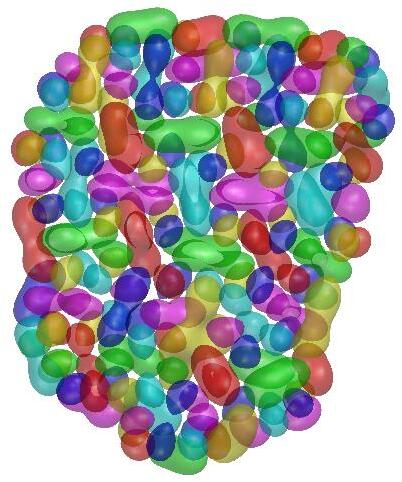}}\\
(as) 5594.28\\
\includegraphics[scale=0.23]{{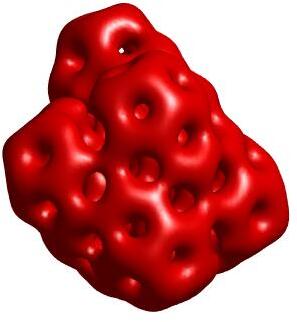}}&
\includegraphics[scale=0.23]{{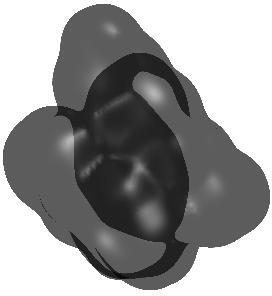}}&
\includegraphics[scale=0.23]{{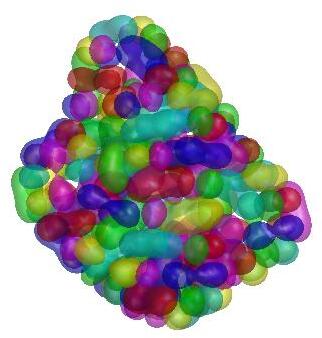}}\\
(au) 5594.58\\
\includegraphics[scale=0.2]{{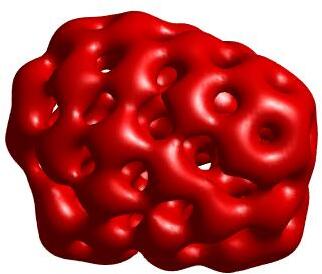}}&
\includegraphics[scale=0.2]{{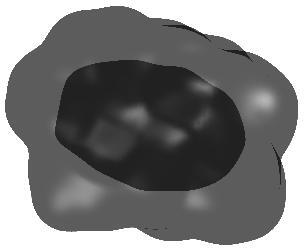}}&
\includegraphics[scale=0.2]{{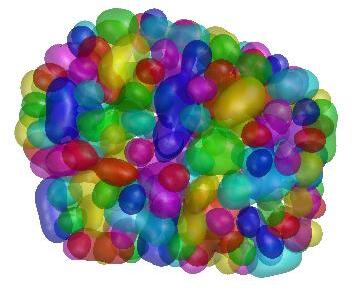}}\\
(av) 5594.96\\
\end{tabular}
\caption{$B=40$ solutions with $\kappa=0.737$ ordered by increasing
  static energy, continued from fig.~\ref{fig:B40rest6} and excluding
  the second row of fig.~\ref{fig:conf_kappaonly_noflowdown}.
}
\label{fig:B40rest7}
\end{figure}

\begin{figure}[!htp]
\centering
\begin{tabular}{lcc}
\includegraphics[scale=0.21]{{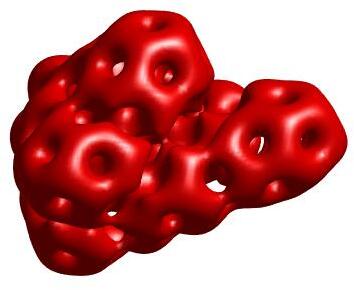}}&
\includegraphics[scale=0.21]{{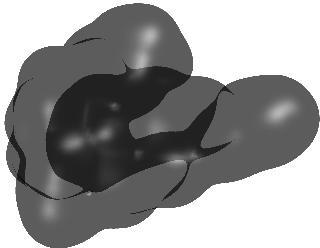}}&
\includegraphics[scale=0.21]{{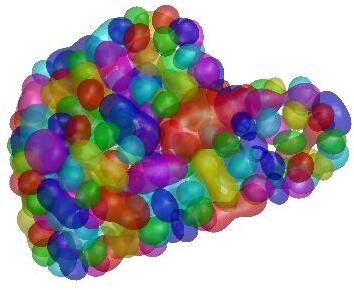}}\\
(aw) 5595.15\\
\includegraphics[scale=0.19]{{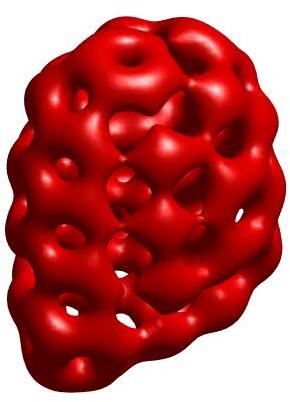}}&
\includegraphics[scale=0.19]{{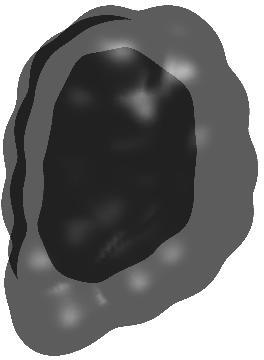}}&
\includegraphics[scale=0.19]{{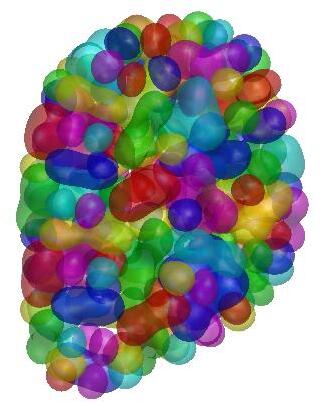}}\\
(ax) 5595.32\\
\includegraphics[scale=0.21]{{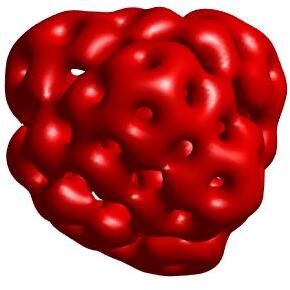}}&
\includegraphics[scale=0.21]{{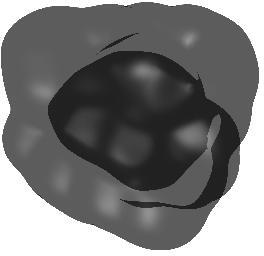}}&
\includegraphics[scale=0.21]{{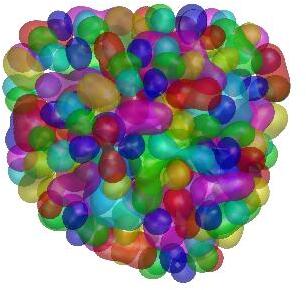}}\\
(ay) 5595.55\\
\includegraphics[scale=0.21]{{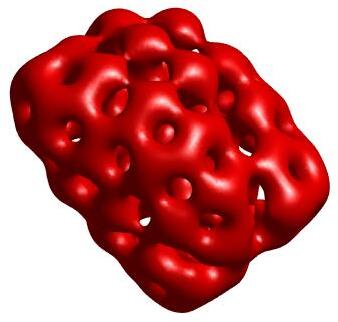}}&
\includegraphics[scale=0.21]{{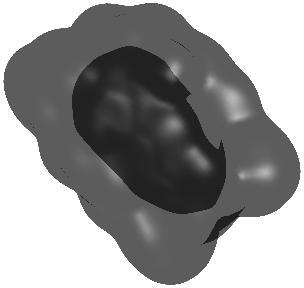}}&
\includegraphics[scale=0.21]{{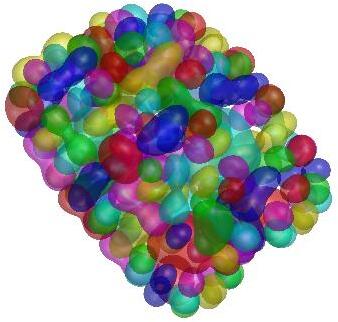}}\\
(az) 5595.61\\
\includegraphics[scale=0.18]{{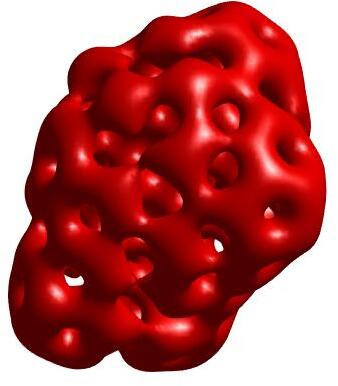}}&
\includegraphics[scale=0.18]{{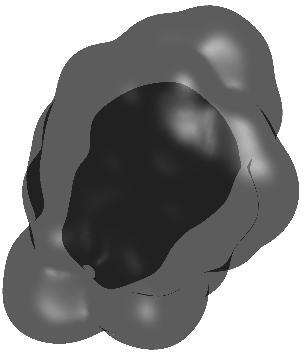}}&
\includegraphics[scale=0.18]{{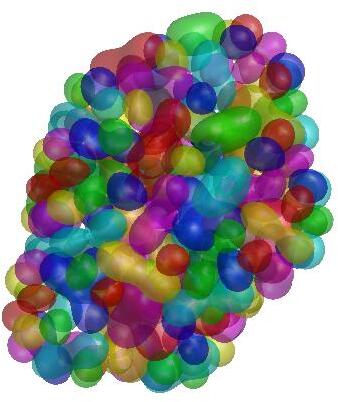}}\\
(ba) 5595.75\\
\end{tabular}
\caption{$B=40$ solutions with $\kappa=0.737$ ordered by increasing
  static energy, continued from fig.~\ref{fig:B40rest7}.
}
\label{fig:B40rest8}
\end{figure}

\begin{figure}[!htp]
\centering
\begin{tabular}{lcc}
\includegraphics[scale=0.19]{{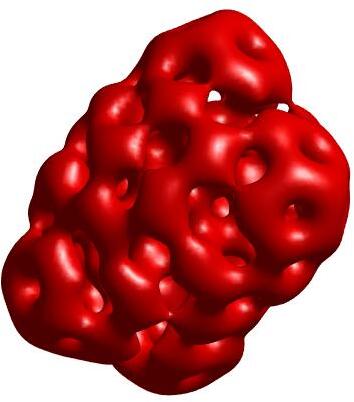}}&
\includegraphics[scale=0.19]{{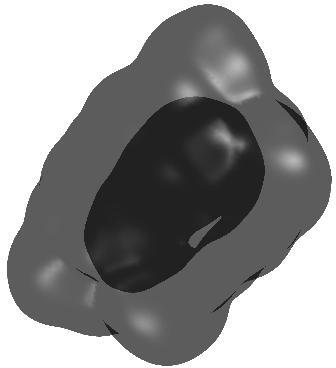}}&
\includegraphics[scale=0.19]{{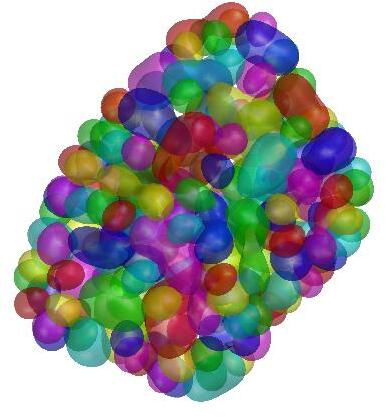}}\\
(bb) 5595.88\\
\includegraphics[scale=0.23]{{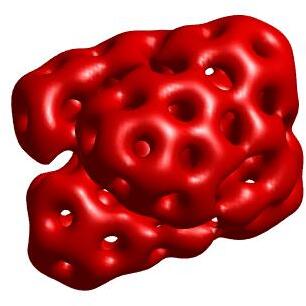}}&
\includegraphics[scale=0.23]{{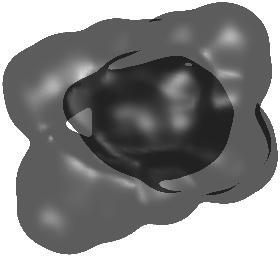}}&
\includegraphics[scale=0.23]{{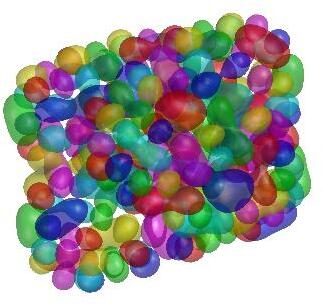}}\\
(bc) 5596.23\\
\includegraphics[scale=0.19]{{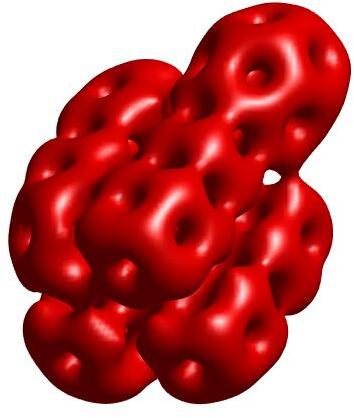}}&
\includegraphics[scale=0.19]{{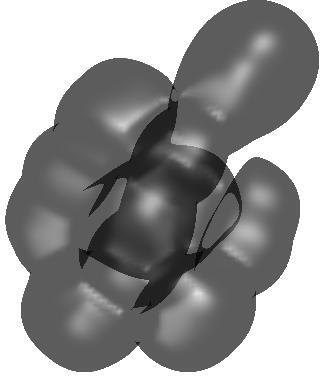}}&
\includegraphics[scale=0.19]{{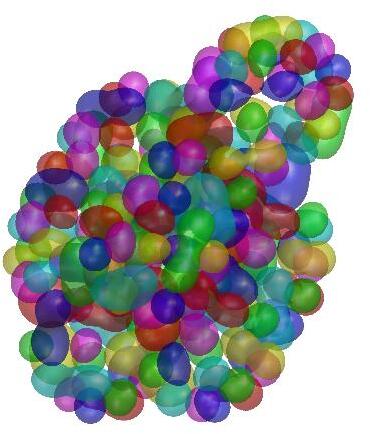}}\\
(be) 5596.32\\
\includegraphics[scale=0.2]{{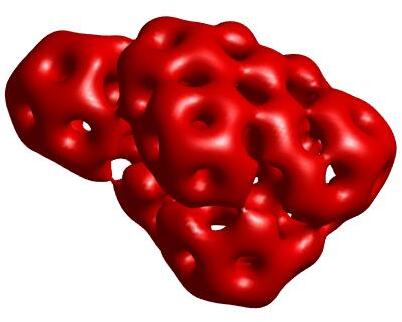}}&
\includegraphics[scale=0.2]{{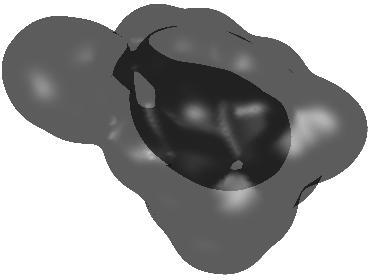}}&
\includegraphics[scale=0.2]{{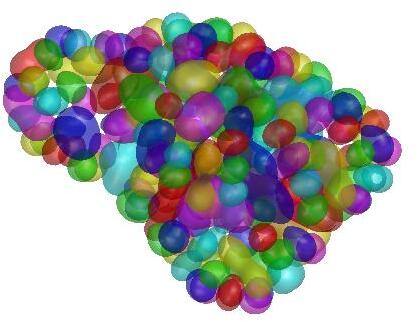}}\\
(bf) 5596.52\\
\includegraphics[scale=0.21]{{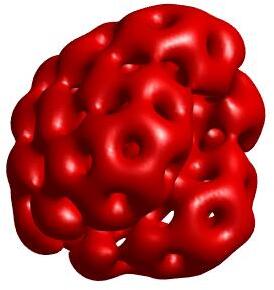}}&
\includegraphics[scale=0.21]{{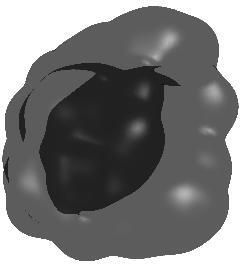}}&
\includegraphics[scale=0.21]{{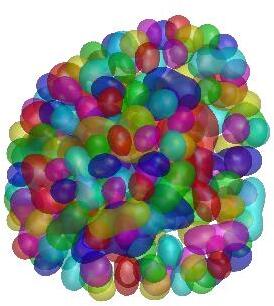}}\\
(bg) 5596.62\\
\end{tabular}
\caption{$B=40$ solutions with $\kappa=0.737$ ordered by increasing
  static energy, continued from fig.~\ref{fig:B40rest8} and excluding
  those of figs.~\ref{fig:conf20+27}(left) and \ref{fig:conf13}.
}
\label{fig:B40rest9}
\end{figure}

\begin{figure}[!htp]
\centering
\begin{tabular}{lcc}
\includegraphics[scale=0.2]{{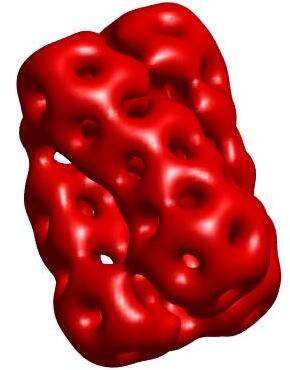}}&
\includegraphics[scale=0.2]{{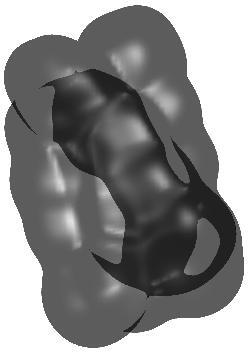}}&
\includegraphics[scale=0.2]{{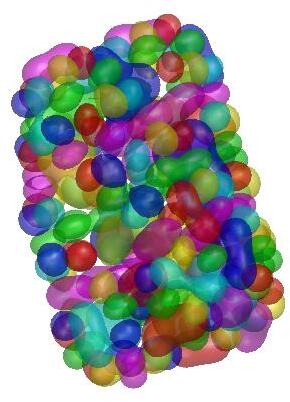}}\\
(bi) 5597.16\\
\includegraphics[scale=0.2]{{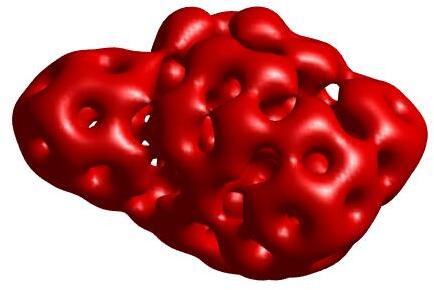}}&
\includegraphics[scale=0.2]{{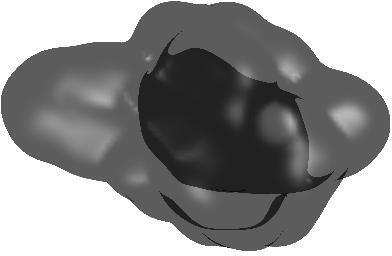}}&
\includegraphics[scale=0.2]{{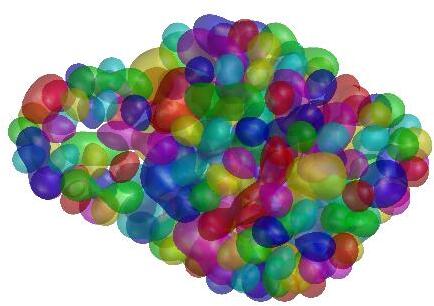}}\\
(bj) 5597.24\\
\includegraphics[scale=0.23]{{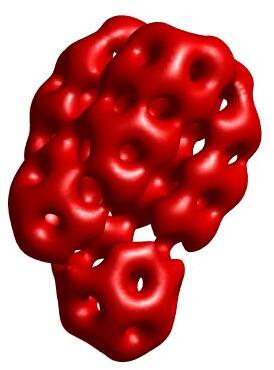}}&
\includegraphics[scale=0.23]{{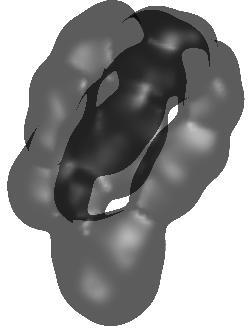}}&
\includegraphics[scale=0.23]{{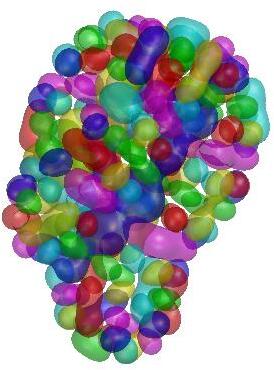}}\\
(bk) 5597.26\\
\includegraphics[scale=0.2]{{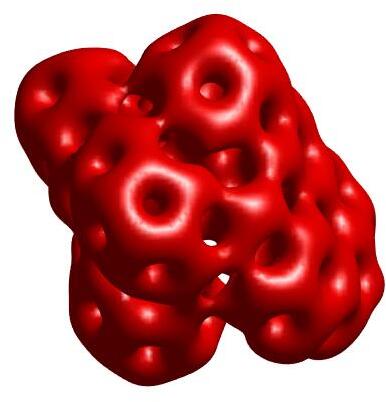}}&
\includegraphics[scale=0.2]{{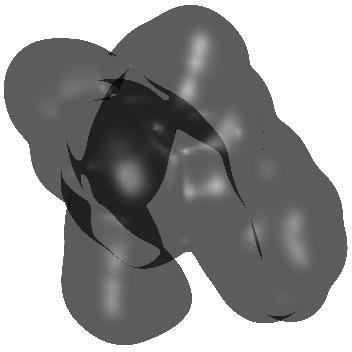}}&
\includegraphics[scale=0.2]{{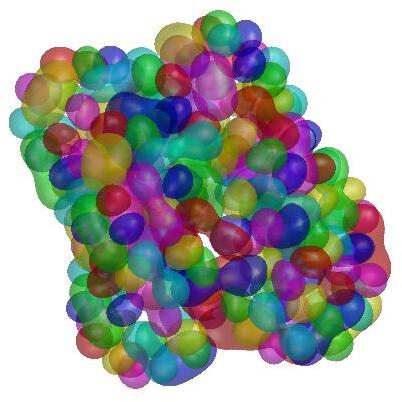}}\\
(bl) 5597.57\\
\includegraphics[scale=0.2]{{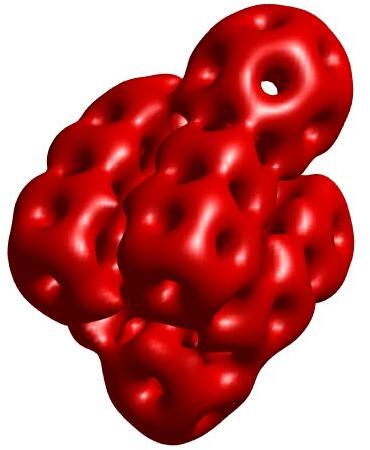}}&
\includegraphics[scale=0.2]{{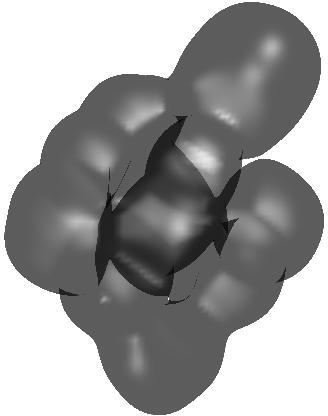}}&
\includegraphics[scale=0.2]{{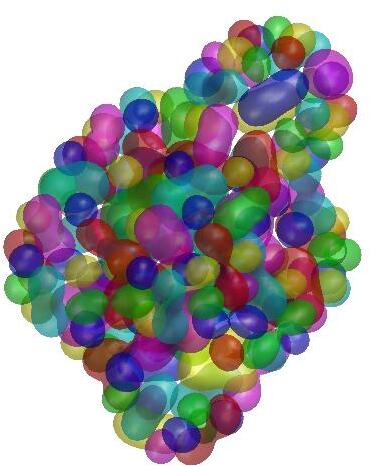}}\\
(bm) 5597.60\\
\end{tabular}
\caption{$B=40$ solutions with $\kappa=0.737$ ordered by increasing
  static energy, continued from fig.~\ref{fig:B40rest9}.
}
\label{fig:B40rest10}
\end{figure}

\begin{figure}[!htp]
\centering
\begin{tabular}{lcc}
\includegraphics[scale=0.22]{{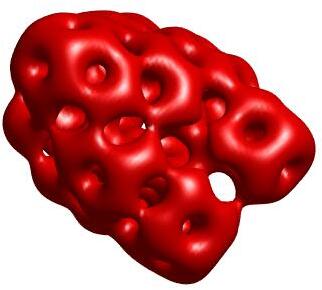}}&
\includegraphics[scale=0.22]{{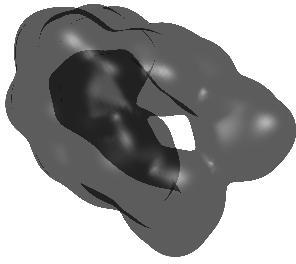}}&
\includegraphics[scale=0.22]{{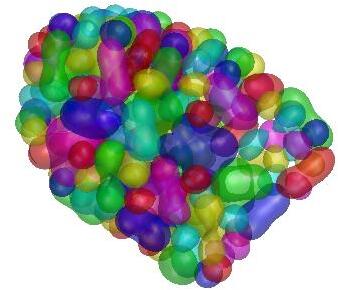}}\\
(bn) 5598.41\\
\includegraphics[scale=0.19]{{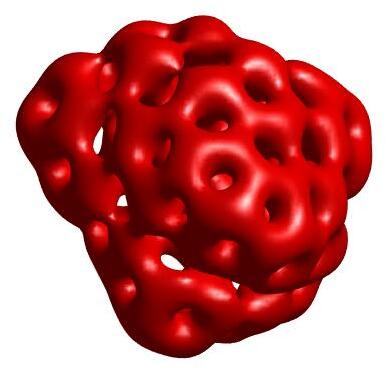}}&
\includegraphics[scale=0.19]{{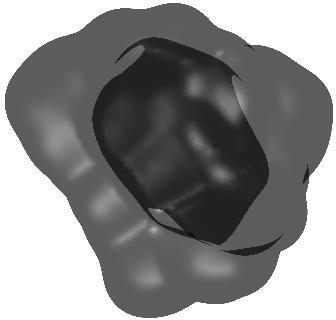}}&
\includegraphics[scale=0.19]{{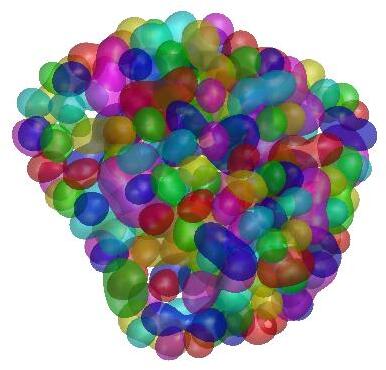}}\\
(bo) 5598.68\\
\includegraphics[scale=0.23]{{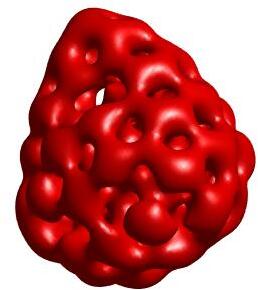}}&
\includegraphics[scale=0.23]{{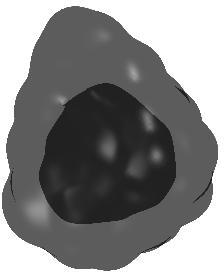}}&
\includegraphics[scale=0.23]{{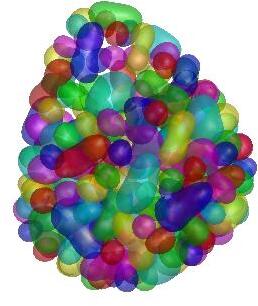}}\\
(bp) 5599.11\\
\includegraphics[scale=0.2]{{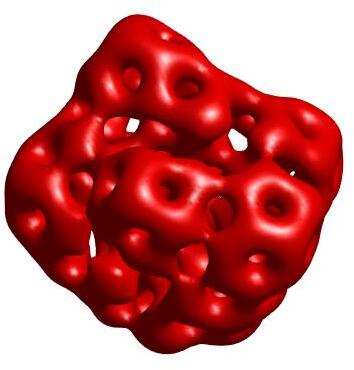}}&
\includegraphics[scale=0.2]{{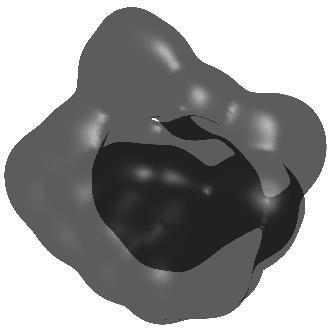}}&
\includegraphics[scale=0.2]{{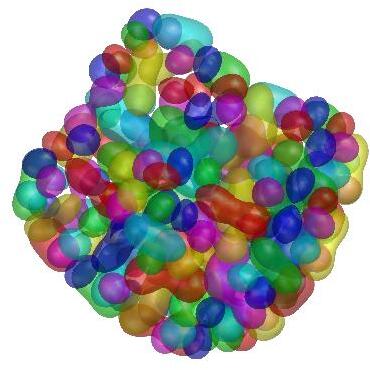}}\\
(bq) 5600.52\\
\includegraphics[scale=0.2]{{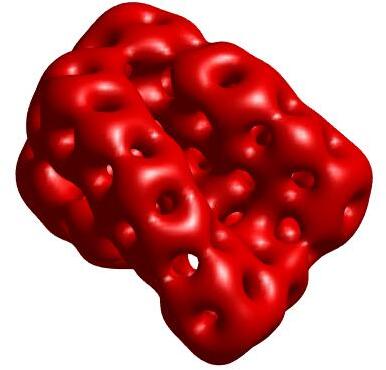}}&
\includegraphics[scale=0.2]{{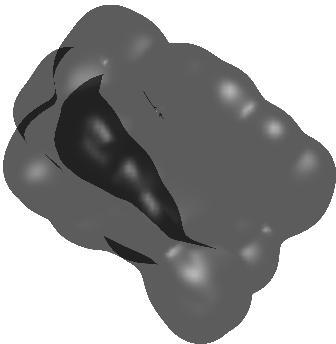}}&
\includegraphics[scale=0.2]{{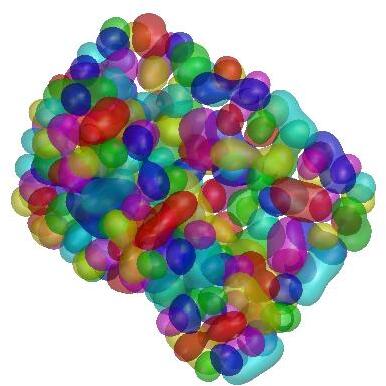}}\\
(br) 5600.55\\
\end{tabular}
\caption{$B=40$ solutions with $\kappa=0.737$ ordered by increasing
  static energy, continued from fig.~\ref{fig:B40rest10}.
}
\label{fig:B40rest11}
\end{figure}

\begin{figure}[!htp]
\centering
\begin{tabular}{lcc}
\includegraphics[scale=0.22]{{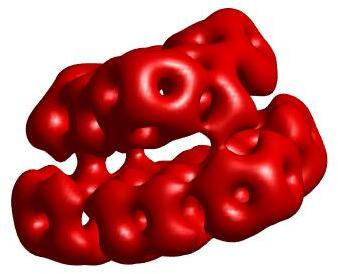}}&
\includegraphics[scale=0.22]{{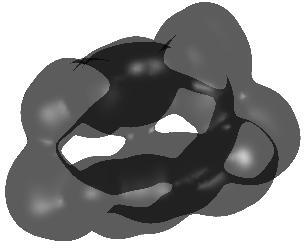}}&
\includegraphics[scale=0.22]{{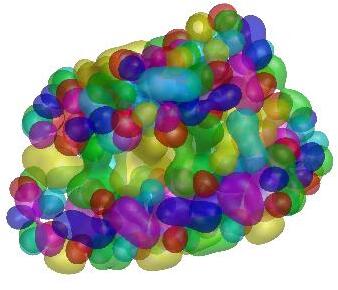}}\\
(bs) 5600.93\\
\includegraphics[scale=0.2]{{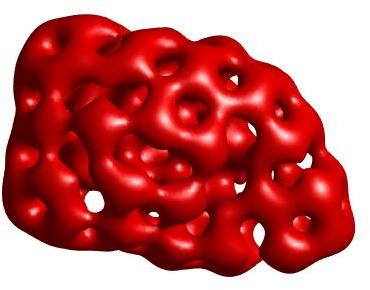}}&
\includegraphics[scale=0.2]{{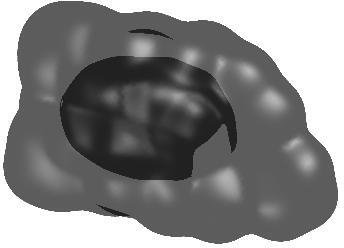}}&
\includegraphics[scale=0.2]{{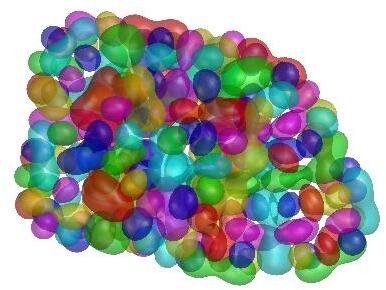}}\\
(bt) 5601.00\\
\includegraphics[scale=0.19]{{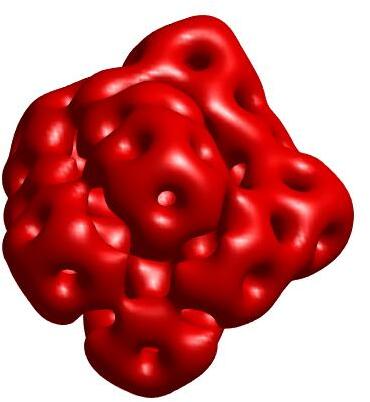}}&
\includegraphics[scale=0.19]{{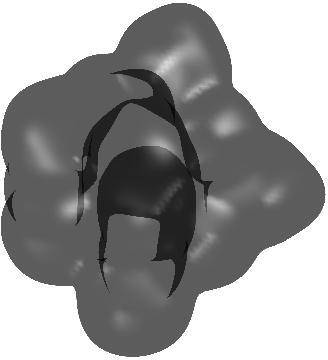}}&
\includegraphics[scale=0.19]{{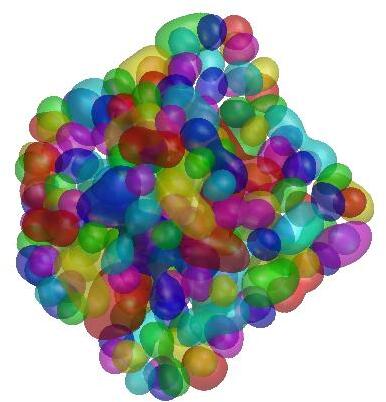}}\\
(bu) 5601.07\\
\includegraphics[scale=0.2]{{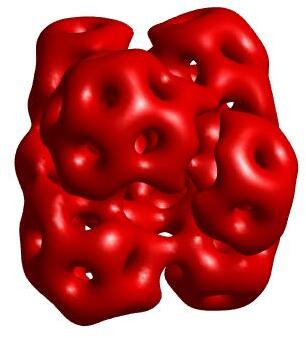}}&
\includegraphics[scale=0.2]{{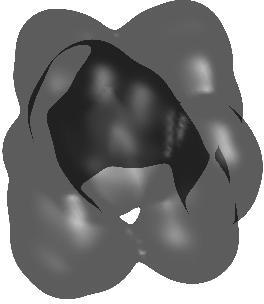}}&
\includegraphics[scale=0.2]{{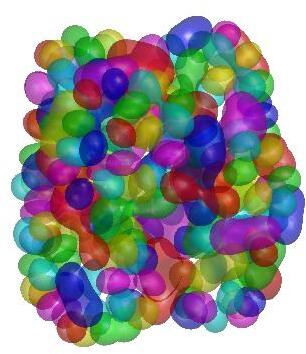}}\\
(bx) 5601.50\\
\includegraphics[scale=0.22]{{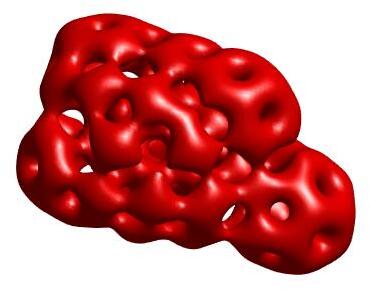}}&
\includegraphics[scale=0.22]{{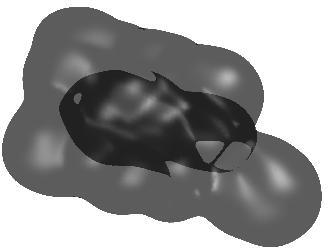}}&
\includegraphics[scale=0.22]{{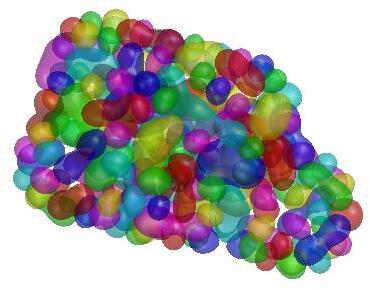}}\\
(by) 5601.75\\
\end{tabular}
\caption{$B=40$ solutions with $\kappa=0.737$ ordered by increasing
  static energy, continued from fig.~\ref{fig:B40rest11} and excluding
  that of fig.~\ref{fig:conf4}.
}
\label{fig:B40rest12}
\end{figure}

\begin{figure}[!htp]
\centering
\begin{tabular}{lcc}
\includegraphics[scale=0.21]{{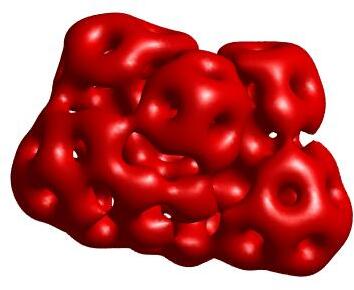}}&
\includegraphics[scale=0.21]{{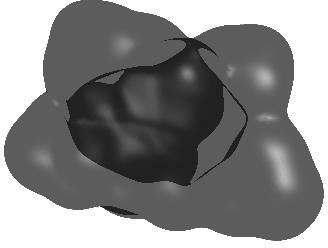}}&
\includegraphics[scale=0.21]{{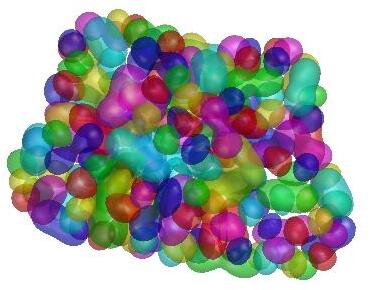}}\\
(bz) 5602.28\\
\includegraphics[scale=0.21]{{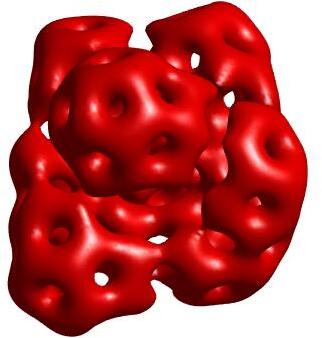}}&
\includegraphics[scale=0.21]{{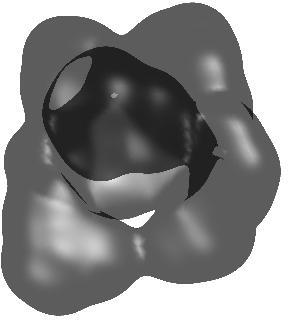}}&
\includegraphics[scale=0.21]{{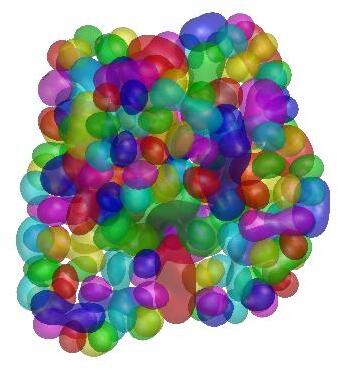}}\\
(cb) 5602.52\\
\includegraphics[scale=0.23]{{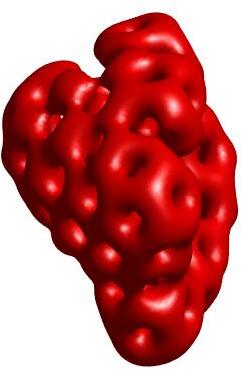}}&
\includegraphics[scale=0.23]{{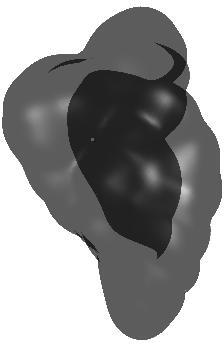}}&
\includegraphics[scale=0.23]{{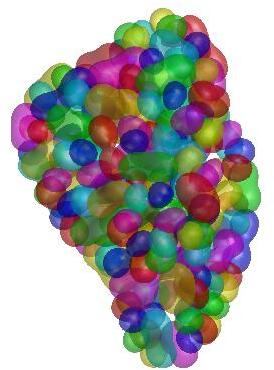}}\\
(cc) 5602.83\\
\includegraphics[scale=0.23]{{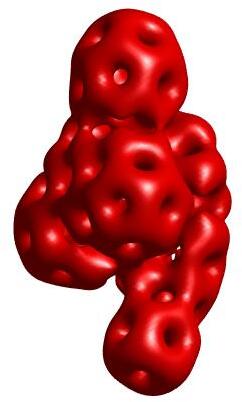}}&
\includegraphics[scale=0.23]{{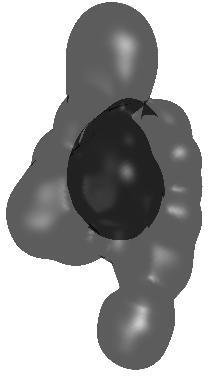}}&
\includegraphics[scale=0.23]{{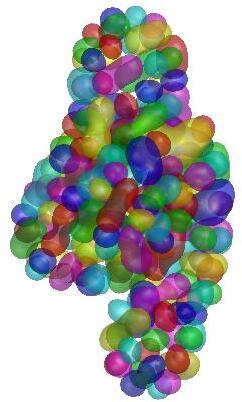}}\\
(cd) 5603.10\\
\includegraphics[scale=0.2]{{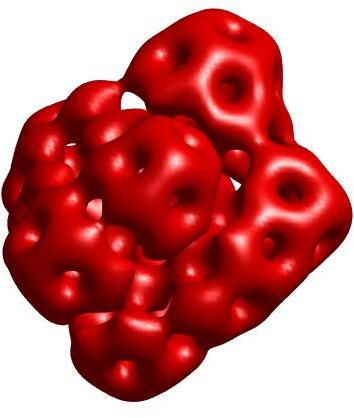}}&
\includegraphics[scale=0.2]{{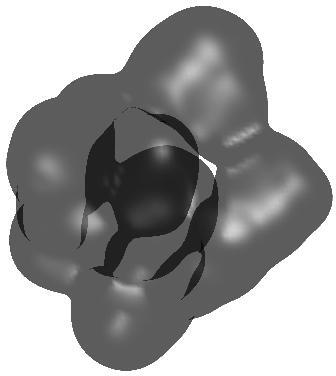}}&
\includegraphics[scale=0.2]{{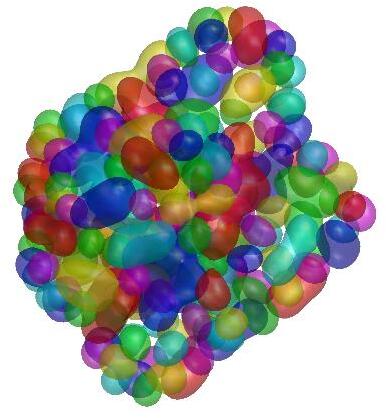}}\\
(ce) 5603.65\\
\end{tabular}
\caption{$B=40$ solutions with $\kappa=0.737$ ordered by increasing
  static energy, continued from fig.~\ref{fig:B40rest12} and excluding
  that of fig.~\ref{fig:conf20+27}(right). 
}
\label{fig:B40rest13}
\end{figure}

\begin{figure}[!htp]
\centering
\begin{tabular}{lcc}
\includegraphics[scale=0.18]{{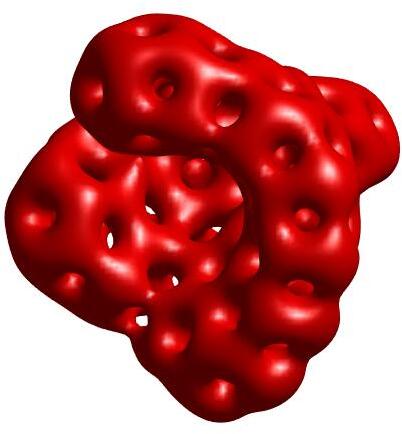}}&
\includegraphics[scale=0.18]{{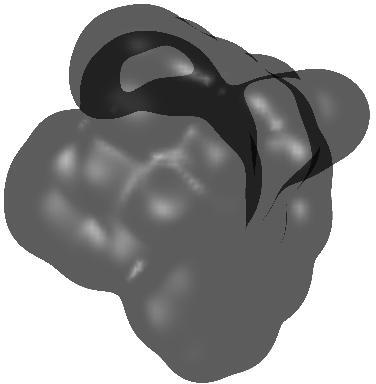}}&
\includegraphics[scale=0.18]{{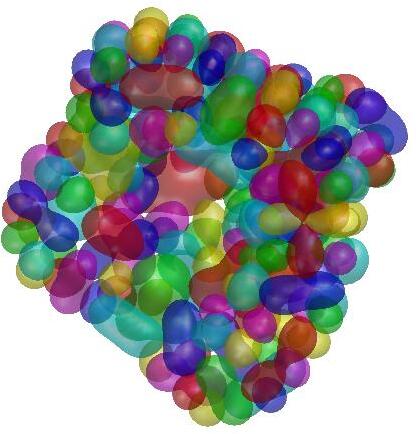}}\\
(cf) 5604.15\\
\includegraphics[scale=0.2]{{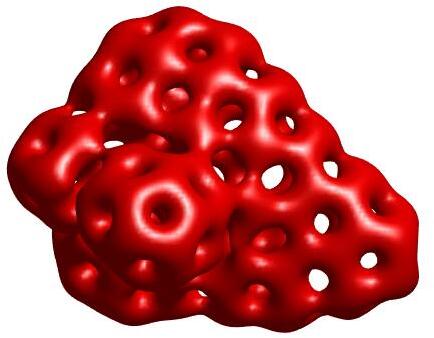}}&
\includegraphics[scale=0.2]{{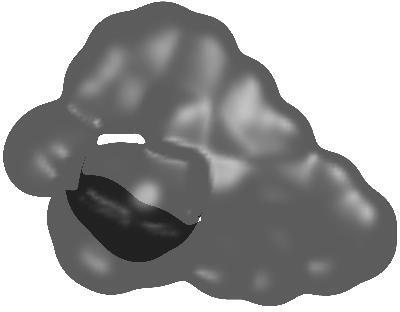}}&
\includegraphics[scale=0.2]{{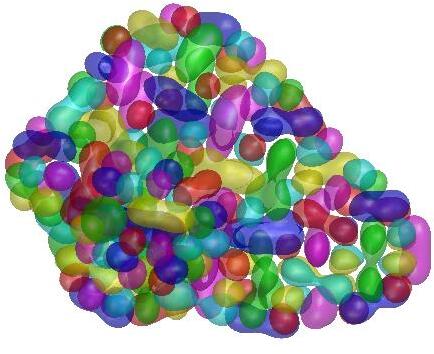}}\\
(cg) 5604.27\\
\includegraphics[scale=0.2]{{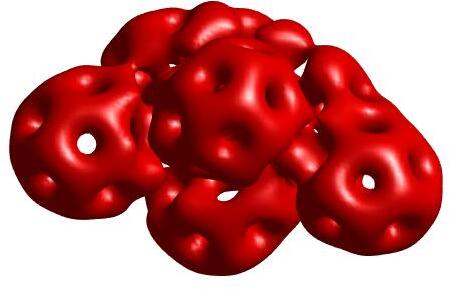}}&
\includegraphics[scale=0.2]{{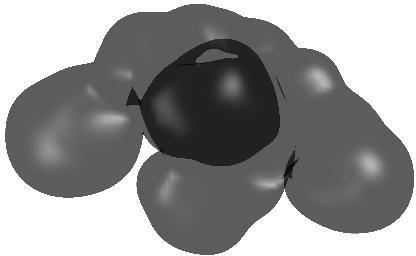}}&
\includegraphics[scale=0.2]{{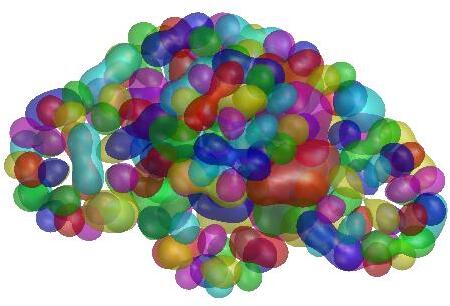}}\\
(ch) 5604.30\\
\includegraphics[scale=0.17]{{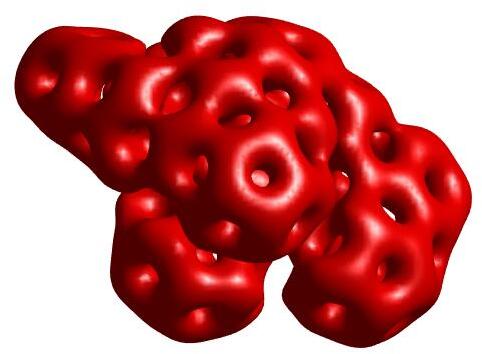}}&
\includegraphics[scale=0.17]{{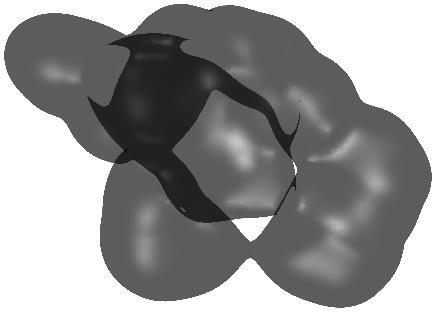}}&
\includegraphics[scale=0.17]{{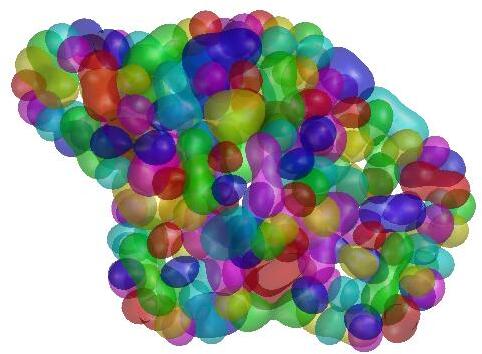}}\\
(ci) 5605.21\\
\includegraphics[scale=0.21]{{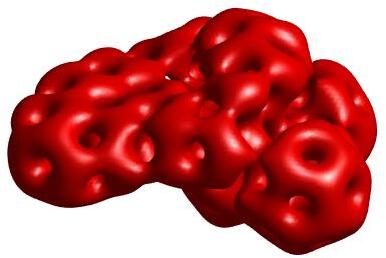}}&
\includegraphics[scale=0.21]{{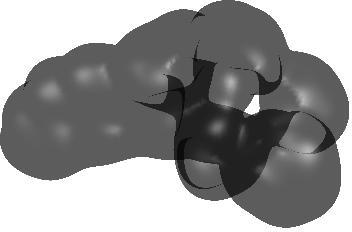}}&
\includegraphics[scale=0.21]{{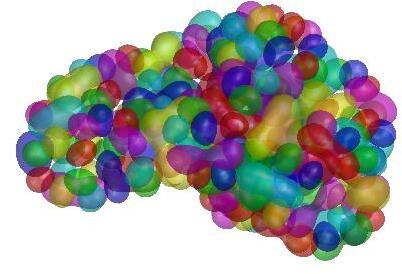}}\\
(cj) 5605.55\\
\end{tabular}
\caption{$B=40$ solutions with $\kappa=0.737$ ordered by increasing
  static energy, continued from fig.~\ref{fig:B40rest13} and excluding
  the second last row of fig.~\ref{fig:conf_kappaonly_noflowdown}. 
}
\label{fig:B40rest14}
\end{figure}

\begin{figure}[!htp]
\centering
\begin{tabular}{lcc}
\includegraphics[scale=0.21]{{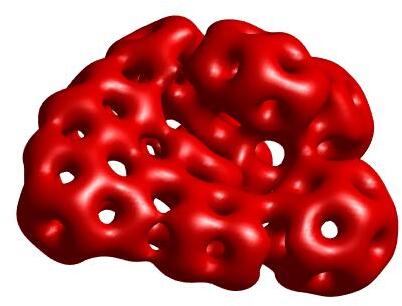}}&
\includegraphics[scale=0.21]{{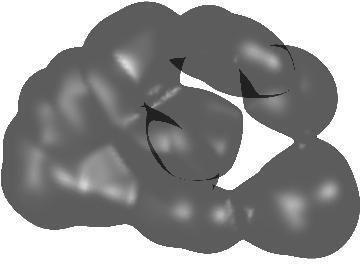}}&
\includegraphics[scale=0.21]{{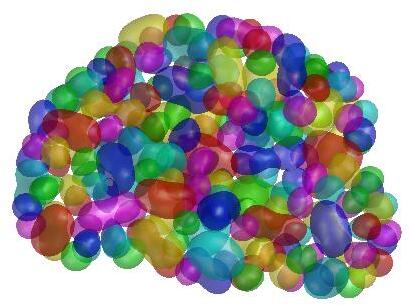}}\\
(cl) 5607.92\\
\includegraphics[scale=0.22]{{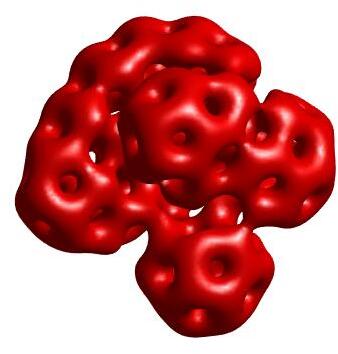}}&
\includegraphics[scale=0.22]{{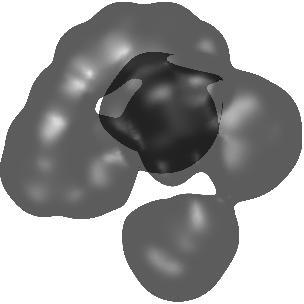}}&
\includegraphics[scale=0.22]{{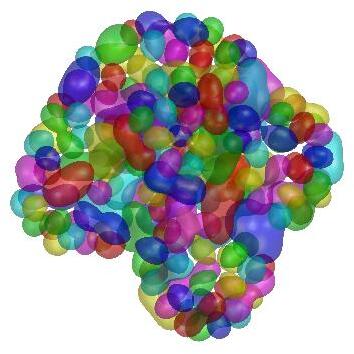}}\\
(cm) 5608.52\\
\includegraphics[scale=0.22]{{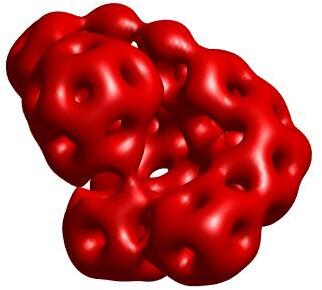}}&
\includegraphics[scale=0.22]{{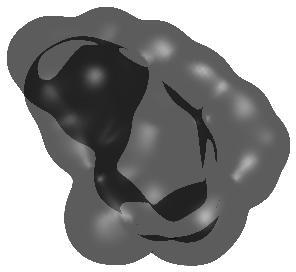}}&
\includegraphics[scale=0.22]{{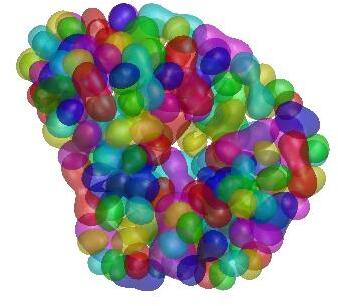}}\\
(cn) 5608.90\\
\includegraphics[scale=0.22]{{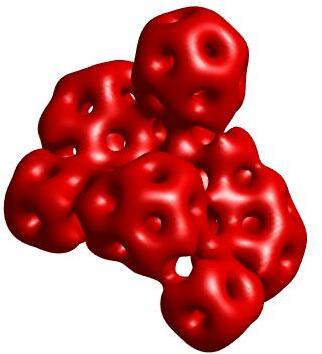}}&
\includegraphics[scale=0.22]{{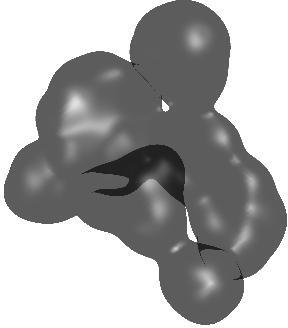}}&
\includegraphics[scale=0.22]{{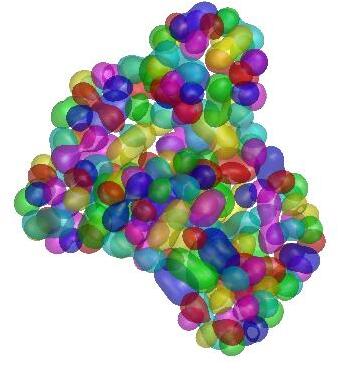}}\\
(cp) 5610.30\\
\includegraphics[scale=0.2]{{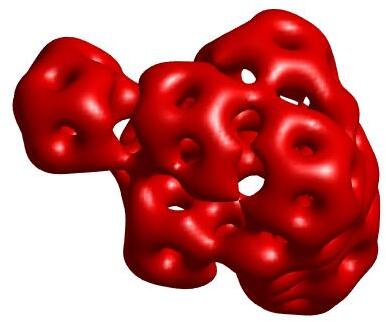}}&
\includegraphics[scale=0.2]{{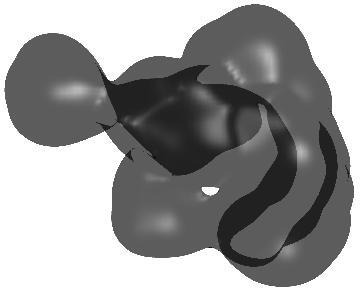}}&
\includegraphics[scale=0.2]{{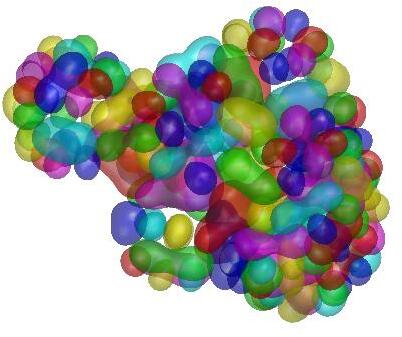}}\\
(cr) 5610.84\\
\end{tabular}
\caption{$B=40$ solutions with $\kappa=0.737$ ordered by increasing
  static energy, continued from fig.~\ref{fig:B40rest14} and excluding
  the last row of fig.~\ref{fig:conf_kappaonly_noflowdown} as well as
  that of fig.~\ref{fig:conf20+27}(right).
}
\label{fig:B40rest15}
\end{figure}

\begin{figure}[!htp]
\centering
\begin{tabular}{lcc}
\includegraphics[scale=0.21]{{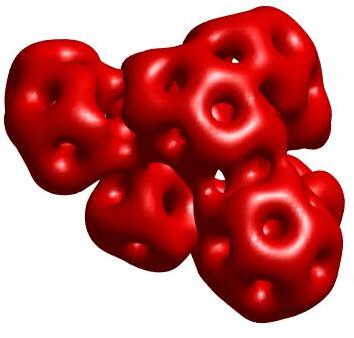}}&
\includegraphics[scale=0.21]{{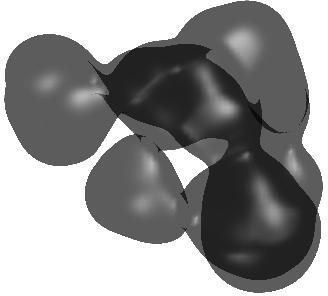}}&
\includegraphics[scale=0.21]{{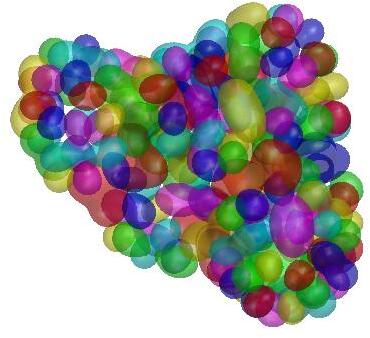}}\\
(cs) 5611.09\\
\includegraphics[scale=0.2]{{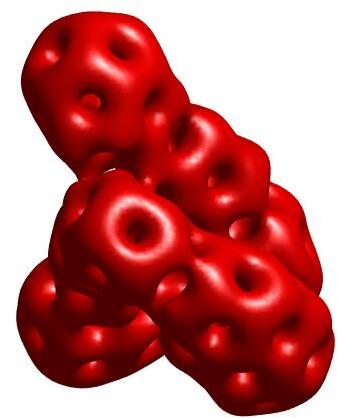}}&
\includegraphics[scale=0.2]{{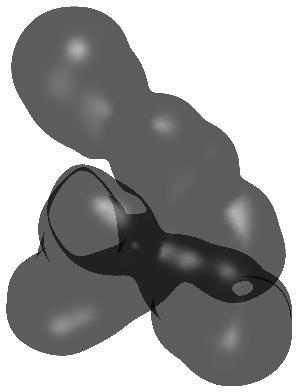}}&
\includegraphics[scale=0.2]{{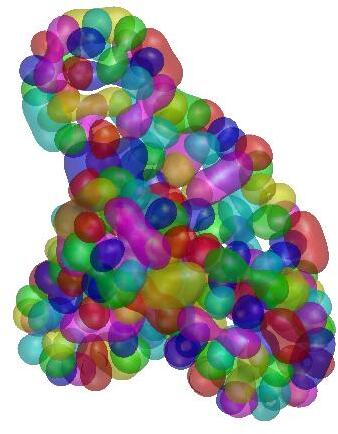}}\\
(ct) 5612.25\\
\includegraphics[scale=0.2]{{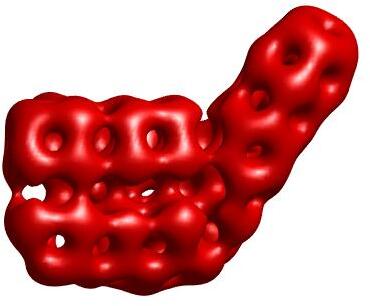}}&
\includegraphics[scale=0.2]{{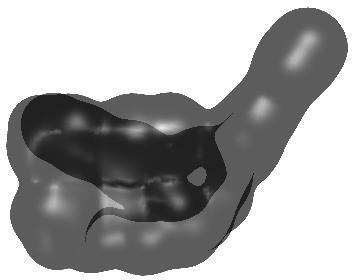}}&
\includegraphics[scale=0.2]{{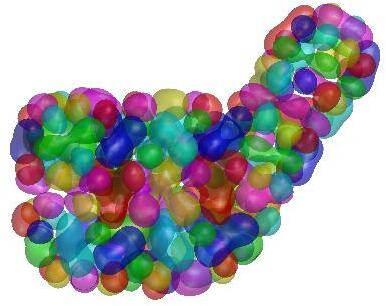}}\\
(cu) 5613.74\\
\includegraphics[scale=0.2]{{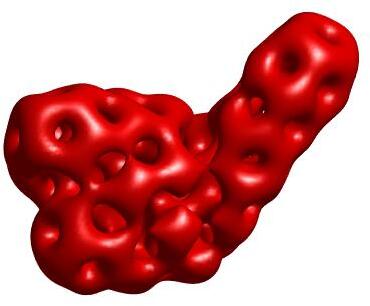}}&
\includegraphics[scale=0.2]{{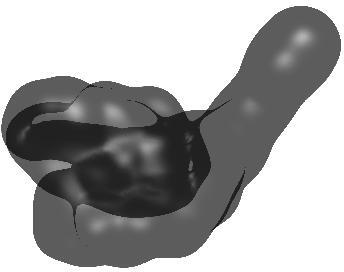}}&
\includegraphics[scale=0.2]{{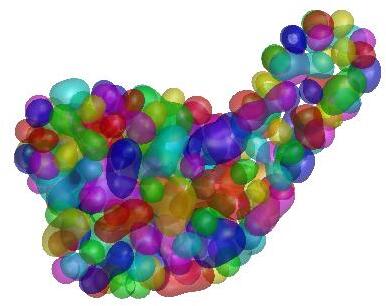}}\\
(cv) 5613.76\\
\end{tabular}
\caption{$B=40$ solutions with $\kappa=0.737$ ordered by increasing
  static energy, continued from fig.~\ref{fig:B40rest15}.
}
\label{fig:B40rest16}
\end{figure}

\end{document}